\documentclass[aps,prb,amsmath,amssymb,twocolumn,letterpaper,superscriptaddress,
         footinbib,bibnotes,showpacs,reprint,balancelastpage,raggedbottom,longbibliography,%
         citeautoscript,floatfix]{revtex4-1}
\usepackage[usenames,dvipsnames]{color}
\usepackage[pdftex,
  breaklinks,             %
  bookmarks = false, %
  pdfpagemode   = UseNone,%
  pdfstartview 	= FitH,		%
  pdfstartpage  = 1,      %
  colorlinks    = true,   %
]{hyperref}
\hypersetup{
    linkcolor=RubineRed,          
    citecolor=ForestGreen,        
    filecolor=Mulberry,      
    urlcolor=RoyalBlue           
}
\usepackage{graphicx}
\usepackage{microtype}
\usepackage{braket}
\usepackage{xspace}
\usepackage{lipsum}
\usepackage{chngcntr}
\usepackage{xfrac}

\def\nsf{Na$_3$ScF$_6$\xspace}



\def\nmf{Na$_3$MnF$_6$}
\def\nsf{Na$_3$ScF$_6$}

\def\mf{MnF$_2$}
\def\vf{VF$_2$}
\def\kmf{KMnF$_3$}

\begin{document}

\title{Assessing exchange-correlation functional performance for structure and property predictions of oxyfluoride compounds from first principles}

\author{Nenian Charles}
	\affiliation{Department\,of\,Materials\,Science\,and\,Engineering,\,Drexel\,University,\,Philadelphia,\,PA\,19104,\,USA} %
	\affiliation{Department\,of\,Materials\,Science\,and\,Engineering,\,Northwestern\,University,\,Evanston,\,IL\,60208,\,USA} %

%
\author{James M.\ Rondinelli}
  \email{jrondinelli@northwestern.edu}
	\affiliation{Department\,of\,Materials\,Science\,and\,Engineering,\,Northwestern\,University,\,Evanston,\,IL\,60208,\,USA} %
\date{\today}

\begin{abstract}
Motivated by the resurgence of electronic and optical property design in ordered fluoride and oxyfluoride compounds, we present a density functional theory (DFT) study on 19 materials with structures, ranging from simple to complex, and variable  oxygen-to-fluorine ratios. We focus on understanding the accuracy of the exchange-correlation potentials ($V_{xc}$) to DFT for the prediction of structural, electronic, and lattice dynamical properties at four different levels of theory, \emph{i.e.}, the local density approximation (LDA), generalized gradient approximation (GGA), metaGGA, and hybrid functional level which includes fractions of exact exchange. We investigate in detail the metaGGA functionals MS2 [Sun \emph{et al}., Phys.\ Rev.\ Lett., \textbf{111}, 106401 (2013)] and SCAN [Sun \emph{et al}., Phys.\ Rev.\ Lett., \textbf{115}, 036402 (2015)], and show that although the metaGGAs show improvements over the LDA and GGA functionals in describing the electronic structure and phonon frequencies, the static structural properties of fluorides and oxyfluorides are often more accurately predicted by the GGA-level functional PBEsol. Results from LDA calculations are unsatisfactory for any compound regardless of oxygen concentration. PBEsol or HSE06 gives good performance in all oxide or all fluoride compounds. For the oxyfluorides, PBEsol  is consistently more accurate for structural properties across all oxygen concentrations, however, we stress the need for detailed property assessment with various functionals for oxyfluorides with variable composition. The ``best'' functional is anticipated to be more strongly dependent on the property of interest. Our study provides useful insights in selecting an $V_{xc}$ that achieves the best performance comprise, enabling more accurate predictive design of functional fluoride-based materials with density functional theory.
\end{abstract}

\pacs{71.15.Mb, 71.20.Ps, 77.80.--e}

\maketitle

\section{Introduction}

Materials researchers are increasingly utilizing 
first-principles simulations to elucidate underlying physical principles, guide compound design, 
and accelerate discovery.
In particular, density functional theory (DFT) calculations have proved the method of choice 
for many researches owing to its speed and scalability compared to traditional many-body approaches. \cite{PhysRevLett.91.146401, Buhl:DFT2006}
DFT benefits from treating a fictitious, Kohn-Sham, system of noninteracting electrons to obtain the interacting electron density.
Although the formalism is exact, there remains no complete and tractable description 
of the exact exchange-correlation (XC) potential ($V_{xc}$) and thus it must be approximated in any practical calculation. \cite{sun2016accurate}
As a consequence, the accuracy of these first-principles simulations depends intimately on the 
suitability of the XC approximation for handling the chemical system and physical property of interest.
Owing to the diversity of bonding interactions present in structurally complex and chemically diverse compounds, various $V_{xc}$ approximations are routinely benchmarked against 
experimental values of representative compounds in different chemical families to assess and calibrate accuracy.\cite{BurkeWhichXC}

In this light, the burgeoning field of anion engineering in crystalline materials, \cite{anion-control} which aims to tune physical 
properties through the incorporation of multiple-anion species of different size, electronegativity, and charge
into the same anionic groups comprising the solid, may present considerable challenges to the predictive capabilities of DFT.
Multianion engineering expands the current set of design strategies for the discovery of functional materials.
Oxynitride compounds display interesting dielectric, magnetoelectic and photocatalytic properties. \cite{OpticalOxyni:2010, Yang/Attfield:2011, Fuertes:2008}
In addition, superconductivity, and ferromagnetic behavior have been reported in novel multianion chalcogenides compounds. \cite{Johnston:2010}
%

Another interesting class of multianion compounds is the transition metal oxyfluorides. \cite{tagkey2016iii}
Oxygen and fluorine substitution presents the opportunity to create materials that possess the 
advantageous properties of both oxide and fluoride compounds. 
Transition metal oxyfluorides have already show improved performance
over their oxide counterparts in electrochemical and solid state lighting applications. \cite{Ceder_FeOF2, Jared:MolliOxyfluo, phosphor:Anant, Ram:Phosphor2010}
Moreover, other novel oxyfluoride compounds display a myriad of technologically useful properties 
ranging from nonlinear optical behavior \cite{Atuchin2012159, Izumi/Poeppelmeier:2005} to superconductivity. \cite{delaCruz/Pengcheng_et_al:2008, Hosono/FeSC:2008}  

From a chemical perspective, the oxide and fluoride ions appear to be quite similar.
For instance, the closeness of their atomic number and ionic radii, \emph{i.e.},  1.35\,\AA\ and 1.29\,\AA\ for O$^{2-}$ and F$^-$, respectively,\cite{Shannon/Prewitt:1969}
allow them to occupy the same sites in a crystal, making anion substitution easy.
However, their differences in charge and electronegativity can lead to distinct physical properties in crystalline solids.
The union of these contrasting traits in oxyfluoride anionic groups, \emph{i.e.}, [$M$O$_x$F$_{6-x}$]$^{n-}$, where $M$ is a (transition) metal center, results in complex bonding interactions compared to 
compounds where only oxygen or fluorine are present.
Specifically, the $2p$ orbitals of the more electronegative fluoride anions are located deeper in the valence band compared to the same $2p$ states of the oxide ions.
When both anions belong to the same coordination sphere, the $M$--O bonds tend to display 
stronger covalent interactions with shorter bond lengths compared to the longer $M$--F bonds. \cite{withers2007cluster}
The result is oxyfluoride anionic groups with varying degrees of covalency within the same coordination sphere.
The efficacy of different DFT $V_{xc}$ approximations in describing oxyfluorides remains to be systematically assessed.

In this work, we evaluate the performance of exchange-correlation functionals at four rungs of Jacob's ladder, \emph{i.e.}, the local density approximation (LDA), generalized gradient approximation (GGA,) as implemented by the PBEsol functional, metaGGA (MS2 and SCAN), and hybrid functional (HSE06) level which includes fractions of exact exchange.
We focus on the accuracy of each $V_{xc}$ potential in predicting the structural, 
electronic and lattice dynamical properties in oxyfluorides.
Specifically, we investigate how the approximations perform as the relative covalency 
increases by increasing oxygen content across 19 known fluoride, oxyfluoride, and oxide materials.
Here we categorize each compound by the composition, $x$, of the local ligand chemistry about the octahedrally coordinated metals forming the anionic groups [$M$O$_x$F$_{6-x}$]$^{n-}$ in the solid.
Owing to the so-called chemical intuition derived from the $\alpha$ parameter found in 
modern metaGGAs such as MS2 and SCAN, we expect them to be optimally suited to capture the subtleties of the oxyfluoride chemistry in the solid state. 
Our main finding is that PBEsol is the most suitable functional for calculating the structural properties regardless of oxygen composition.
In addition, we find that the inclusion of the kinetic energy in the metaGGAs MS2 and SCAN introduces functional dependencies, 
which deviate from trends previously reported for LDA, GGA and hybrid functionals. 


\section{Methods and Materials Suite}

\subsection{Computational Details}
We performed density functional theory calculations 
as implemented in the Vienna
{\it Ab initio} Simulation Package ({\sc vasp}) 
\cite{Kresse/Furthmuller:1996a,Kresse/Joubert:1999} 
with the projector augmented wave (PAW) method \cite{Blochl:1994} 
to treat the interactions between the core and valence electrons. 
The atomic reference configurations are $2s^22p^4$ for the O, and $2s^22p^5$ for F in all compounds studied. 
The PAW atomic configurations of the cations, appearing in each compound, are tabulated in Ref.~\onlinecite{Supplmental_Note:Benchmark}.
We also performed a series of convergence test using different Monkhorst-Pack $k$-point meshes \cite{Monkhorst/Pack:1976} and
planewave cutoffs to find the appropriate values for each chemistry. \cite{Supplmental_Note:Benchmark}
Full structural optimizations are performed until the Hellmann-Feynman forces are less than 5 meV \AA$^{-1}$.
The phonon modes are analyzed using the PHONOPY package. \cite{Phonopy:2008}

Here we note that Fuchs et al.\ showed that the distinct behavior of the 
core-valence exchange-correlation contributions in the LDA and GGA formalisms requires pseudopotentials to be tailored to calculations at each level of theory. \cite{Fuchs_PAW_LDAvGGA}
In particular, differences in core-valence overlap substantially affect the binding properties of solids. \cite{Fuchs_PAW_LDAvGGA}
With this understanding, we perform the LDA and PBEsol calculations  with the respective {\sc vasp} LDA and PBE optimized PAWs.
Because the hybrid-$V_{xc}$ HSE06 is a PBE-based functional, we also use PBE optimized PAWs in these calculations.
At present, {\sc vasp} does not support PAWs specifically optimized for metaGGA functionals;  however the newest PAWs distributed with the {\sc vasp.5.X} package include information on the kinetic energy density of the core-electrons that can be utilized in metaGGA calculations.
Therefore, we also use PBE-optimized PAWs for all calculations using metaGGA exchange-correlation potentials.

\subsection{$V_{xc}$ Performance Trends in Solids}
DFT benchmarks of various solid-state materials have been successful in identifying the general \emph{habits} of 
exchange-correlation functionals at different rungs of Jacob's ladder. \cite{Jacobsladder}
The local density approximation (LDA) and generalized gradient approximation (GGA)  
are at the first and second rung of the ladder, respectively.
The XC energy density in the LDA is designed to mimic the density of the uniform electron gas. \cite{Kohn/Sham:1965} 
LDA works well for simple metallic solids, however, it also has a tendency 
to overestimate cohesive energies of ionic solids and transition metal complexes. \cite{Kresse_STO-HSE} 
The failure of LDA to treat complex 
bonding interactions results in equilibrium volumes that are too small for most solids.

GGA improves on LDA by incorporating gradient corrections of the density. \cite{Perdew/Wang:1986, Becke:GGA1992, PhysRevB.46.6671}
One of the most widely used GGA among materials researcher is based on the parameterization by Perdew, Burke, and  Ernzerhof (PBE). \cite{Perdew/Burke/Ernzerhof:1996, Perspec:Burke12}
Although it corrects the overbinding errors of the LDA, 
PBE tend to give bond lengths and cell volumes that are too large. 
For solids the overestimation of cell volumes is significantly improved by functional such as, PBEsol, WC and SG4. \cite{PBEsol:2008, Wu/Cohen:2006, SG4_XC:2016}

\begin{figure*}[ht] 
\centering
\includegraphics[width=1.7\columnwidth]{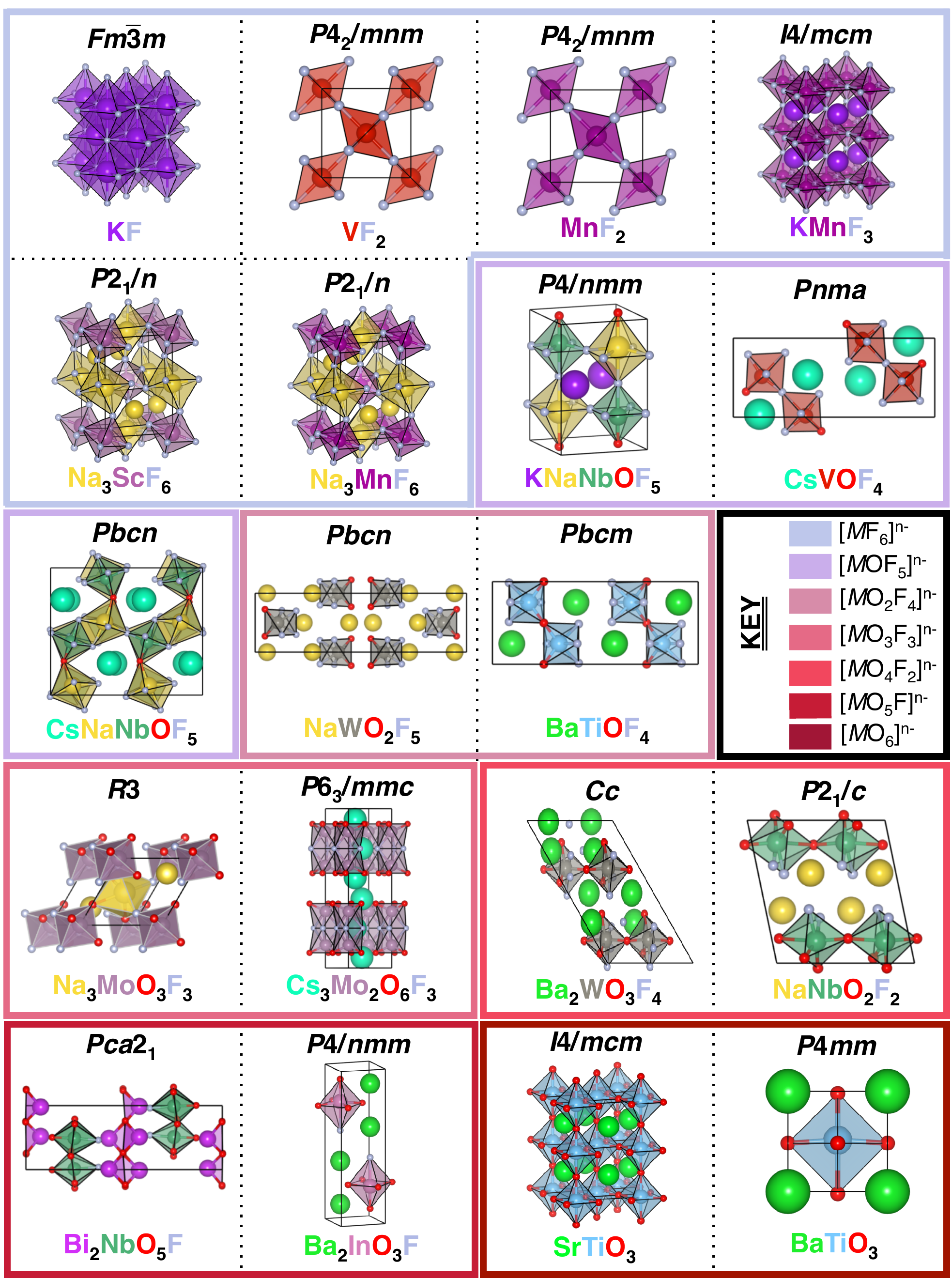}\vspace{-0.75\baselineskip}
\caption{%
Crystal structures of the 19 compounds appearing in this study. The structures are grouped based on the oxygen/fluorine ratio in the [$M$O$_x$F$_{6-x}$]$^{n-}$ anionic group.
Each anionic group can be differentiated by the colors bordering the structures, as indicated in the KEY (black outlined panel), and this color scheme is used in all subsequent figures to differentiate anionic groups.}
\label{fig:all_strs}
\end{figure*}

The semilocal metaGGAs form the third rung of Jacob's density functional ladder and include the kinetic energy density of the Kohn-Sham orbitals as an additional ingredient.
MetaGGAs strive to be more accurate than the GGA for a wider range of applications and chemical systems. 
Indeed, metaGGAs such as TPSS, have proved to be more accurate than PBE in predicting the properties of bulk solids and surfaces. \cite{PhysRevB.69.075102, PhysRevLett.115.036402}
However, TPSS often fails to capture the van der Waals' bonding in layered compounds like graphite. \cite{Haas/Tran/Blaha:2009}
The inconsistencies of early metaGGAs were due in part to the
inability of the dimensionless variables built from the kinetic energy to distinguish between different types of
chemical bonding. \cite{Jianwei_orbitalOverlap, MS2_paper, SunKEdepend:2013, sun2016accurate}
One way modern metaGGAs attempt to circumvent this problem is by tuning the kinetic energy density through a dimensionless parameter $\alpha = [1/$ELF $- 1]^{1/2}$ (ELF is defined as the electron localization function, cf.\ Refs.\ \onlinecite{silvi1994classification, Becke:ELF1990}). 
The $\alpha$ parameter characterizes the extent of orbital overlap throughout the crystal  and thus is expected to 
enable modern metaGGAs to \emph{intuitively} distinguish between a variety of bonding situations. \cite{Jianwei_orbitalOverlap}
In addition, $\alpha$ is essential for describing the asymptotic properties of the exchange energy density and exchange potential. \cite{PhysRevB.91.035126, PhysRevB.93.115127}
However, including the $\alpha$ parameter is not essential to functional development and to the overall accuracy of modern metaGGAs because accurate metaGGAs can be constructed without the so-called $\alpha$ parameter, for example, M06-L \cite{M06L:2006} and the recent TM metaGGA. \cite{TM:xc_2016} 
Although the modern metaGGAs like MS2 \cite{MS2_paper} and SCAN \cite{PhysRevLett.115.036402} show great promise, they have not been as 
extensively investigated as LDA, GGAs, or the hybrid functionals, which appear at the next rung. \cite{BlahaDFTbenchmark:2016, PhysRevB.69.075102}

Hybrid functionals mix semilocal XC (often from the GGA) with some contribution of exact exchange. \cite{Becke:1993}
The goal of hybrid $V_{xc}$ functionals is to produce more accurate thermochemical, magnetic and electronic properties by mimicking the effects of static correlation. \cite{BurkeWhichXC, Becke:1993, PhysRevB.84.115114, Perdew_et_al:1996}
The most popular hybrids, B3LYP \cite{Becke:1993} and HSE06 \cite{HSE06:2005} 
include approximately 20\% to 25\% exact exchange; however, studies have shown that this 
value is not unique and can be tuned to improve the accuracy of the desired property, such as equilibrium geometry or electronic band gap. \cite{PhysRevB.83.035119, garciabenchmark2012, FranchiniHSE:2013, Perdew_et_al:1996}
Notably, hybrid functionals overcome a major shortcoming of the non-empirical semilocal functionals,
in that they \emph{cannot} provide accurate band gaps beyond the exact Kohn-Sham gap.
For semilocal functionals this often results in a significant underestimation of the electronic band gap in solids.
In fact, correlated transition metal compounds are often gapless at these levels of theory.
DFT calculations using hybrid functionals often predict band gaps values that are much closer to experiment, 
marking a significant improvement over semilocal functionals.
However, the inclusion of exact exchange comes at significantly greater  
computational cost compared to the approximations at lower rungs, 
particularly in DFT codes based on planewave basis sets.

\subsection{Compounds and Crystal Data}
\autoref{fig:all_strs} shows the crystal structures for the 19 compounds investigated, which 
belong to either the fluoride, oxyfluoride or oxide families.
The materials can be further delineated by considering the ratio of 
oxygen $x$ to fluorine $6-x$ within the octahedral anionic 
groups comprising the structure, \emph{i.e.}, [$M$O$_x$F$_{6-x}$]$^{n-}$, where $x$ is an integer ranging from  0 to 6.
We note that all phases examined have been experimentally synthesized and 
both the cation and anion sublattices are fully ordered with no cation-site or anion-site intermixing. 
All crystallographic data for the DFT optimized structures are provided in the Supplemental Material.\cite{Supplmental_Note:Benchmark}

For the fluorides with $M$F$_6$ ($x$ = 0) octahedra, we study six compounds belonging to four crystal families.
KF is a simple binary fluoride with the $AB$ rock-salt structure. \cite{KF_exp_str}
VF$_2$ \cite{Costa:VF2} and MnF$_2$ \cite{MnF2:Benchmark} exhibit the $AB_2$ rutile structure. 
KMnF$_3$ \cite{Scatturin:KMnF3} exhibits the $ABX_3$ perovskite structure whereas \nsf\,\cite{Carlson1998116} and \nmf\,\cite{Carlson/Norrestam:1998} are ordered double perovskites with the cryolite structure---Na occupies both the $A$ and $B$ positions in the $A_2B_2$X$_6$ structure. \cite{hawthorne1975refinement}

By inspection of \autoref{fig:all_strs}, we find that  
ordered\cite{Pinlac/Poeppelmeier:2011,CsVOF4:1972, Marvel/Poeppelmeier:2007, Na2WOF:1982, 
BaTiOF4:1992, Brink2003450, Cs3Mo2O6F3:1980, Ba2WO3F4:1985, Bi2NbO5F:2007, Ba2InO3F:1995} oxyfluorides ($x$ = 1 \dots 5) 
crystallize in a wide variety of structure types with corner, edge and face connectivity among polyhedral units.
As an illustration of our schema to differentiate the oxyfluorides using the local 
[$M$O$_x$F$_{6-x}$]$^{n-}$ anionic groups rather than usual chemical formula, consider 
Ba$_2$WO$_3$F$_4$ and NaNbO$_2$F$_2$. 
Although these compounds have different chemical formulas, they both exhibit 
$M$O$_4$F$_2$ octahedral units, \emph{i.e.}, [WO$_4$F$_2$]$^{4-}$ and  [NbO$_4$F$_2$]$^{5-}$, and thus are grouped together in 
 \autoref{fig:all_strs}. 
 
Last, we  study two ternary oxides with $M$O$_6$ ($x$ = 6) octahedra: 
The perovskites SrTiO$_3$ \cite{I4mcm:STO1996} and BaTiO$_3$ \cite{BTO:1992} 
with distorted tetragonal configurations, \emph{i.e.}, nonpolar $I4/mcm$ (with out-of-phase TiO$_6$ octahedral rotations and polar $P4mm$ (with an electric polarization), respectively.
All 19 compounds are electronic insulators, with 15 of the 19 compounds investigated  nonmagnetic $d^0$ insulators.
The antiferromagnetic compounds VF$_2$ ($d^3$), MnF$_2$ ($d^5$), and KMnF$_3$ ($d^5$) are Slater insulators. \cite{Tressaud1982, Slater:Insulator1951}
Here we simulate  VF$_2$, MnF$_2$, and KMnF$_3$ with C-type magnetic order. \cite{C-type:KMnF31972}
\nmf\,($d^4$) is known experimentally to be insulating, however, the magnetic properties have not been throughly explored. \cite{Carlson/Norrestam:1998}
In our previous work, we established that \nmf\ exhibits a weak energetic preference  for ferromagnetic spin order over an antiferromagnetic spin configuration. \cite{Nenian_PRB_NMF_2014} 
Because these four magnetic compounds are Slater insulators, we are able to assess the performance of each XC functional without requiring a Hubbard-$U$ correction to ensure insulating behavior. \cite{Sawatzky:1994, Liechtenstein/Anisimov/Zaanen:1995}

\section{Results}
\subsection{Structure Analysis}

\begin{figure*}[ht] 
\centering
\includegraphics[width=1.75\columnwidth]{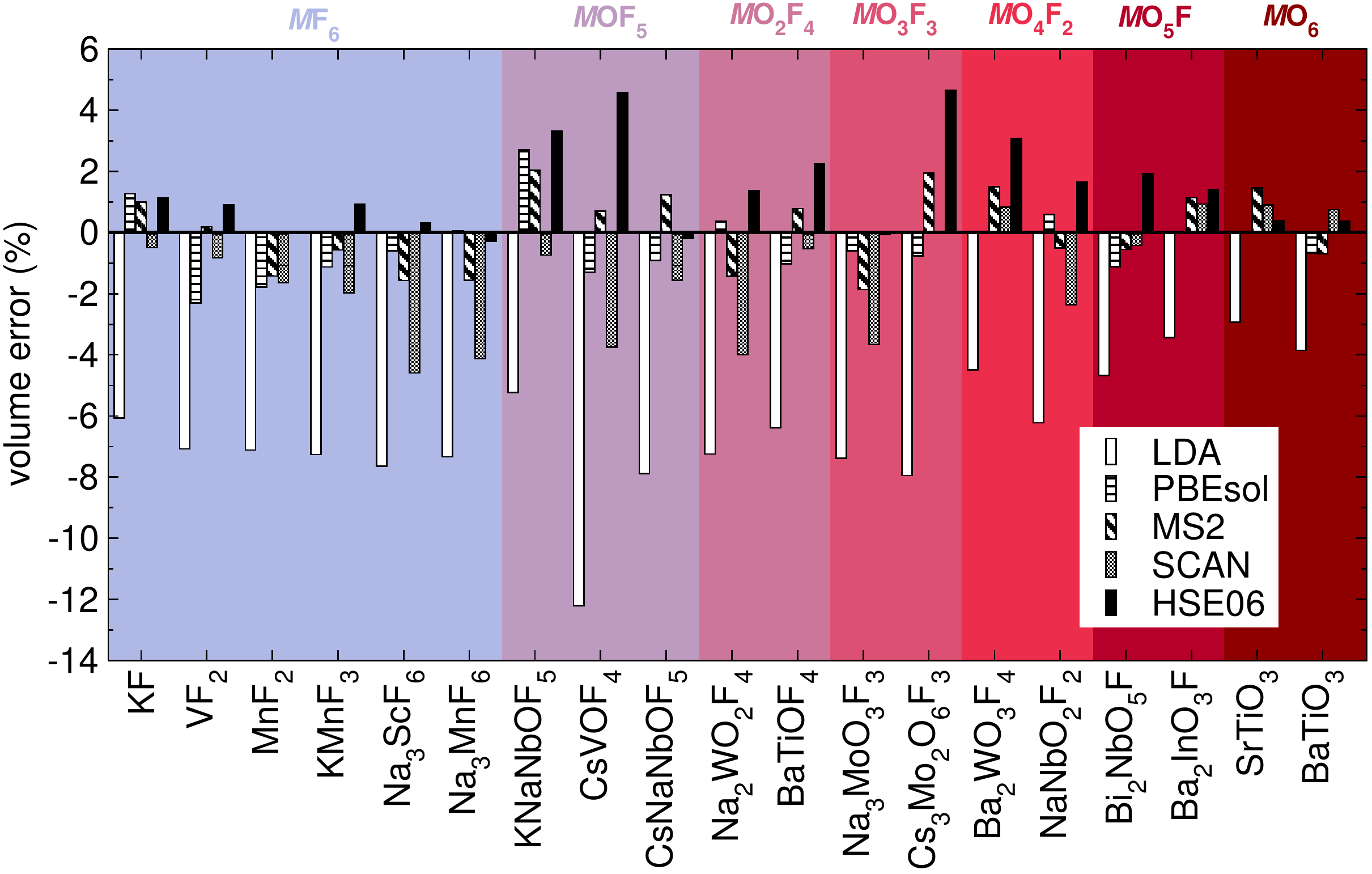}
\caption{%
Percent error in DFT computed cell volumes relative to experiment obtained using different exchange-correlation potentials as indicated by the different shaded bars.
The $M$O$_x$F$_{6-x}$ octahedra present in each compound are shown on the upper abscissa,  differentiated by shaded regions.}
\label{fig:vol_str}
\end{figure*}

\subsubsection{Cell Volume}
\autoref{fig:vol_str} shows the percentage error (\% error) in the cell volume $\Omega$  obtained from DFT with the different XC functionals relative to the experimentally (exp) measured structures at room temperature, which is obtained as  
$\%\,\mathrm{error} = ({\Omega_\mathrm{XC} - \Omega_\mathrm{exp}})/{\Omega_\mathrm{exp}}\times100$.
First, we observe the well-known overbinding of LDA for all compounds. 
The underestimation in the cell volume is larger than 3\% across all chemistries, with the exception of SrTiO$_3$ where the discrepancy is 2.3\%.

In contrast, cell optimizations with XC functionals at rungs higher than LDA show increased accuracy (errors $\lesssim$ 3\%).
The GGA PBEsol performs well, consistently predicting cell volumes to within 2\% error, with a slight tendency towards overbinding (\autoref{fig:vol_str}).
Our results also show that the hybrid functional HSE06, appearing at the highest rung of Jacob's ladder, more frequently overestimates the cell parameters than the other functionals.
This result is consistent with what has been previously reported for perovskite oxides studied using either HSE06 or the unscreened hybrid PBE0 . \cite{garciabenchmark2012, Kresse_STO-HSE}
We find that HSE06 gives the best overall performance for the ionic all-fluoride $M$F$_6$ ($\sim\!$ 1\%) and more covaleny all-oxide $M$O$_6$ ($\sim\!$ 0.5\%) compounds. 
Surprisingly, our results suggest that the degree of unit-cell overestimation is often greater for the oxyfluorides with $M$O$_x$F$_{6-x}$, $x$ = 1 \dots 5, where errors $>$1.5\% are found.
At the metaGGA level, we investigate the MS2 and SCAN functionals (\autoref{fig:vol_str}). 
First, our structural optimizations with MS2 reveals comparable and sometimes better accuracy than PBEsol ($\sim\!$ 2\%) particularly for the all-fluoride $M$F$_6$ compounds.
Although MS2 shows a strong tendency to underestimate the volumes of the fluorides, a consistent trend is less clear across the oxyfluorides and oxide compounds.
Next, our results indicate that SCAN produces smaller cell volumes across all chemistries 
and is often less accurate than MS2 particularly in predicting $\Omega$ of the more ionic compounds ($M$O$_x$F$_{6-x}$, $x$ = 0, 1, 2) with an error $\sim\!$ 3\%. 
These results are inconsistent with a recent study reported by Sun et al. \cite{PRL_SCAN} and we conjecture two factors may be responsible for the poorer performance herein: 
\begin{itemize}
\item Finite temperature effects, \emph{i.e.}, the experimental structures for which we assess the XC functional accuracy are measured at room temperature and thus must be zero-point corrected. 
This is germane to all volume calculated with any XC functional.
\item Maladapted PBE PAWs resulting in inaccurate core-valence exchange and correlation for the metaGGA functionals.
\end{itemize}

First, we investigate the contribution of thermal expansion for KF, Na$_3$ScF$_6$ and BaTiO$_3$ to address inaccuracies derived from missing finite-temperature contributions in \autoref{fig:vol_str}. 
From the measured coefficient of thermal expansion (CTE) for BaTiO$_3$, 
we approximate that the material will contract by 0.19\% at 0 K. \cite{He2004135} 
For KF and Na$_3$ScF$_6$, we compute the CTE with a self-consistent quasiharmonic approach at the PBEsol level. \cite{Huang201684} 
We observe that the effect of zero point volume correction is much larger for the fluorides 
with volume changes of approximately 1.3\% and 1.7\% for KF and Na$_3$ScF$_6$ respectively.
This result suggests the effects of thermal expansion may be more significant for compounds comprised of $M$F$_6$ rather than $M$O$_6$ octahedra. 
Although the zero-point correction improves the calculated metaGGA volumes compared 
to experiment, it alone does not fully account for the observed discrepancies, particularly in the case of Na$_3$ScF$_6$.

\begin{table}[t]
\begin{ruledtabular}
\centering
\caption{\label{tab:PAWdepend}The cell-volume error (\%) for 
 KF, Na$_3$ScF$_6$ and BaTiO$_3$ optimized with DFT using different PBE cation PAW valence configurations from three different exchange-correlation functionals: PBEsol, MS2,  and SCAN. The computed volumes are compared to the experimentally measured volumes at room temperature.} 
\begin{tabular}{llccc}

Solid				&PAW 				& PBEsol 		& MS2 		& SCAN 	 \\
\hline
KF				& K, $3p^64s^1$		& 1.26		& 1.00		& -0.50		\\
				& K, $3s^23p^64s^1$	& 1.13		& -0.19		& -3.10		\\
\hline
\hline	
Na$_3$ScF$_6$	&Na, $3s^1$			&			&			&	\\
				&Sc, $3d^24s^1$		& -4.98		& -6.38		& -8.73	\\
\hline
				&Na, $2p^63s^1$		&			&			&	\\
				&Sc, $3d^24s^1$		& -1.69		& -3.38		& -5.46	\\
\hline
				&Na, $3s^1$			&			&			&	\\
				&Sc, $3s^23p^63d^24s^1$	& -4.25		& -5.12		& -7.85	\\
\hline
				&Na, $2p^63s^1$			&			&			&	\\
				&Sc, $3s^23p^63d^24s^1$	& -0.60		& -1.57		& -4.57	\\
\hline
\hline
BaTiO$_3$		&Ba, $5s^25p^66s^2$	&			&			&	\\
				&Ti, $3p^63d^34s^1$ & -0.68		& -0.70		& -0.75	\\
\hline
				&Ba, $5s^25p^66s^2$	&			&			&	\\
				&Ti, $3s^23p^63d^34s^1$ & -0.87		& -0.60		& -0.62	\\
\end{tabular}
\end{ruledtabular}
\end{table}


Next, we investigate the equilibrium volume dependence on the electronic configuration of the PAW pseudopotential (all based on PBE exchange and correlation) 
for three XC functionals. 
Here we vary the cations' valence electronic configuration in the KF, Na$_3$ScF$_6$ and BaTiO$_3$ compounds as a means 
to perturb the core-valence exchange-correlation interaction owing to the unavailability of PAWs constructed specifically for metaGGA functionals.

\autoref{tab:PAWdepend} presents the volume errors for PBEsol, MS2, and SCAN as a function of cation PAW choice.
We observe the computed BaTiO$_3$ volumes show little sensitivity to the form of the Ti PAW.
All three XC functionals result in volume errors $<$ 1\% compared to experiment. 
In contrast, a change in the PAW electronic configuration has a pronounced effect on the equilibrium volume in both the  $M$F$_6$ containing compounds KF and Na$_3$ScF$_6$.
Moreover, the computed volumes obtained from SCAN and MS2 show a much larger variance than PBEsol when more or fewer electrons are included as valence. 
These results suggest that the optimized PBE PAWs may not be ``totally'' transferable to modern metaGGAs like MS2 and SCAN.
We believe that a more in-depth study following that of Fuchs et al. \cite{Fuchs_PAW_LDAvGGA} which compares the role of pseudopotential 
choice on the accuracy metaGGAs may be necessary before ultimately drawing schematic conclusions on its performance across many material classes.

\begin{figure}[t] 
\centering
\includegraphics[width=1\columnwidth]{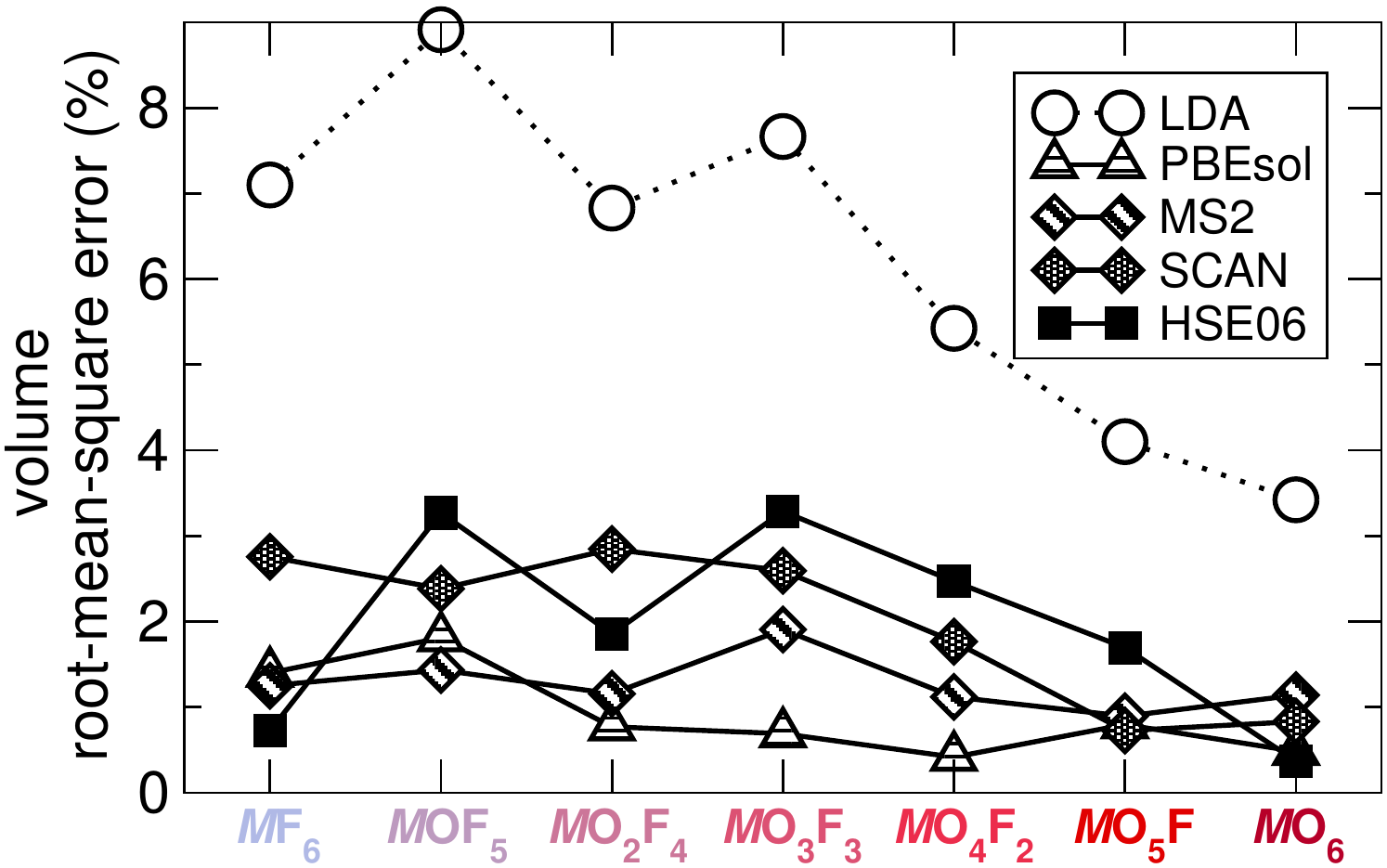}\vspace{-0.75\baselineskip}
\caption{%
RMS error in volume with different $V_{xc}$ potentials as a function of increasing oxygen content, 
$x$ = 0 (left) to $x$ = 6 (right) within the $M$O$_x$F$_{6-x}$ anionic units.}
\label{fig:rms_vol}
\end{figure}

To understand the performance of the XC functionals as the relative covalent bonding active in the compounds 
evolves from the ionic fluorides to the more covalent oxides, we plot the root-mean-squared (RMS)
error in the cell volume for each material in our test suite with respect to anionic composition $M$O$_x$F$_{6-x}$, ($x$ = 0 \dots 6) ratio (\autoref{fig:rms_vol}).
Our statistical analysis confirms the previously stated results, \emph{i.e.}, the  LDA is the least accurate and HSE06 performs the best for fluorides and oxides.
However, it also reveals other less obvious trends described next.
Notably, we observe that the unit cell  volumes predicted by all functionals become more accurate as the relative covalency 
of the material increases, \emph{i.e.}, with increasing oxygen content $x$ in the anionic group.
LDA shows the most noticeable improvement across the series, with the RMS error changing from 7\,\% and 9\,\% for $M$O$_x$F$_{6-x}$ with $x$ = 0 and 1, respectively,  to 3.1\,\% for the oxides ($x$ = 6).
Interestingly, we observe that across all O-to-F ratios, the PBEsol functional provides the 
most consistent performance and is overall the most accurate for the intermediate anion ratios exhibited by the $M$O$_x$F$_{6-x}$, $x$ = 2, 3 and 4 oxyfluorides.
Last, we find that between the metaGGAs, MS2 exhibits better overall performance than SCAN (\autoref{fig:rms_vol}).
We observe that the RMS-error of SCAN is significantly better for the oxides compared to the fluorides, which suggests 
that it may be better suited to describe the more pronounced covalent bonding interactions oxides and chalcogenides. 

\begin{figure*}[t] 
\centering
\includegraphics[width=1.75\columnwidth]{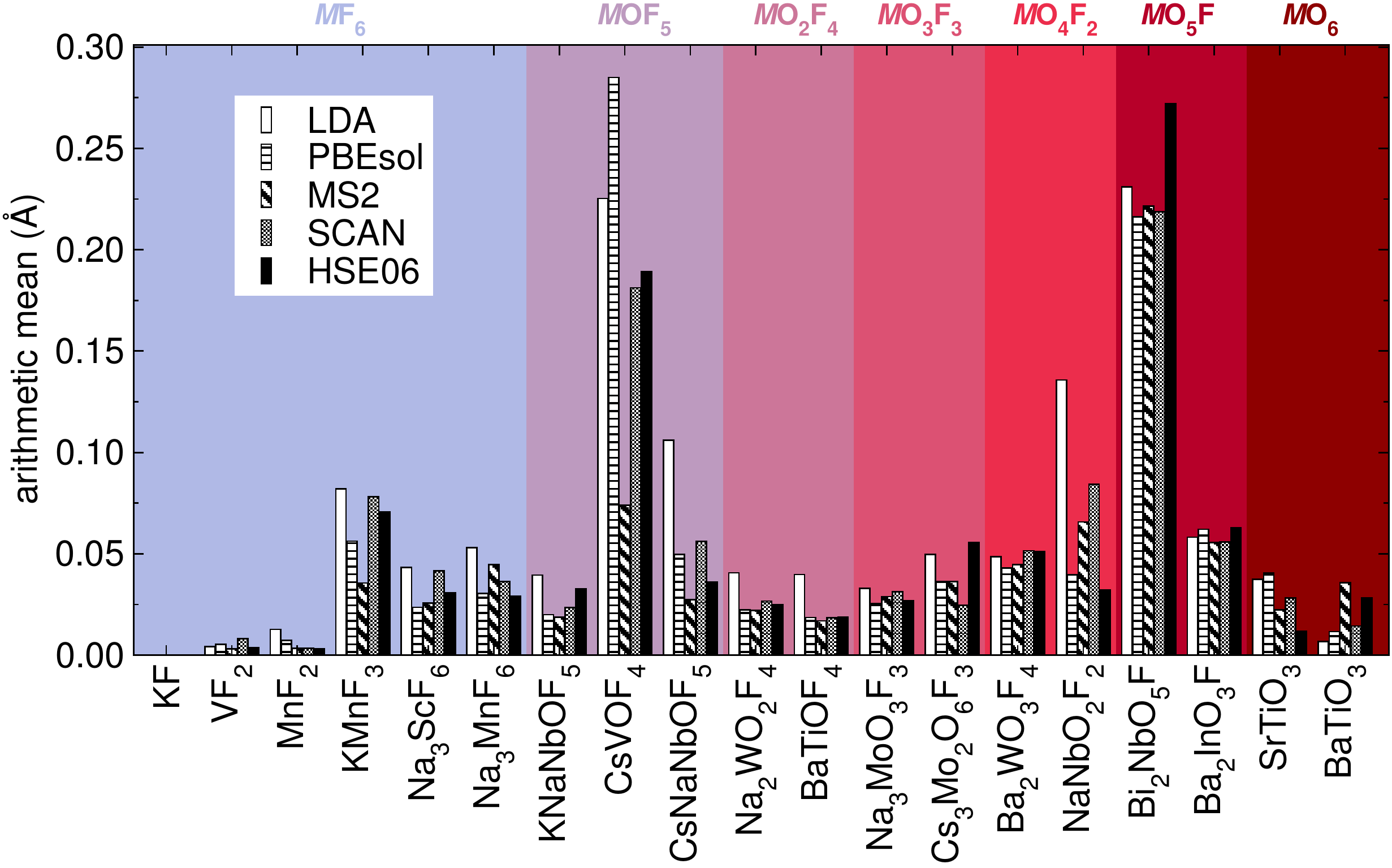}\vspace{-0.75\baselineskip}
\caption{%
Error in internal coordinates of DFT computed structures relative to experiment obtained using different XC functionals indicated by shaded bars.
The anionic group present in each compound is shown on the upper abscissa differentiated by shaded regions. An arithmetic mean of $d_{av}$ = 0\,\AA\ indicates perfect agreement, which occurs for KF because by symmetry there are no free internal positions.}
\label{fig:inter_str}
\end{figure*}

\subsubsection{Internal Coordinates} 
Next, we assess the effectiveness of each XC functional at accurately predicting the internal ionic positions compared to those reported in the experimental structures. 
In \autoref{fig:inter_str} we report the structural arithmetic mean introduced in Ref.~\onlinecite{delaFlor:to5129} to 
assess the likeness of two structures, and
defined as $ d_{av} = (\sum_i m_{i}u_{i})/n$, 
where $i$ is the index of the occupied Wyckoff site, $m_i$ is the multiplicity of the Wyckoff orbit in the unit cell, 
$u_i$ represents the atomic displacement {amplitude required to map one structure onto another}, and $n$ is the total number of atoms in the unit cell.
Note, that an arithmetic mean of $d_{av}$ = 0~\AA\,corresponds to perfect agreement between two structures. 
In this definition, the cell strain contribution is omitted. 

\paragraph*{Fluorides.} The structure of KF has $Fm\bar{3}m$ symmetry and no free internal parameters for the occupied Wyckoff positions.
As a result, the arithmetic mean for KF is zero.
\vf\,and \mf\, exhibit the rutile structure ($P4_2/mnm$) and thus have two positional degrees of freedom for the fluoride ion located on the $4f$ Wyckoff site. 
We find that the predicted equilibrium atomic positions obtained from all $V_{xc}$ functionals generally matches closely with the experimental rutile structures.
For example in \vf\, we find that the error is quite similarly at all levels of theory (\autoref{fig:inter_str}); 
however, for \mf\ we observe that the predicted internal coordinates are systematically improved as we climb Jacob's ladder.

The perovskite structured \kmf, \nsf, and \nmf\ compounds exhibit corner-connected $M$F$_6$ octahedral units that rotate about several crystal directions. 
The main distortion in \kmf\ is an out-of-phase rotation of the fluorine octahedra about the [001] direction (of 2.6$^\circ$), which corresponds to two free internal positions for the fluoride anion ($8h$ Wyckoff site). 
\autoref{fig:inter_str} shows that the LDA gives the poorest performance (angle of 8.2$^\circ$) in predicting the internal positions 
followed by SCAN (8.0$^\circ$), HSE06 (7.4$^\circ$), PBEsol (6.5$^\circ$), and MS2 (5.1$^\circ$).

The primary distortions in the double perovskites \nsf\ and \nmf\ are out-of-phase fluorine octahedral rotations about the [110] direction ($\phi$) and an in-phase rotations about [001] ($\theta$). 
From \autoref{fig:inter_str} we observe that in general all functionals perform slightly better for the double perovskites than \kmf. 
The amplitude of the octahedral rotations in the double perovskites (\nsf: $\phi = 21.1^\circ$, $\theta = 13.3^\circ$ and \nmf: $\phi = 19.4^\circ$, $\theta = 12.5^\circ$) 
are much larger compared to the amplitude of the rotation in \kmf.
We find that the amplitude of the in-phase and out-of-phase rotation angles are  within  
$1-2^\circ$ (cf. Ref.~\onlinecite{Supplmental_Note:Benchmark}) of the experimental values for all functionals studied.

The increased accuracy of the predicted rotations is likely responsible for the improved performance of the XC functionals for the double perovskite over \kmf. 
For these ternary fluorides, the LDA functional remains the least accurate at predicting the internal coordinates. 
For \nsf\ we note that the arithmetic mean obtained for the non-hybrids PBEsol and MS2 perform slightly better than HSE06; however, the metaGGA SCAN is the worse performer among the functionals for \nsf.
On the other hand,  PBEsol and HSE06 are found to be the most accurate for \nmf.
Mode-crystallographic analysis between the ground state monoclinic double perovskite structures and the ideal high-symmetry parent $Fm\bar{3}m$ lattice 
reveals that \nsf\ and \nmf\ have complex atomic distortions (see Ref.~\onlinecite{Nenian_PRB_NMF_2014} for full description).
From the comparison the symmetry-adapted mode amplitudes for \nsf\ and \nmf\ (see Ref.~\onlinecite{Supplmental_Note:Benchmark}) 
we observe that discrepancy in the internal coordinates is predominately due to the inaccuracy in predicting secondary distortions as opposed to octahedral rotations. 

\paragraph*{Oxyfluorides.} The oxyfluoride compounds studied here belong to a diverse set of structure types, with different polyhedral connectivity. 
Our results do not reveal a clear trend in the accuracy of the calculated internal coordinates from difference $V_{xc}$ potentials as a function of increasing oxygen content in the anionic unit.
However, from \autoref{fig:inter_str} we can conclude that the LDA is most often the least accurate functional in predicting the internal coordinates regardless of oxygen content and thus should be avoided.

One anomalous result we find is the significant arithmetic means, corresponding to possible large errors in the optimized in atomic positions, for 
the compounds CsVOF$_4$ (with [VOF$_5$]$^{n-}$) and Bi$_2$NbO$_5$F (with [NbO$_5$F]$^{6-}$). 
To understand the origin of this behavior, we compare the global instability index (GII) of the experimental structures and the DFT optimized geometries for each material.
The GII judges the ``plausibility'' of a crystal structure based on the deviation of 
the bond valence sums of the constituent ions from their formal valence averaged over all atoms in the unit cell. \cite{BondValenceBookPoeppe, Adams2004281}
Using this metric, crystal structures that are characterized by GII $>$ 0.2 valence units (v.u.) either show evidence of strained bonds or an incorrectly reported structure.

For CsVOF$_4$ we calculate a GII value of 0.203 v.u.\ for the experimentally reported structure, which is close to the boundary of a material exhibiting large bond strains. 
Further analysis of the experimental structure reveals that the large GII is due to compressed V--O and V--F bonds in the [VOF$_5$]$^{n-}$ units 
compared to the bond valence parameters.
Supplementary Figure 1 shows that the DFT optimized structures with the exception of that obtained from the LDA, 
reduce the GII to $<$ 0.14 v.u. (see Ref.~\onlinecite{Supplmental_Note:Benchmark}).
Therefore we contend that the large discrepancies for CsVOF$_4$ observed in \autoref{fig:inter_str} result from the comparison with an experimental structure with strained bonds.
We observe that the cell optimizations with PBEsol, HSE06 and the metaGGAs 
increase the [VOF$_5$]$^{n-}$ polyhedral volumes and reduce the strain on the bonds observed in the experimental structure.
Our results suggest that the internal coordinates of the CsVOF$_4$ experimental structure may require reassessment.

We calculate GII=0.260 for the experimental Bi$_2$NbO$_5$F structure. 
All functionals besides the LDA are found to reduce the strain in the stretched Bi--O bonds of the experimental structure\cite{Supplmental_Note:Benchmark}
However, unlike CsVOF$_4$, 0.2 $<$ GII  $<$ 0.26 v.u.\ for all structures obtained with the different XC functionals except from the LDA.
This result also suggests  the internal coordinates of Bi$_2$NbO$_5$F  may need to be revisited. 

\paragraph*{Oxides.}  The tetragonal perovskite oxides SrTiO$_3$  and BaTiO$_3$ are found in space groups $I4/mcm$ and $P4mm$ respectively. 
For SrTiO$_3$ the experimental out-of-phase TiO$_6$ rotation angle is $\approx$2.1$^{\circ}$ about the [001] direction. 
LDA and PBEsol both predict the out-of-phase angle to be $\sim\,$5$^{\circ}$, whereas the metaGGAs preform slightly better, \emph{i.e.}, MS2 gives  3.7$^{\circ}$ and SCAN gives 4.1$^{\circ}$.
HSE06 performs most accurately for SrTiO$_3$, providing an equilibrium rotation angle of 2.9$^{\circ}$. 
This result is consistent with the findings of Aramburu and co-workers, who proposed that the subtle energy variations associated with octahedral rotations are a more stringent test of the predictive capabilities of exchange-correlation functionals compared to strain tensors in structural complex materials. \cite{garciabenchmark2012}
They argue that the accuracy of the hybrid functionals derives from 
the pseudo-Jahn-Teller vibronic contribution to octahedral tilting in semicovalent solids. \cite{GarciaFernandezetal:2010}
Specifically, the larger band gaps predicted by hybrid functionals like HSE06 reduce the covalent interactions between occupied and unoccupied orbitals, which manifest through electron-lattice interactions, thus resulting in smaller predicted rotation angles.\cite{garciabenchmark2012}

\begin{figure}[t] 
\centering
\includegraphics[width=0.9\columnwidth]{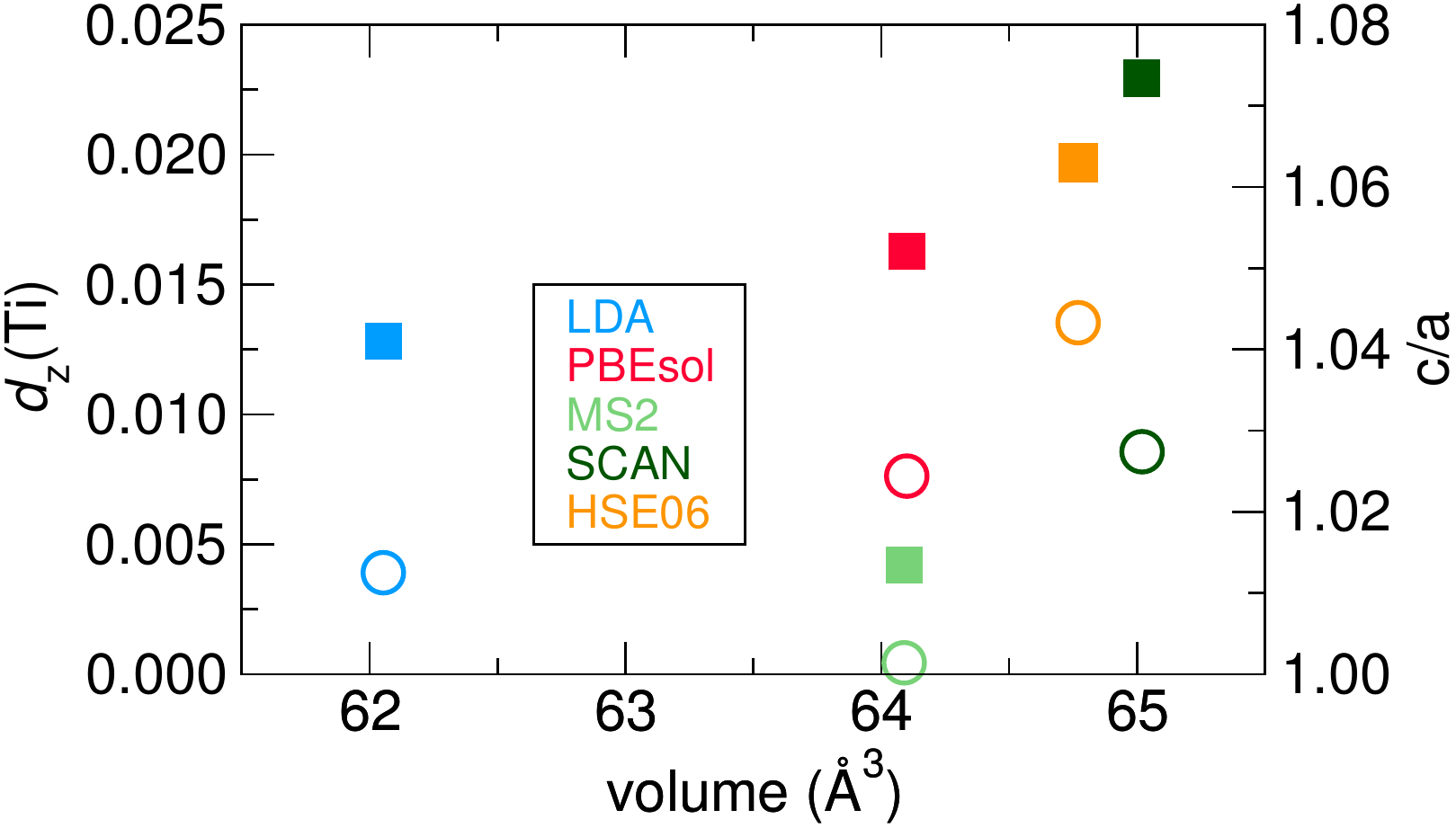}\vspace{-0.75\baselineskip}
\caption{%
The Ti ion displacement $d_z(\mathrm{Ti})$ along the [001] direction (filled symbols) and corresponding axial ratios (empty symbols) in tetragonal $P4/mmm$ BaTiO$_3$ as a function of optimized unit cell volumes obtained with difference exchange-correlation potentials. 
Experimentally $d_z(\mathrm{Ti})\!\sim\!0.015$ (Ref.~\onlinecite{Shirane/BTO:1957}).
}
\label{fig:bto_dist}
\end{figure}

\autoref{fig:inter_str} shows that $d_{av}$ of the polar oxide BaTiO$_3$ increases as we climb Jacob's ladder. 
To understand this trend, we examine the ferroelectric displacements in BaTiO$_3$ and focus on the Ti off centering.
Prior studies have established that the amplitude of the ferroelectric distortion in BaTiO$_3$ is strongly linked to the equilibrium cell volume as determined by the selected exchange-correlation functional, \emph{e.g.}, LDA, GGA, or hybrid. \cite{Kresse_STO-HSE, Bilc/Ghosez_et_al:2008, Wu/CohenSingh:2004} 
Variable cell and atomic relaxations with $V_{xc}$ functionals that favor larger unit cell volumes predict larger ferroelectric displacements, 
thus suggesting no direct functional dependence of the ferroelectric distortion in BaTiO$_3$, but rather that the dependency originates from indirect errors in the cell volume. 

In \autoref{fig:bto_dist}, we plot the displacement of the 
Ti ion from the $(\sfrac{1}{2}, \sfrac{1}{2}, \sfrac{1}{2})$ position with Ba at the origin and the $c/a$ lattice parameter axial ratio as a function of optimized equilibrium volume for each XC functional. 
We observe that the $V_{xc}$-volume dependence reported in previous studies is maintained by the LDA, PBEsol, SCAN and HSE06 functionals and 
that the $c/a$ ratio obeys a similar trend to the Ti ion displacement as a function of optimized cell volume  for LDA, PBEsol and HSE06, \emph{i.e.} $c/a$ increase with larger cell volume.
However, the metaGGA functionals, MS2 and SCAN, deviate from these trends.
First, MS2 predicts an optimized cell volume for BaTiO$_3$ that is nearly identical to PBEsol yet exhibits a Ti displacement that is nearly three times smaller. We can understand this by examining the $c/a$ ratio obtained from MS2. The lattice tetragonality for BaTiO$_3$ is nearly quenched with MS2 (\autoref{fig:bto_dist}).
In the case of SCAN we observe that although the functional predicts the largest optimized cell volume, the $c/a$ ratio obtained is only slightly larger than the PBEsol value.
These results strongly suggests that the inclusion of the kinetic energy density in metaGGAs introduces a direct functional dependence for the ferroelectric distortion in BaTiO$_3$.
Based on these finding, we recommend that when these modern metaGGAs are used to study and predict new ferroelectric compounds a detailed structural analysis is essential.

\subsection{Electronic Properties}

\begin{figure*}[t] 
\centering
\includegraphics[width=1.75\columnwidth]{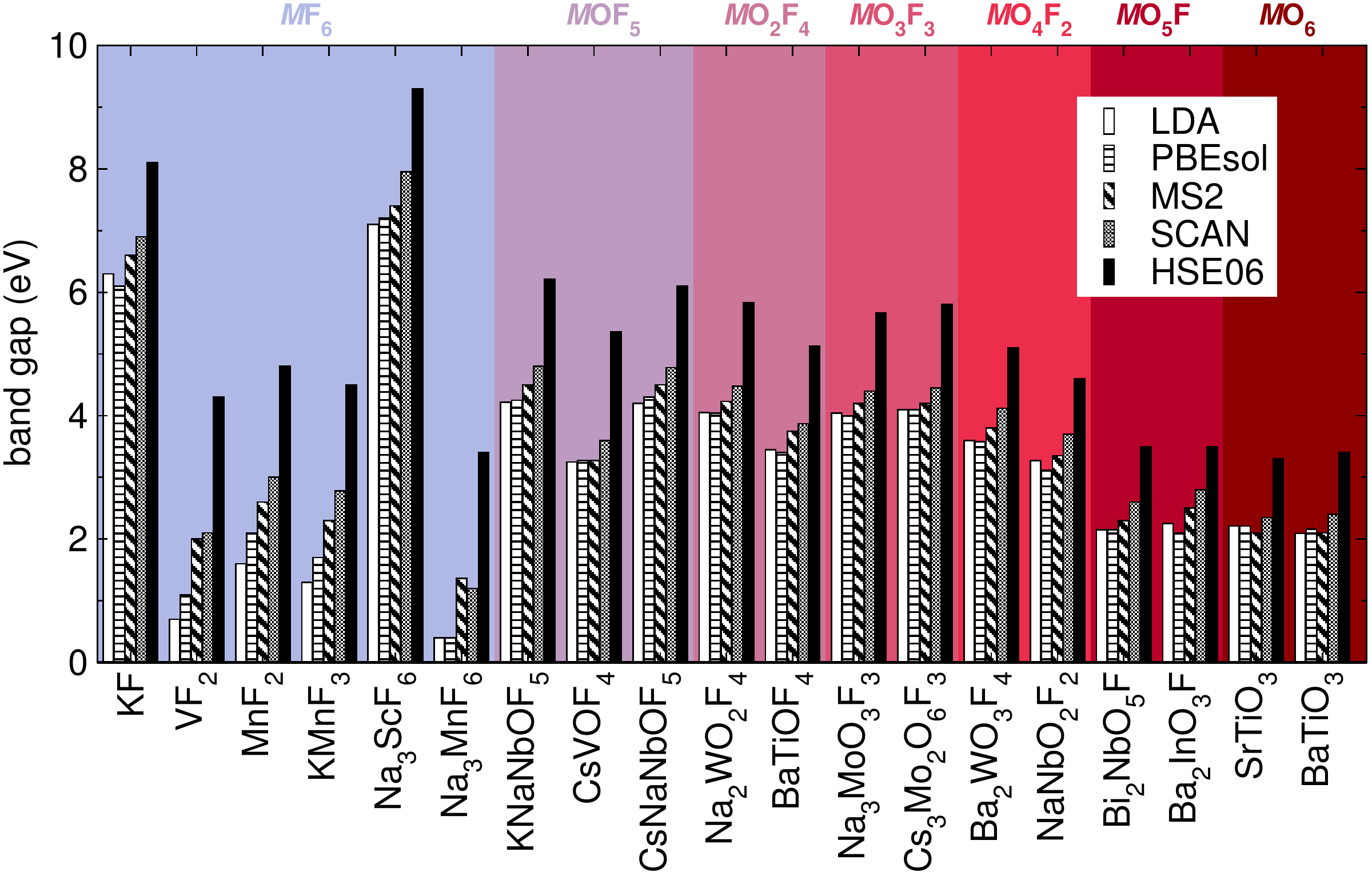}\vspace{-0.75\baselineskip}
\caption{%
Electronic band gap evolution with variable oxygen content $x$ in the $M$O$_x$F$_{6-x}$ octahedra from different exchange-correlation functionals (shaded bars).
The anionic group present in each compound is shown on the upper abscissa differentiated by shaded regions.
Experimental measured gap for KF is 10.8 eV \cite{fox2001optical}. 
SrTiO$_3$ has a measured direct $\Gamma$--$\Gamma$ band gap of 3.75 eV and an indirect $R$--$\Gamma$ gap of 3.25 eV. \cite{Benthem/French:2001} The experimental band gap of BaTiO$_3$ is 3.4 eV. \cite{BTOBandgap:1970}
}
\label{fig:BandGaps}
\end{figure*}

We now compute the electronic density-of-states (DOS) for each compound optimized with each XC functional. The calculated electronic gaps are reported in \autoref{fig:BandGaps}.
Owing to a shortage of experimental data on the band gaps of most of the compounds in this study, we are unable to provide a direct assessment of the accuracy of the each functional against experiment.
Qualitatively, we anticipate the relative size of the band gap to decrease as fluorine is replaced with the less electronegative oxide anions.
Indeed, \autoref{fig:BandGaps} demonstrates that this trend is observed for the $d^0$ compounds regardless of the choice of $V_{xc}$.

First we find that the DFT band gap increases as one climbs Jacob's ladder of exchange-correlation functionals (\autoref{fig:BandGaps}).
As expected, the LDA and PBEsol give the smallest band gaps and HSE06 the largest for all compounds regardless of oxygen content $x$ in the $M$O$_x$F$_{6-x}$ octahedra.
In addition, both metaGGAs MS2 and SCAN show larger band gaps than the LDA and GGA as observed in other binary semiconductors including GaAs and CaO, \cite{Zeng-hui:metaGGAgaps2016} with SCAN always producing larger band gaps than MS2.

\subsection{Lattice Dynamical Properties}

Now we investigate the $V_{xc}$-dependent lattice vibrational frequencies. 
We compute the phonons at the optimized theoretical volumes with each level of theory and report the transverse optical frequencies.
Given the high computational cost of accurate phonon calculations and the scarcity of 
experimental vibrational data for most compounds in this study, 
we focus on four representative materials that sample the range of O/F concentrations observed in the material suite: 
KMnF$_3$ ($x$ = 0), 
KNaNbOF$_5$ ($x$ = 1), 
Na$_2$W$_2$F$_4$ ($x$ = 2), 
and Na$_3$MoO$_3$F$_3$ ($x$ = 3).
With regards to the oxide only containing compounds ($x$ = 6), BaTiO$_3$ and SrTiO$_3$, we note that a prior study by Kresse et al.\ has already established the phonon functional dependence and therefore we do not repeat it here. \cite{Kresse_STO-HSE}

\begin{figure}[th] 
\centering
\includegraphics[width=1\columnwidth]{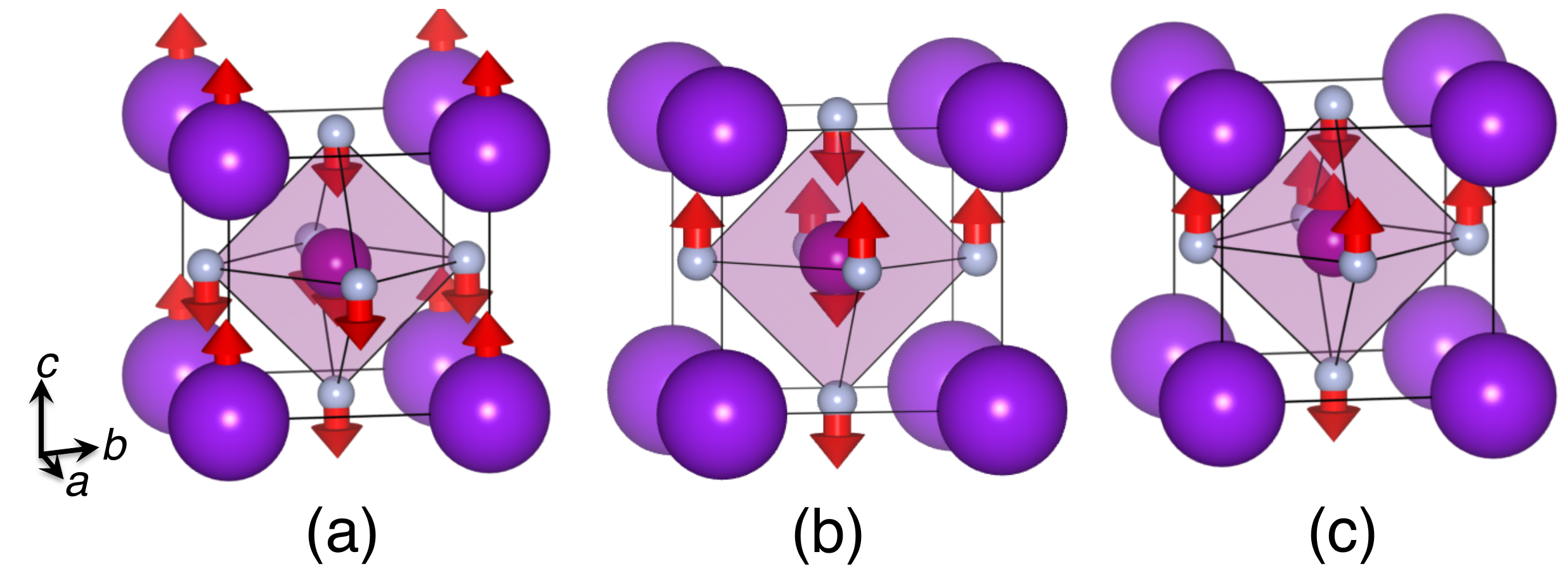}\vspace{-0.75\baselineskip}
\caption{%
Eigendisplacements (red arrows) for the IR-active transverse optical (TO) modes (a) TO1, (b) TO2 and (c) TO3 in 
cubic $Pm\bar{3}m$ KMnF$_3$.
\label{fig:kmf_modes}
}
\end{figure}

\begin{figure}[h] 
\centering
\includegraphics[width=1\columnwidth]{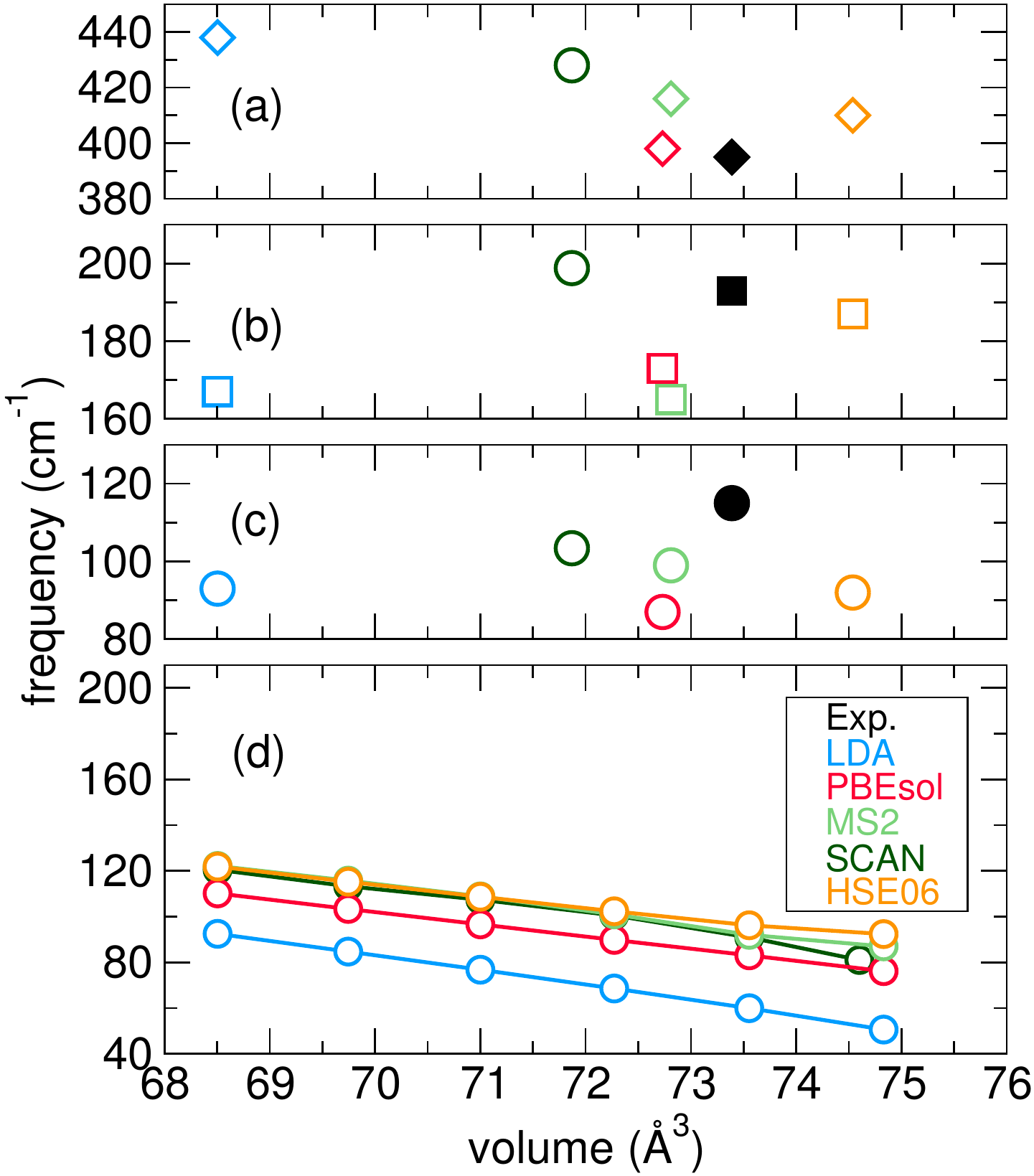}\vspace{-0.75\baselineskip}
\caption{%
Infrared frequencies computed different cell volumes with the LDA, PBEsol, MS2, SCAN and HSE06 functionals for cubic KMnF$_3$. 
In panels (a)-(c) the computed frequencies of the TO3 mode (a), TO2 mode (b) and TO1 mode (c) are plotted at respective equilibrium volume for each functional.
In panel (d) we plot the evolution of the TO1 mode frequency as a function of cell volume for each functional.
The empty diamond, square, circular symbols represent the TO3, TO2 and TO1 modes respectively.}
\label{fig:vol_phon_kmf}
\end{figure}

The fluoroperovskite KMnF$_3$ is cubic ($Pm\bar{3}m$) at room temperature. \cite{KMnF3_revised_str} 
Here we investigate the $\Gamma$ point phonon frequencies of the cubic phase, which like other perovskites consists of four triply degenerate transverse optical (TO)  modes. Three of the TO modes are infrared (IR) active (\autoref{fig:kmf_modes}). \cite{Rondinelli/Spanier:2013}
In \autoref{fig:vol_phon_kmf} we plot the computed IR mode frequencies \emph{vs.}\ volume at four rungs of DFT alongside available experimental data. \cite{KMnF3_Phonondata_cubic}
We find that all  XC functionals perform well, providing calculated TO mode frequencies close to experiment at the respective equilibrium volume [\autoref{fig:kmf_modes}(a)-(c)].
Specifically, the TO1 mode is underestimated by every functional [\autoref{fig:vol_phon_kmf}(a)].
While the TO2 mode is underestimated by every functional besides SCAN [\autoref{fig:vol_phon_kmf}(b)].
The frequency of the highest experimental IR mode frequency nearly coincides with that obtained from PBEsol [\autoref{fig:vol_phon_kmf}(c)].

In \autoref{fig:vol_phon_kmf}(d) we plot the evolution of the lowest frequency IR active mode as a function of the cubic cell volume.
Our results demonstrate that the TO1 mode decreases almost linearly with increasing cell volume for all functionals investigated.
We find that given the same cell volume, LDA always predicts a softer TO1 mode frequency followed by PBEsol.
The largest frequencies are predicted by the metaGGAs and HSE06.
In fact, \autoref{fig:vol_phon_kmf}(d) shows that for the same cell volume, the TO1 mode obtained from the metaGGA and hybrid functionals are nearly identical.
The trend of higher phonon frequencies obtained from hybrid-$V_{xc}$ functional calculations has been reported previously and is attributed to stiffer bonds owing to the inclusion of exact exchange. \cite{Kresse_STO-HSE}
Interestingly, despite having no exact exchange added, and predicting smaller optimized cell volumes, the metaGGAs behave nearly identically to HSE06 with respect to the mode frequencies. 

\begin{figure}[t] 
\centering
\includegraphics[width=1\columnwidth]{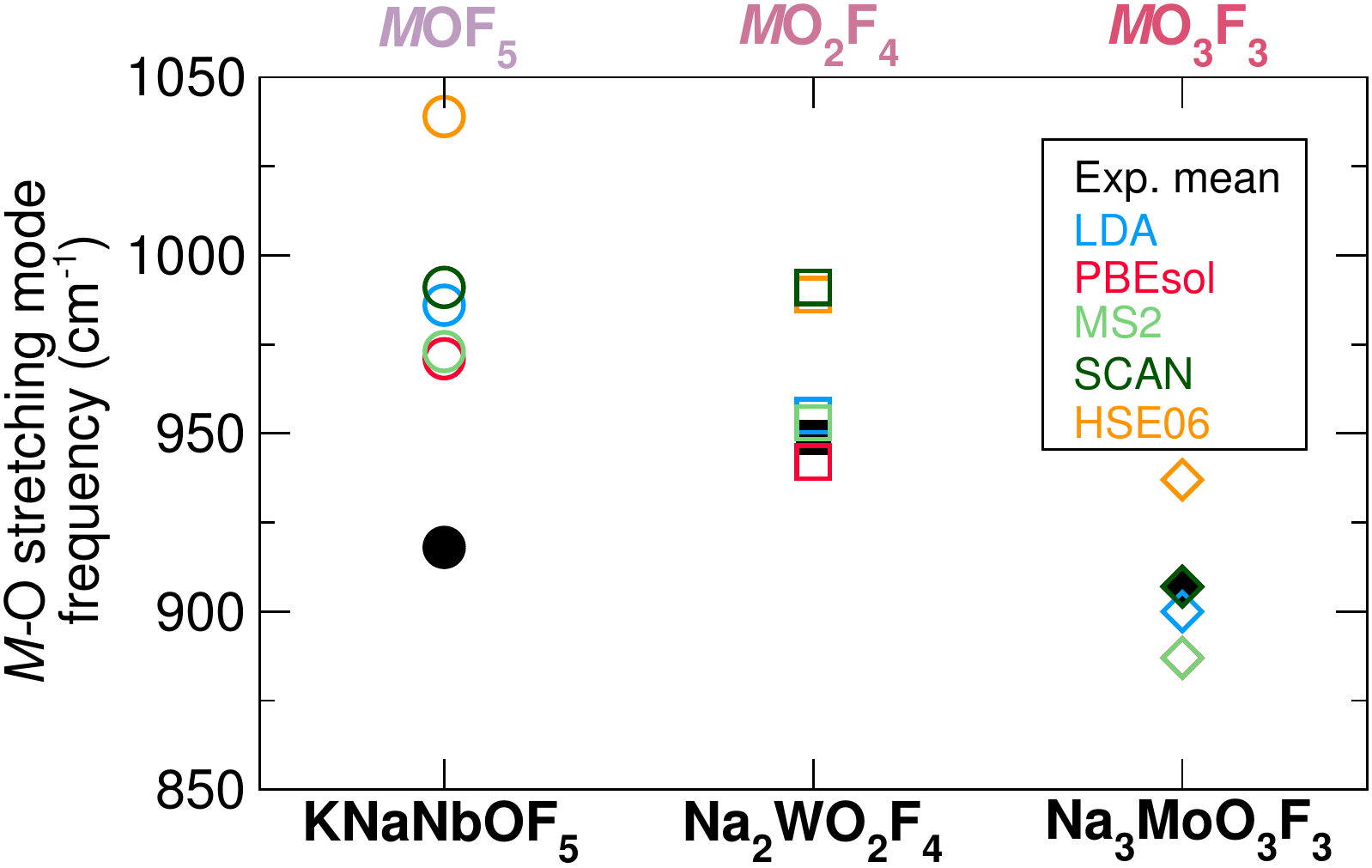}
\caption{%
Infrared frequencies for the $M$--O stretching mode in KNaNbOF$_5$, 
Na$_2$W$_2$F$_4$ and Na$_3$MoO$_3$F$_3$ compared to the experimental mean of related compounds. 
Experimental phonon frequencies taken from Refs.~\onlinecite{{Krylov201432, Voit2006, doi:10.1021/ic50008a029, OxF6-x_Vibrational_spec, Atuchin2012159, Krylov2012, ZAAC19693660107}}.
Note that the PBEsol result overlaps with MS2 and SCAN matches the experimental mean for Na3MoO3F3.
}
\label{fig:oxyfly_phonons}
\end{figure}

\begin{figure*}[ht] 
\centering
\includegraphics[width=1.8\columnwidth]{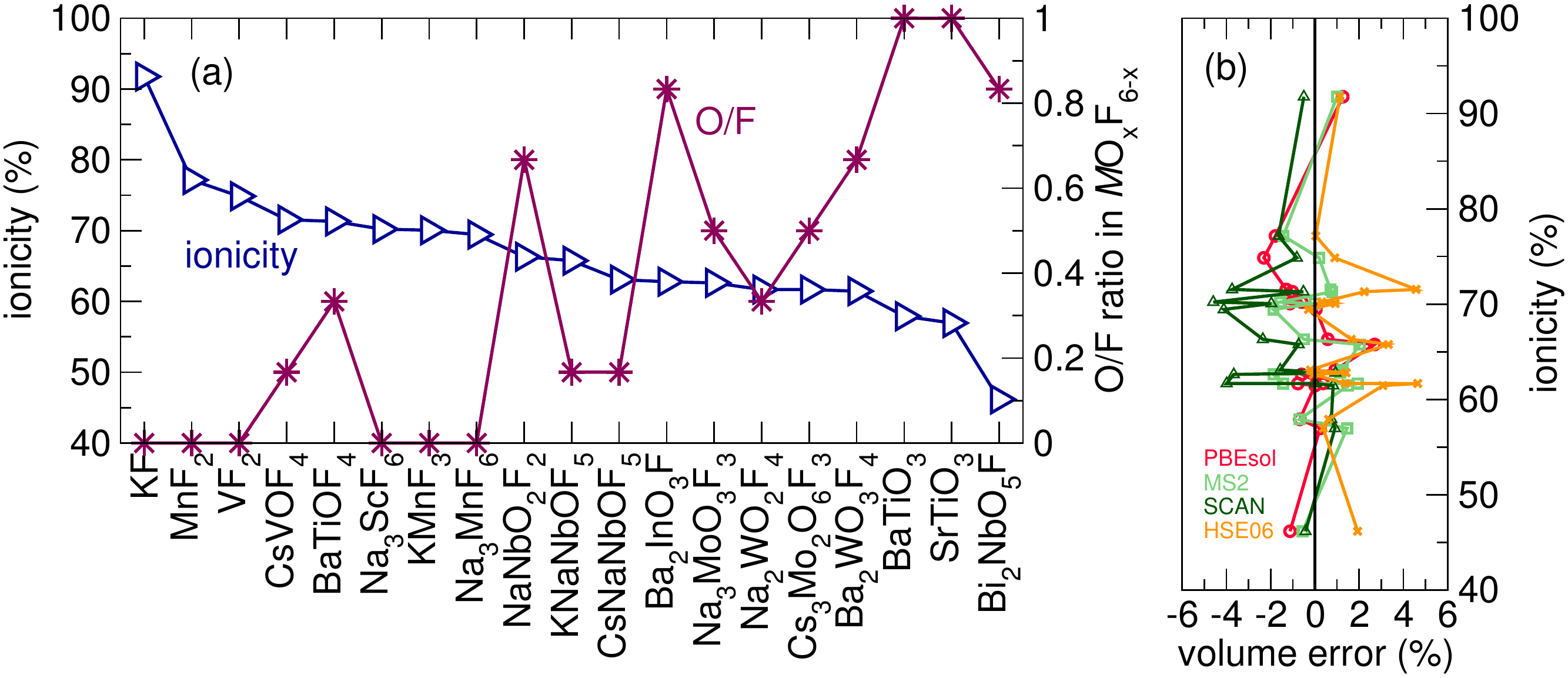}\vspace{-0.75\baselineskip}
\caption{%
(a) Calculated ionicities and O/F ratio in the $M$O$_x$F$_{6-x}$ anionic group for the 19 compounds explored. The materials are arranged in order of decreasing ionicity.  The electronegativity differences (A--B interactions) considered in the calculation of the ionicity of the solids were cation--anion and cation--cation. Anion-anion interactions are ignored. 
(b) Volume error in percentage as a function of ionicity. For clarity LDA is excluded owing to its large \% errors compared to the other functionals.
}
\label{fig:percent_ion}
\end{figure*}

We were unable to find experimental vibrational data for the ordered oxyfluoride compounds to use for a direct comparison with our calculations.  
However, we note that vibrational spectroscopy measurements are 
routinely used to help characterize local anion order in $M$O$_x$F$_{6-x}$ complexes, 
especially in cases where the compounds appears disordered to X-ray diffraction. \cite{Voit2006, Krylov201432}
In this regard, the $M$--O stretching mode is particularly important in identifying the nature of anion order 
in the octahedral environment, for example, when deciphering $cis$ vs.\ $trans$ configurations for $M$O$_2$F$_4$ units or $fac$ vs.\ $mer$ configurations for $M$O$_3$F$_3$ units. 
Our literature survey of [NbOF$_5$]$^{2-}$, [WO$_2$F$_4$]$^{2-}$ and [MoO$_3$F$_3$]$^{3-}$ anions revealed that the IR active
frequency of the M--O stretching mode appears in the 800 -- 1000 cm$^{-1}$ range
and was relatively insensitive to the choice of counter cations. 
For that reason, we investigate the functional dependence of the $M$--O mode 
frequency for the compounds KNaNbOF$_5$, Na$_2$WO$_2$F$_4$ and Na$_3$MoO$_3$F$_3$.  
The DFT frequencies are compared to the statistical mean, $\sigma$, of the experimentally measured $M$--O 
frequencies for similar chemistries consisting of 
[NbOF$_5$]$^{2-}$, $cis$-[WO$_2$F$_4$]$^{2-}$ and $fac$-[MoO$_3$F$_3$]$^{3-}$ octahedra. 

\autoref{fig:oxyfly_phonons} shows that for these three oxyfluorides, the LDA, PBEsol,  and MS2 functionals predict the 
$M$--O stretching mode with comparable accuracy compared to the experimental (literature) mean. 
Specifically, we observe that the mode frequencies computed by semilocal functionals  are very similar, with LDA consistently predicting the largest frequency for all three compounds.
The calculated PBEsol frequencies are slightly smaller than those obtained from MS2 for the compounds KNaNbOF$_5$ ($x$ = 1) and Na$_2$WO$_2$F$_4$ ($x$ = 2), whereas the values are nearly identical for Na$_3$MoO$_3$F$_3$ ($x$ = 3).
SCAN, on the other hand, predicts stiffer mode frequencies compared to the other semilocal functionals.  
In addition, we observe that $M$--O mode frequencies calculated with the hybrid HSE06 are often harder than the semilocal functional predictions.
We also note that the overestimation the $M$--O mode frequency by HSE06 decreases as the percent overestimation of the cell volume is reduced (cf.\  \autoref{fig:vol_str}).
The latter suggests that the increase frequency observed in our HSE06 calculations is strongly tied to the degree of overestimation of the cell volume.

\begin{table}[t]
\begin{ruledtabular}
\centering
\caption{\label{tab:StatPhononErr}
The mean absolute error (Mean AE) and maximum absolute error (Max AE) from experiment of vibrational frequencies for the 6 modes 
(3 IR modes of KMnF$_3$ and $M$--O stretching mode of KNaNbOF$_5$, Na$_2$WO$_2$F$_4$, Na$_3$MoO$_3$F$_3$) measured in this study.
All values are in cm$^{-1}$.
Error = theory $-$ experiment.
$
\mathrm{Mean\,AE} = 1/n \sum^n_{i=1} |\mathrm{Error}|.\, 
$
} 
\begin{tabular}{lcc}

Method			&Mean AE 	& Max AE 	 \\
\hline
LDA				& 28.7			& 68 		\\
PBEsol			& 21.7			& 53		\\	
MS2				& 24.1			& 55		\\
SCAN			& 27.6			& 73		\\
HSE06			& 39.4			& 121		\\
\end{tabular}
\end{ruledtabular}
\end{table}

In \autoref{tab:StatPhononErr} we summarize the deviation of the six computed phonon frequencies from experiment. 
We observe that PBEsol has the lowest mean absolute error of all functionals while HSE06 has the highest. 
Between the metaGGA functionals, MS2 has a lower mean absolute error than SCAN. 
As we have already seen for KMnF$_3$, if given identical cell volumes then the metaGGAs predict very similar mode frequencies to HSE06; 
therefore, this analysis further suggests that the major source of these discrepancies is due to the error in the optimized cell volume.

Our results are consistent with previous reports on the sensitivity of the phonon mode frequencies with XC functionals in crystalline solids. \cite{Kresse_STO-HSE, Bilc/Ghosez_et_al:2008, sun2016accurate}
For the $M$O$_x$F$_{6-x}$ ($x$ = 0, 1, 2, 3) fluorides and oxyfluorides studied here, we  observe ($i$) that although there is some dependence of the phonon frequency on the choice of exchange-correlation functional  the accuracy of the computed phonon modes is most significantly affected by cell volume and  ($ii$) the lower rung functionals can be as accurate as the metaGGA and hybrid functionals.

\section{Discussion}

For the first time, our benchmark study systematically investigates the performance of different $V_{xc}$ approximations within DFT 
at four different levels against oxygen content in a variety of oxyfluoride compounds.
With the evolution of the oxygen-to-fluorine (O/F) ratio in these compounds, the implicit assumption is that the 
balance of covalency (or ionicity) of the solids would also evolve proportionally.
Although the terms \emph{ionic `bond'} and \emph{covalent bond} can be defined,
decades of debate has proved that there exists
no unambiguous scale with which ionicity can be quantified, particularly as a bulk property. \cite{IonicityReview:83, DUFFY1986145, Prin/ion:1994}
Nevertheless, various ionicity scales have proved to provide useful 
approximations in a variety of applications. \cite{AntonioCovalency, Cohen:1992, DUFFY1986145, PhysRev.182.891, Whiteside:2011} 
Perhaps one of the most useful and transferable ionicity formalisms was introduced by Pauling. \cite{PaulingElectro}
His approach of using electronegativity differences as a measure of bond polarizabilities \cite{Shannon/Prewitt:1969} between ions has proved accurate in describing the bonding in simple metal fluorides and oxide crystals. \cite{IonicityReview:83}
Here we qualitatively assess the effect of increasing oxygen content in this materials suite by defining \emph{\% ionicity of a solid} as the weighted average ionicity obtained from all atomic pair electronegativity differences in the crystal as follows:  
$
\mathrm{\%\,ionicity} = (\sum w_\mathrm{AB}f_\mathrm{AB})/W,\, 
$
 where $f_\mathrm{AB}$ is the ionicity of the bond between ions A and B according to Pauling's electronegativity scale. \cite{PaulingElectro}
The weight of the interaction $w_\mathrm{AB}$ is computed by adding the site multiplicity of ions A and B, and 
$W=\sum_i w_i$ is the sum of all the interaction weights. 
Using this approximation, 
we observe that decreasing ionicity does not directly correlate with a change in oxygen/fluorine ratio [\autoref{fig:percent_ion}(a)].
However, the general behavior of this metric shows that the fluorides and oxides cluster at the high- and low-ionicity 
ends of the scale, respectively, with the oxyfluorides located between the extremes.

\autoref{fig:percent_ion}(b) also shows that the volume error does not evolve 
monotonically as a function of \% ionicity with any of the functionals studied.
Although this behavior might arguably reflect the diverse collection of structure types studied, 
it is significant to note that no clear trends emerge within the hierarchy imposed by Jacob's ladder of exchange-correlation functionals.
Generally, PBEsol shows a remarkable degree of accuracy as a function of \% ionicity. 
However, no overall improvement in the accuracy of the cell volume is obtained by using more sophisticated functionals (MS2, SCAN, or HSE06). 
Furthermore, \autoref{fig:percent_ion}(b) demonstrates that the evolution of volume as a function of  \% ionicity varies 
significantly for functionals at the same level of theory, \emph{e.g.}, the metaGGAs MS2 and SCAN.

\begin{figure}[t] 
\centering
\includegraphics[width=1\columnwidth]{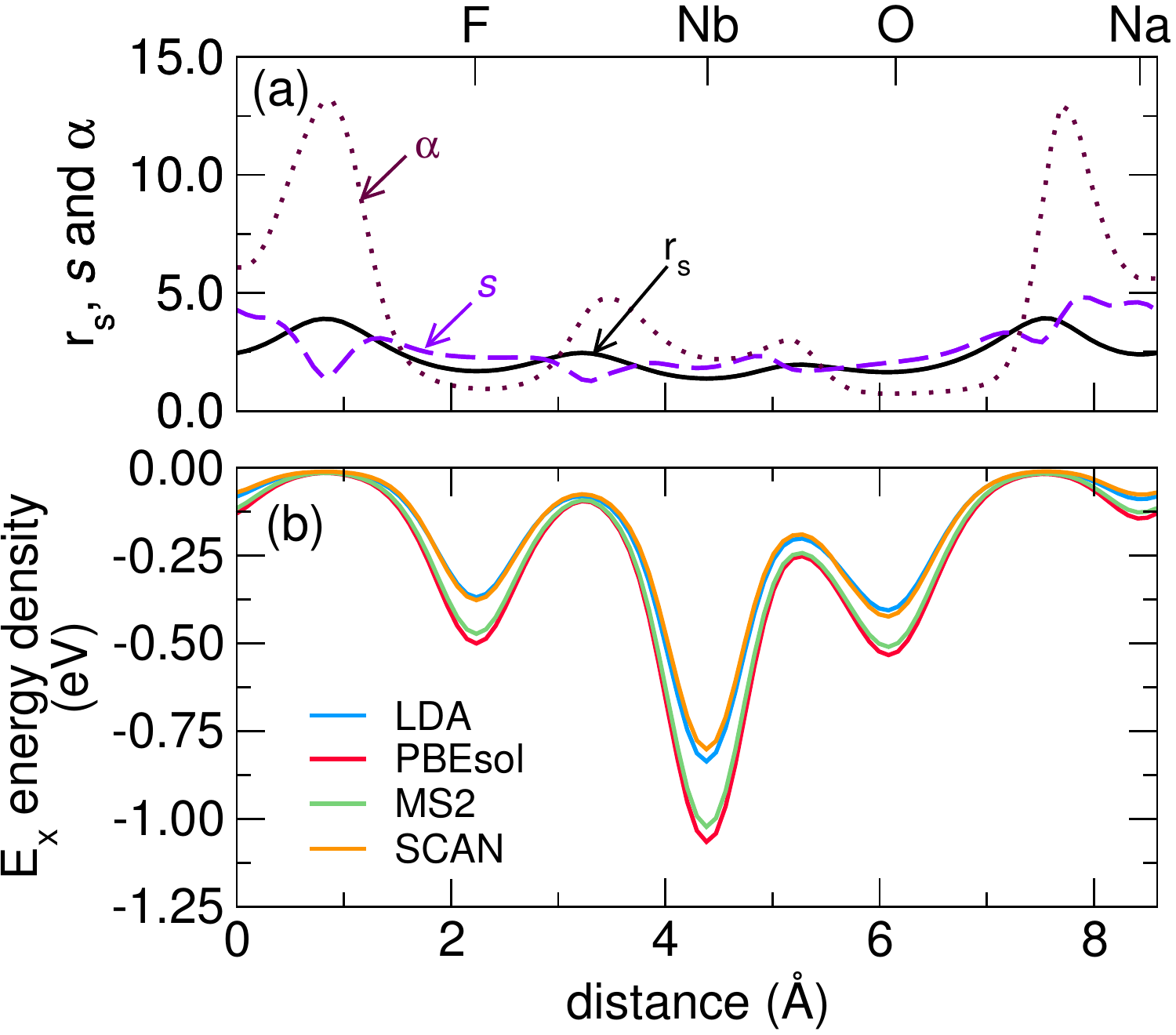}\vspace{-0.75\baselineskip}
\caption{%
(a) The computed r$_s$, $s$ and $\alpha$ distributions along a line in the [001] direction 
within the (100) plane and near the F--Nb--O--Na bond in the compound KNaNbOF$_5$.
(b) The exchange energy densities along the same path for the functionals
LDA, PBEsol, and MS2 and SCAN. 
The charge density and ELF are computed with PBEsol at the experimental lattice constants.}
\label{fig:s_alpha}
\end{figure}

A semilocal functional can be defined in terms of its enhancement
factors $F_{xc}$, \cite{BlahaDFTbenchmark:2016} which is typically expressed as a function of
the Wigner-Seitz radius, $r_s = [3/(4\pi n)]^{1/3}$, the reduced density gradient, \cite{SunKEdepend:2013} $s = |\nabla n|/2 (3 \pi^2 )^{1/3} n^{4/3}$ 
and dimensionless parameter $\alpha$. 
The Wigner-Seitz radius is related to the electron density, $n$, the parameter \emph{s} measures how fast the charge density varies on the scale of the 
local Fermi wavelength $2 \pi /k_F$ where $k_F = (3 \pi^2 n)^{1/3}$, \cite{Xiao:Sun2013SiO, Perdew/Burke/Wang:1996} 
and $\alpha$ is an ingredient of the metaGGAs MS2 and SCAN that characterizes the extent of orbital overlap. \cite{MS2_paper, Xiao:Sun2013SiO}

The spacial distributions of $r_s$, $s$, $\alpha$ along the [001] direction in the oxyfluoride KNaNbOF$_5$ ($x$ = 1) allows us to identify important regions in the solid [\autoref{fig:s_alpha}(a)]. 
[The positions of the ions along the [001] direction are indicated on the upper abscissa of \autoref{fig:s_alpha}(a).]
Specifically, the regions of core-valence separation is characterized by larger values of $s$, \cite{CoreVal:Blaha:2009} while the regions of single-bonding 
and the overlap of closed shells can be identified where $\alpha < 1$ and $\alpha \gg 1$, respectively. \cite{MS2_paper}

Prior studies have established that a strong correlation exists between predicted cell volumes and the parameters 
$r_s$, $s$ and $\alpha$ for semilocal functionals. \cite{CoreVal:Blaha:2009, Fuchs_PAW_LDAvGGA, BlahaDFTbenchmark:2016, PhysRevB.79.155107}
Thus, to better understand the distinct tendencies of the semilocal functionals observed in this work, 
we plot the exchange ($E_x$) energy density for LDA, PBEsol, MS2 and SCAN [\autoref{fig:s_alpha}(b)]. 
\footnote{The exchange energy density, is $n(\mathbf{r})\times E_x (n)$ in LDA or $n(\mathbf{r})\times E_x (n)\times F_x(s,\alpha)$ 
in GGA or meta-GGA. $E_x(n)$ is the exchange energy per particle of LDA, which is a function of $r_s$. }
Interestingly from the evolution of the $E_x$ energy density [\autoref{fig:s_alpha}(b)], we observe that the metaGGA SCAN 
behaves very similar to LDA while the other metaGGA, MS2 bears closer resemblance to PBEsol.
This result is consistent with the trends observed for the functional dependence of the cell volume in the compounds studied,
\emph{i.e.} LDA and SCAN show a strong tendency to underpredict cell volumes particularly for more ionic compounds 
while the absolute errors of the volumes predicted by PBEsol and MS2 are very similar.

Moreover, we observe that the major difference in the $E_x$ energy density between the 
LDA and SCAN results and the PBEsol and MS2 results occurs in the region near the core of the ions.
It has been shown that for solids the overlap between the core and
the valence electrons of the constituent atoms is a particularly important region for the prediction of the cell volume. \cite{CoreVal:Blaha:2009, Fuchs_PAW_LDAvGGA}
Indeed, the difference between each semilocal functional is dominated 
by the analytical form chosen to represent the enhancement factor, but we conjecture that subtle differences in the core-valence regions likely produces the distinct behavior in the structural properties predicted by the functionals with similar $F_x$ and motivates further 
detailed analyses.

\section{Summary}

We assessed the accuracy of exchange-correlation potentials for DFT  found at 
the first four rungs of Jacob's ladder with an emphasis on predicting the structural, electronic, 
and lattice dynamical properties in oxyfluorides compounds. The performance of the functionals was found to vary with oxygen content in the octahedral $M$O$_x$F$_{6-x}$ anionic groups.
A summary of our main conclusions is as follows:
\begin{itemize}
\item \textbf{Structural properties:}
Overall, LDA and PBEsol are consistently the least and most accurate 
functional for computation of the cell volume and internal coordinates across all materials in our suite, respectively.
The modern metaGGAs, MS2 and SCAN, also show satisfactory performance 
but do not exhibit the same consistent improvement as the PBEsol functional as the O/F ratio changes. For example, MS2 almost completely removes the tetragonality of BaTiO$_3$.
It is significant to note that we observe a large variance in the predicted cell 
volume for the metaGGAs based on the valence configuration of the cation pseudopotentials. 
We suspect that the discrepancies observed in the structural properties 
for the metaGGAs may be in part due to the poor transferability of the GGA PAWs used in our study.
We suggest a more in-depth pseudopotential study to clarify these observations.
The hybrid $V_{xc}$ HSE06 is indeed most accurate for the fluoride ($x$ = 0) and oxide ($x$ = 6) compounds, however, it shows a strong tendency to overestimate cell volumes for the oxyfluorides (intermediate integer values of $x$).

\item \textbf{Electronic properties:}
All of the compounds in this study are known  experimentally to be insulating. 
Generally, the smallest band gaps are predicted by LDA and PBEsol.
The metaGGAs MS2 and SCAN both improve the DFT gap slightly 
over the functionals at the first two rungs.
HSE06 consistently predicts the largest band gaps.

\item \textbf{Lattice dynamical properties:}
We computed the phonons for four compounds: 
KMnF$_3$ ($x$ = 0), KNaNbOF$_5$ ($x$ = 1), 
Na$_2$WO$_2$F$_4$ ($x$ = 2), and Na$_3$MoO$_3$F$_3$ ($x$ = 3).
Generally, our results indicate that the phonon frequencies show a strong volume dependence for all XC functionals.
\end{itemize}

Based on the observations in this study, our general recommendations for DFT calculations of related compounds are as follows:
($i$) LDA consistently underperforms and as a result we do not recommend it for this family of materials.
($ii$) For all fluorides ($x$ = 0) and oxides ($x$ = 6) compounds PBEsol or HSE06 are most accurate; and
($iii$) for intermediate oxygen compositions ($x$ = 1 \dots 5), PBEsol is consistently accurate for structural properties.
In light of this, we also anticipate that functionals which contain \emph{ingredients} based on similar construction principles as PBEsol such as the revised TPSS functional \cite{PhysRevLett.103.026403} (revTPSS) to yield satisfactory results;
however, we advise careful testing of each functional relative to the property being predicted.

\begin{acknowledgments}
N.C.\ and J.M.R.\ acknowledge support from the National Science Foundation (NSF, DMR-1454688) and are grateful to Jianwei Sun for many useful discussions and guidance in using the SCAN functional. 
N.C.\ also thanks Liang-Fang Huang and Bing Xiao for useful comments.
DFT calculations were performed on the high-performance computing 
facilities available at the Center for Nanoscale Materials (CARBON Cluster) 
at Argonne National Laboratory, supported by the U.S.\ DOE, 
Office of Basic Energy Sciences (BES), DE-AC02-06CH11357.
\end{acknowledgments}

\bibliography{7.10.15_rondo_nenian_refs}

\begin{thebibliography}{120}%
\makeatletter
\providecommand \@ifxundefined [1]{%
 \@ifx{#1\undefined}
}%
\providecommand \@ifnum [1]{%
 \ifnum #1\expandafter \@firstoftwo
 \else \expandafter \@secondoftwo
 \fi
}%
\providecommand \@ifx [1]{%
 \ifx #1\expandafter \@firstoftwo
 \else \expandafter \@secondoftwo
 \fi
}%
\providecommand \natexlab [1]{#1}%
\providecommand \enquote  [1]{``#1''}%
\providecommand \bibnamefont  [1]{#1}%
\providecommand \bibfnamefont [1]{#1}%
\providecommand \citenamefont [1]{#1}%
\providecommand \href@noop [0]{\@secondoftwo}%
\providecommand \href [0]{\begingroup \@sanitize@url \@href}%
\providecommand \@href[1]{\@@startlink{#1}\@@href}%
\providecommand \@@href[1]{\endgroup#1\@@endlink}%
\providecommand \@sanitize@url [0]{\catcode `\\12\catcode `\$12\catcode
  `\&12\catcode `\#12\catcode `\^12\catcode `\_12\catcode `\%12\relax}%
\providecommand \@@startlink[1]{}%
\providecommand \@@endlink[0]{}%
\providecommand \url  [0]{\begingroup\@sanitize@url \@url }%
\providecommand \@url [1]{\endgroup\@href {#1}{\urlprefix }}%
\providecommand \urlprefix  [0]{URL }%
\providecommand \Eprint [0]{\href }%
\providecommand \doibase [0]{http://dx.doi.org/}%
\providecommand \selectlanguage [0]{\@gobble}%
\providecommand \bibinfo  [0]{\@secondoftwo}%
\providecommand \bibfield  [0]{\@secondoftwo}%
\providecommand \translation [1]{[#1]}%
\providecommand \BibitemOpen [0]{}%
\providecommand \bibitemStop [0]{}%
\providecommand \bibitemNoStop [0]{.\EOS\space}%
\providecommand \EOS [0]{\spacefactor3000\relax}%
\providecommand \BibitemShut  [1]{\csname bibitem#1\endcsname}%
\let\auto@bib@innerbib\@empty
\bibitem [{\citenamefont {Tao}\ \emph {et~al.}(2003)\citenamefont {Tao},
  \citenamefont {Perdew}, \citenamefont {Staroverov},\ and\ \citenamefont
  {Scuseria}}]{PhysRevLett.91.146401}%
  \BibitemOpen
  \bibfield  {author} {\bibinfo {author} {\bibfnamefont {Jianmin}\ \bibnamefont
  {Tao}}, \bibinfo {author} {\bibfnamefont {John~P.}\ \bibnamefont {Perdew}},
  \bibinfo {author} {\bibfnamefont {Viktor~N.}\ \bibnamefont {Staroverov}}, \
  and\ \bibinfo {author} {\bibfnamefont {Gustavo~E.}\ \bibnamefont
  {Scuseria}},\ }\bibfield  {title} {\enquote {\bibinfo {title} {Climbing the
  density functional ladder: Nonempirical meta-generalized gradient
  approximation designed for molecules and solids},}\ }\href {\doibase
  10.1103/PhysRevLett.91.146401} {\bibfield  {journal} {\bibinfo  {journal}
  {Phys. Rev. Lett.}\ }\textbf {\bibinfo {volume} {91}},\ \bibinfo {pages}
  {146401} (\bibinfo {year} {2003})}\BibitemShut {NoStop}%
\bibitem [{\citenamefont {B{\"u}hl}\ and\ \citenamefont
  {Kabrede}(2006)}]{Buhl:DFT2006}%
  \BibitemOpen
  \bibfield  {author} {\bibinfo {author} {\bibfnamefont {Michael}\ \bibnamefont
  {B{\"u}hl}}\ and\ \bibinfo {author} {\bibfnamefont {Hendrik}\ \bibnamefont
  {Kabrede}},\ }\bibfield  {title} {\enquote {\bibinfo {title} {Geometries of
  transition-metal complexes from density-functional theory},}\ }\href
  {\doibase 10.1021/ct6001187} {\bibfield  {journal} {\bibinfo  {journal}
  {Journal of Chemical Theory and Computation}\ }\textbf {\bibinfo {volume}
  {2}},\ \bibinfo {pages} {1282--1290} (\bibinfo {year} {2006})},\ \bibinfo
  {note} {pMID: 26626836},\ \Eprint
  {http://arxiv.org/abs/http://dx.doi.org/10.1021/ct6001187}
  {http://dx.doi.org/10.1021/ct6001187} \BibitemShut {NoStop}%
\bibitem [{\citenamefont {Sun}\ \emph {et~al.}(2016)\citenamefont {Sun},
  \citenamefont {Remsing}, \citenamefont {Zhang}, \citenamefont {Sun},
  \citenamefont {Ruzsinszky}, \citenamefont {Peng}, \citenamefont {Yang},
  \citenamefont {Paul}, \citenamefont {Waghmare}, \citenamefont {Wu} \emph
  {et~al.}}]{sun2016accurate}%
  \BibitemOpen
  \bibfield  {author} {\bibinfo {author} {\bibfnamefont {Jianwei}\ \bibnamefont
  {Sun}}, \bibinfo {author} {\bibfnamefont {Richard~C}\ \bibnamefont
  {Remsing}}, \bibinfo {author} {\bibfnamefont {Yubo}\ \bibnamefont {Zhang}},
  \bibinfo {author} {\bibfnamefont {Zhaoru}\ \bibnamefont {Sun}}, \bibinfo
  {author} {\bibfnamefont {Adrienn}\ \bibnamefont {Ruzsinszky}}, \bibinfo
  {author} {\bibfnamefont {Haowei}\ \bibnamefont {Peng}}, \bibinfo {author}
  {\bibfnamefont {Zenghui}\ \bibnamefont {Yang}}, \bibinfo {author}
  {\bibfnamefont {Arpita}\ \bibnamefont {Paul}}, \bibinfo {author}
  {\bibfnamefont {Umesh}\ \bibnamefont {Waghmare}}, \bibinfo {author}
  {\bibfnamefont {Xifan}\ \bibnamefont {Wu}},  \emph {et~al.},\ }\bibfield
  {title} {\enquote {\bibinfo {title} {Accurate first-principles structures and
  energies of diversely bonded systems from an efficient density functional},}\
  }\href {\doibase http://dx.doi.org/10.1038/nchem.2535} {\bibfield  {journal}
  {\bibinfo  {journal} {Nature Chemistry}\ } (\bibinfo {year} {2016}),\
  http://dx.doi.org/10.1038/nchem.2535}\BibitemShut {NoStop}%
\bibitem [{\citenamefont {Rappoport}\ \emph {et~al.}(2006)\citenamefont
  {Rappoport}, \citenamefont {Crawford}, \citenamefont {Furche},\ and\
  \citenamefont {Burke}}]{BurkeWhichXC}%
  \BibitemOpen
  \bibfield  {author} {\bibinfo {author} {\bibfnamefont {Dmitrij}\ \bibnamefont
  {Rappoport}}, \bibinfo {author} {\bibfnamefont {Nathan R.~M.}\ \bibnamefont
  {Crawford}}, \bibinfo {author} {\bibfnamefont {Filipp}\ \bibnamefont
  {Furche}}, \ and\ \bibinfo {author} {\bibfnamefont {Kieron}\ \bibnamefont
  {Burke}},\ }\enquote {\bibinfo {title} {Approximate density functionals:
  Which should {I} choose?}}\ in\ \href {\doibase 10.1002/0470862106.ia615}
  {\emph {\bibinfo {booktitle} {Encyclopedia of Inorganic Chemistry}}}\
  (\bibinfo  {publisher} {John Wiley and Sons, Ltd},\ \bibinfo {year}
  {2006})\BibitemShut {NoStop}%
\bibitem [{ani(2014)}]{anion-control}%
  \BibitemOpen
  \bibfield  {title} {\enquote {\bibinfo {title} {Anion-controlled new
  inorganic materials},}\ \ }(\bibinfo  {publisher} {American Chemical
  Society},\ \bibinfo {year} {2014})\BibitemShut {NoStop}%
\bibitem [{\citenamefont {Logvinovich}\ \emph {et~al.}(2010)\citenamefont
  {Logvinovich}, \citenamefont {Ebbinghaus}, \citenamefont {Reller},
  \citenamefont {Marozau}, \citenamefont {Ferri},\ and\ \citenamefont
  {Weidenkaff}}]{OpticalOxyni:2010}%
  \BibitemOpen
  \bibfield  {author} {\bibinfo {author} {\bibfnamefont {Dmitry}\ \bibnamefont
  {Logvinovich}}, \bibinfo {author} {\bibfnamefont {Stefan}\ \bibnamefont
  {Ebbinghaus}}, \bibinfo {author} {\bibfnamefont {Armin}\ \bibnamefont
  {Reller}}, \bibinfo {author} {\bibfnamefont {Ivan}\ \bibnamefont {Marozau}},
  \bibinfo {author} {\bibfnamefont {Davide}\ \bibnamefont {Ferri}}, \ and\
  \bibinfo {author} {\bibfnamefont {Anke}\ \bibnamefont {Weidenkaff}},\
  }\bibfield  {title} {\enquote {\bibinfo {title} {Synthesis, crystal structure
  and optical properties of {LaNbON$_2$}�},}\ }\href {\doibase
  10.1002/zaac.201000067} {\bibfield  {journal} {\bibinfo  {journal}
  {Zeitschrift f{\"u}r anorganische und allgemeine Chemie}\ }\textbf {\bibinfo
  {volume} {636}},\ \bibinfo {pages} {905--912} (\bibinfo {year}
  {2010})}\BibitemShut {NoStop}%
\bibitem [{\citenamefont {Yang}\ \emph {et~al.}(2011)\citenamefont {Yang},
  \citenamefont {Or\`o-Sol\`e}, \citenamefont {Rodgers}, \citenamefont {Jorge},
  \citenamefont {Fuertes},\ and\ \citenamefont
  {Attfield}}]{Yang/Attfield:2011}%
  \BibitemOpen
  \bibfield  {author} {\bibinfo {author} {\bibfnamefont {Minghui}\ \bibnamefont
  {Yang}}, \bibinfo {author} {\bibfnamefont {Judith}\ \bibnamefont
  {Or\`o-Sol\`e}}, \bibinfo {author} {\bibfnamefont {Jennifer~A.}\ \bibnamefont
  {Rodgers}}, \bibinfo {author} {\bibfnamefont {Ana~Bel\`en}\ \bibnamefont
  {Jorge}}, \bibinfo {author} {\bibfnamefont {Amparo}\ \bibnamefont {Fuertes}},
  \ and\ \bibinfo {author} {\bibfnamefont {J.~Paul}\ \bibnamefont {Attfield}},\
  }\bibfield  {title} {\enquote {\bibinfo {title} {Anion order in perovskite
  oxynitrides},}\ }\href {http://dx.doi.org/10.1038/nchem.908} {\bibfield
  {journal} {\bibinfo  {journal} {Nat.\ Chem.}\ }\textbf {\bibinfo {volume}
  {3}},\ \bibinfo {pages} {47--52} (\bibinfo {year} {2011})}\BibitemShut
  {NoStop}%
\bibitem [{\citenamefont {Jorge}\ \emph {et~al.}(2008)\citenamefont {Jorge},
  \citenamefont {Or\`o-Sol\`e}, \citenamefont {Bea}, \citenamefont {Mufti},
  \citenamefont {Palstra}, \citenamefont {Rodgers}, \citenamefont {Attfield},\
  and\ \citenamefont {Fuertes}}]{Fuertes:2008}%
  \BibitemOpen
  \bibfield  {author} {\bibinfo {author} {\bibfnamefont {A.~Bel�n}\
  \bibnamefont {Jorge}}, \bibinfo {author} {\bibfnamefont {Judith}\
  \bibnamefont {Or\`o-Sol\`e}}, \bibinfo {author} {\bibfnamefont {Ana~M.}\
  \bibnamefont {Bea}}, \bibinfo {author} {\bibfnamefont {Nandang}\ \bibnamefont
  {Mufti}}, \bibinfo {author} {\bibfnamefont {Thomas T.~M.}\ \bibnamefont
  {Palstra}}, \bibinfo {author} {\bibfnamefont {Jennifer~A.}\ \bibnamefont
  {Rodgers}}, \bibinfo {author} {\bibfnamefont {J.~Paul}\ \bibnamefont
  {Attfield}}, \ and\ \bibinfo {author} {\bibfnamefont {Amparo}\ \bibnamefont
  {Fuertes}},\ }\bibfield  {title} {\enquote {\bibinfo {title} {Large coupled
  magnetoresponses in {EuNbO$_2$N}},}\ }\href {\doibase 10.1021/ja804139g}
  {\bibfield  {journal} {\bibinfo  {journal} {Journal of the American Chemical
  Society}\ }\textbf {\bibinfo {volume} {130}},\ \bibinfo {pages}
  {12572--12573} (\bibinfo {year} {2008})},\ \bibinfo {note} {pMID: 18759396},\
  \Eprint {http://arxiv.org/abs/http://dx.doi.org/10.1021/ja804139g}
  {http://dx.doi.org/10.1021/ja804139g} \BibitemShut {NoStop}%
\bibitem [{\citenamefont {Johnston}(2010)}]{Johnston:2010}%
  \BibitemOpen
  \bibfield  {author} {\bibinfo {author} {\bibfnamefont {David~C.}\
  \bibnamefont {Johnston}},\ }\bibfield  {title} {\enquote {\bibinfo {title}
  {The puzzle of high temperature superconductivity in layered iron pnictides
  and chalcogenides},}\ }\href {\doibase 10.1080/00018732.2010.513480}
  {\bibfield  {journal} {\bibinfo  {journal} {Advances in Physics}\ }\textbf
  {\bibinfo {volume} {59}},\ \bibinfo {pages} {803--1061} (\bibinfo {year}
  {2010})}\BibitemShut {NoStop}%
\bibitem [{\citenamefont {Tressaud}\ and\ \citenamefont
  {Poeppelmeier}(2016)}]{tagkey2016iii}%
  \BibitemOpen
  \bibinfo {editor} {\bibfnamefont {Alain}\ \bibnamefont {Tressaud}}\ and\
  \bibinfo {editor} {\bibfnamefont {Kenneth}\ \bibnamefont {Poeppelmeier}},\
  eds.,\ \href {\doibase http://dx.doi.org/10.1016/B978-0-12-801639-8.01001-1}
  {\emph {\bibinfo {title} {Photonic and Electronic Properties of Fluoride
  Materials}}}\ (\bibinfo  {publisher} {Elsevier},\ \bibinfo {address}
  {Boston},\ \bibinfo {year} {2016})\BibitemShut {NoStop}%
\bibitem [{\citenamefont {Chevrier}\ \emph {et~al.}(2013)\citenamefont
  {Chevrier}, \citenamefont {Hautier}, \citenamefont {Ong}, \citenamefont
  {Doe},\ and\ \citenamefont {Ceder}}]{Ceder_FeOF2}%
  \BibitemOpen
  \bibfield  {author} {\bibinfo {author} {\bibfnamefont {Vincent~L.}\
  \bibnamefont {Chevrier}}, \bibinfo {author} {\bibfnamefont {Geoffroy}\
  \bibnamefont {Hautier}}, \bibinfo {author} {\bibfnamefont {Shyue~Ping}\
  \bibnamefont {Ong}}, \bibinfo {author} {\bibfnamefont {Robert~E.}\
  \bibnamefont {Doe}}, \ and\ \bibinfo {author} {\bibfnamefont {Gerbrand}\
  \bibnamefont {Ceder}},\ }\bibfield  {title} {\enquote {\bibinfo {title}
  {First-principles study of iron oxyfluorides and lithiation of {FeOF}},}\
  }\href {\doibase 10.1103/PhysRevB.87.094118} {\bibfield  {journal} {\bibinfo
  {journal} {Phys. Rev. B}\ }\textbf {\bibinfo {volume} {87}},\ \bibinfo
  {pages} {094118} (\bibinfo {year} {2013})}\BibitemShut {NoStop}%
\bibitem [{\citenamefont {Incorvati}\ \emph {et~al.}(2016)\citenamefont
  {Incorvati}, \citenamefont {Wan}, \citenamefont {Key}, \citenamefont {Zhou},
  \citenamefont {Liao}, \citenamefont {Fuoco}, \citenamefont {Holland},
  \citenamefont {Wang}, \citenamefont {Prendergast}, \citenamefont
  {Poeppelmeier},\ and\ \citenamefont {Vaughey}}]{Jared:MolliOxyfluo}%
  \BibitemOpen
  \bibfield  {author} {\bibinfo {author} {\bibfnamefont {Jared~T.}\
  \bibnamefont {Incorvati}}, \bibinfo {author} {\bibfnamefont {Liwen~F.}\
  \bibnamefont {Wan}}, \bibinfo {author} {\bibfnamefont {Baris}\ \bibnamefont
  {Key}}, \bibinfo {author} {\bibfnamefont {Dehua}\ \bibnamefont {Zhou}},
  \bibinfo {author} {\bibfnamefont {Chen}\ \bibnamefont {Liao}}, \bibinfo
  {author} {\bibfnamefont {Lindsay}\ \bibnamefont {Fuoco}}, \bibinfo {author}
  {\bibfnamefont {Michael}\ \bibnamefont {Holland}}, \bibinfo {author}
  {\bibfnamefont {Hao}\ \bibnamefont {Wang}}, \bibinfo {author} {\bibfnamefont
  {David}\ \bibnamefont {Prendergast}}, \bibinfo {author} {\bibfnamefont
  {Kenneth~R.}\ \bibnamefont {Poeppelmeier}}, \ and\ \bibinfo {author}
  {\bibfnamefont {John~T.}\ \bibnamefont {Vaughey}},\ }\bibfield  {title}
  {\enquote {\bibinfo {title} {Reversible magnesium intercalation into a
  layered oxyfluoride cathode},}\ }\href {\doibase
  10.1021/acs.chemmater.5b02746} {\bibfield  {journal} {\bibinfo  {journal}
  {Chemistry of Materials}\ }\textbf {\bibinfo {volume} {28}},\ \bibinfo
  {pages} {17--20} (\bibinfo {year} {2016})},\ \Eprint
  {http://arxiv.org/abs/http://dx.doi.org/10.1021/acs.chemmater.5b02746}
  {http://dx.doi.org/10.1021/acs.chemmater.5b02746} \BibitemShut {NoStop}%
\bibitem [{\citenamefont {Setlur}\ \emph {et~al.}(2010)\citenamefont {Setlur},
  \citenamefont {Radkov}, \citenamefont {Henderson}, \citenamefont {Her},
  \citenamefont {Srivastava}, \citenamefont {Karkada}, \citenamefont {Kishore},
  \citenamefont {Kumar}, \citenamefont {Aesram}, \citenamefont {Deshpande},
  \citenamefont {Kolodin}, \citenamefont {Grigorov},\ and\ \citenamefont
  {Happek}}]{phosphor:Anant}%
  \BibitemOpen
  \bibfield  {author} {\bibinfo {author} {\bibfnamefont {Anant~A.}\
  \bibnamefont {Setlur}}, \bibinfo {author} {\bibfnamefont {Emil~V.}\
  \bibnamefont {Radkov}}, \bibinfo {author} {\bibfnamefont {Claire~S.}\
  \bibnamefont {Henderson}}, \bibinfo {author} {\bibfnamefont {Jae-Hyuk}\
  \bibnamefont {Her}}, \bibinfo {author} {\bibfnamefont {Alok~M.}\ \bibnamefont
  {Srivastava}}, \bibinfo {author} {\bibfnamefont {Nagaveni}\ \bibnamefont
  {Karkada}}, \bibinfo {author} {\bibfnamefont {M.~Satya}\ \bibnamefont
  {Kishore}}, \bibinfo {author} {\bibfnamefont {N.~Prasanth}\ \bibnamefont
  {Kumar}}, \bibinfo {author} {\bibfnamefont {Danny}\ \bibnamefont {Aesram}},
  \bibinfo {author} {\bibfnamefont {Anirudha}\ \bibnamefont {Deshpande}},
  \bibinfo {author} {\bibfnamefont {Boris}\ \bibnamefont {Kolodin}}, \bibinfo
  {author} {\bibfnamefont {Ljudmil~S.}\ \bibnamefont {Grigorov}}, \ and\
  \bibinfo {author} {\bibfnamefont {Uwe}\ \bibnamefont {Happek}},\ }\bibfield
  {title} {\enquote {\bibinfo {title} {Energy-efficient, high-color-rendering
  {LED} lamps using oxyfluoride and fluoride phosphors},}\ }\href {\doibase
  10.1021/cm100960g} {\bibfield  {journal} {\bibinfo  {journal} {Chemistry of
  Materials}\ }\textbf {\bibinfo {volume} {22}},\ \bibinfo {pages} {4076--4082}
  (\bibinfo {year} {2010})},\ \Eprint
  {http://arxiv.org/abs/http://dx.doi.org/10.1021/cm100960g}
  {http://dx.doi.org/10.1021/cm100960g} \BibitemShut {NoStop}%
\bibitem [{\citenamefont {Im}\ \emph {et~al.}(2010)\citenamefont {Im},
  \citenamefont {Brinkley}, \citenamefont {Hu}, \citenamefont {Mikhailovsky},
  \citenamefont {DenBaars},\ and\ \citenamefont {Seshadri}}]{Ram:Phosphor2010}%
  \BibitemOpen
  \bibfield  {author} {\bibinfo {author} {\bibfnamefont {Won~Bin}\ \bibnamefont
  {Im}}, \bibinfo {author} {\bibfnamefont {Stuart}\ \bibnamefont {Brinkley}},
  \bibinfo {author} {\bibfnamefont {Jerry}\ \bibnamefont {Hu}}, \bibinfo
  {author} {\bibfnamefont {Alexander}\ \bibnamefont {Mikhailovsky}}, \bibinfo
  {author} {\bibfnamefont {Steven~P.}\ \bibnamefont {DenBaars}}, \ and\
  \bibinfo {author} {\bibfnamefont {Ram}\ \bibnamefont {Seshadri}},\ }\bibfield
   {title} {\enquote {\bibinfo {title}
  {{Sr$_{2.975-x}$Ba$_x$Ce$_{0.025}$AlO$_4$F}: a highly efficient
  green-emitting oxyfluoride phosphor for solid state white lighting},}\ }\href
  {\doibase 10.1021/cm100010z} {\bibfield  {journal} {\bibinfo  {journal}
  {Chemistry of Materials}\ }\textbf {\bibinfo {volume} {22}},\ \bibinfo
  {pages} {2842--2849} (\bibinfo {year} {2010})},\ \Eprint
  {http://arxiv.org/abs/http://dx.doi.org/10.1021/cm100010z}
  {http://dx.doi.org/10.1021/cm100010z} \BibitemShut {NoStop}%
\bibitem [{\citenamefont {Atuchin}\ \emph {et~al.}(2012)\citenamefont
  {Atuchin}, \citenamefont {Isaenko}, \citenamefont {Kesler}, \citenamefont
  {Lin}, \citenamefont {Molokeev}, \citenamefont {Yelisseyev},\ and\
  \citenamefont {Zhurkov}}]{Atuchin2012159}%
  \BibitemOpen
  \bibfield  {author} {\bibinfo {author} {\bibfnamefont {V.V.}\ \bibnamefont
  {Atuchin}}, \bibinfo {author} {\bibfnamefont {L.I.}\ \bibnamefont {Isaenko}},
  \bibinfo {author} {\bibfnamefont {V.G.}\ \bibnamefont {Kesler}}, \bibinfo
  {author} {\bibfnamefont {Z.S.}\ \bibnamefont {Lin}}, \bibinfo {author}
  {\bibfnamefont {M.S.}\ \bibnamefont {Molokeev}}, \bibinfo {author}
  {\bibfnamefont {A.P.}\ \bibnamefont {Yelisseyev}}, \ and\ \bibinfo {author}
  {\bibfnamefont {S.A.}\ \bibnamefont {Zhurkov}},\ }\bibfield  {title}
  {\enquote {\bibinfo {title} {Exploration on anion ordering, optical
  properties and electronic structure in {K$_3$WO$_3$F$_3$} elpasolite},}\
  }\href {\doibase http://dx.doi.org/10.1016/j.jssc.2011.12.037} {\bibfield
  {journal} {\bibinfo  {journal} {Journal of Solid State Chemistry}\ }\textbf
  {\bibinfo {volume} {187}},\ \bibinfo {pages} {159 -- 164} (\bibinfo {year}
  {2012})}\BibitemShut {NoStop}%
\bibitem [{\citenamefont {Izumi}\ \emph {et~al.}(2005)\citenamefont {Izumi},
  \citenamefont {Kirsch}, \citenamefont {Stern},\ and\ \citenamefont
  {Poeppelmeier}}]{Izumi/Poeppelmeier:2005}%
  \BibitemOpen
  \bibfield  {author} {\bibinfo {author} {\bibfnamefont {Heather~K.}\
  \bibnamefont {Izumi}}, \bibinfo {author} {\bibfnamefont {Janet~E.}\
  \bibnamefont {Kirsch}}, \bibinfo {author} {\bibfnamefont {Charlotte~L.}\
  \bibnamefont {Stern}}, \ and\ \bibinfo {author} {\bibfnamefont {Kenneth~R.}\
  \bibnamefont {Poeppelmeier}},\ }\bibfield  {title} {\enquote {\bibinfo
  {title} {{Examining the Out-of-Center Distortion in the [NbOF$_5$]$^{2-}$
  Anion}},}\ }\href@noop {} {\bibfield  {journal} {\bibinfo  {journal}
  {Inorganic Chemistry}\ }\textbf {\bibinfo {volume} {44}},\ \bibinfo {pages}
  {884--895} (\bibinfo {year} {2005})}\BibitemShut {NoStop}%
\bibitem [{\citenamefont {de~la Cruz}\ \emph {et~al.}(2008)\citenamefont {de~la
  Cruz}, \citenamefont {Huang}, \citenamefont {Lynn}, \citenamefont {Li},
  \citenamefont {II}, \citenamefont {Zarestky}, \citenamefont {Mook},
  \citenamefont {Chen}, \citenamefont {Luo}, \citenamefont {Wang},\ and\
  \citenamefont {Dai}}]{delaCruz/Pengcheng_et_al:2008}%
  \BibitemOpen
  \bibfield  {author} {\bibinfo {author} {\bibfnamefont {Clarina}\ \bibnamefont
  {de~la Cruz}}, \bibinfo {author} {\bibfnamefont {Q.}~\bibnamefont {Huang}},
  \bibinfo {author} {\bibfnamefont {J.~W.}\ \bibnamefont {Lynn}}, \bibinfo
  {author} {\bibfnamefont {Jiying}\ \bibnamefont {Li}}, \bibinfo {author}
  {\bibfnamefont {W.~Ratcliff}\ \bibnamefont {II}}, \bibinfo {author}
  {\bibfnamefont {J.~L.}\ \bibnamefont {Zarestky}}, \bibinfo {author}
  {\bibfnamefont {H.~A.}\ \bibnamefont {Mook}}, \bibinfo {author}
  {\bibfnamefont {G.~F.}\ \bibnamefont {Chen}}, \bibinfo {author}
  {\bibfnamefont {J.~L.}\ \bibnamefont {Luo}}, \bibinfo {author} {\bibfnamefont
  {N.~L.}\ \bibnamefont {Wang}}, \ and\ \bibinfo {author} {\bibfnamefont
  {Pengcheng}\ \bibnamefont {Dai}},\ }\bibfield  {title} {\enquote {\bibinfo
  {title} {{Magnetic order close to superconductivity in the iron-based layered
  LaO$_{1-x}$F$_x$FeAs systems}},}\ }\href@noop {} {\bibfield  {journal}
  {\bibinfo  {journal} {Nature}\ }\textbf {\bibinfo {volume} {453}},\ \bibinfo
  {pages} {899--902} (\bibinfo {year} {2008})}\BibitemShut {NoStop}%
\bibitem [{\citenamefont {Kamihara}\ \emph {et~al.}(2008)\citenamefont
  {Kamihara}, \citenamefont {Watanabe}, \citenamefont {Hirano},\ and\
  \citenamefont {Hosono}}]{Hosono/FeSC:2008}%
  \BibitemOpen
  \bibfield  {author} {\bibinfo {author} {\bibfnamefont {Yoichi}\ \bibnamefont
  {Kamihara}}, \bibinfo {author} {\bibfnamefont {Takumi}\ \bibnamefont
  {Watanabe}}, \bibinfo {author} {\bibfnamefont {Masahiro}\ \bibnamefont
  {Hirano}}, \ and\ \bibinfo {author} {\bibfnamefont {Hideo}\ \bibnamefont
  {Hosono}},\ }\bibfield  {title} {\enquote {\bibinfo {title} {{Iron-Based
  Layered Superconductor LaO$\left[_{1-x}\textrm{F}_x\right]$FeAs ($x =
  0.05-0.12$) with $T_c$ = 26 K}},}\ }\href@noop {} {\bibfield  {journal}
  {\bibinfo  {journal} {Journal of the American Chemical Society}\ }\textbf
  {\bibinfo {volume} {130}},\ \bibinfo {pages} {3296--3297} (\bibinfo {year}
  {2008})}\BibitemShut {NoStop}%
\bibitem [{\citenamefont {Shannon}\ and\ \citenamefont
  {Prewitt}(1970)}]{Shannon/Prewitt:1969}%
  \BibitemOpen
  \bibfield  {author} {\bibinfo {author} {\bibfnamefont {R.~D.}\ \bibnamefont
  {Shannon}}\ and\ \bibinfo {author} {\bibfnamefont {C.~T.}\ \bibnamefont
  {Prewitt}},\ }\bibfield  {title} {\enquote {\bibinfo {title} {{Revised values
  of effective ionic radii}},}\ }\href {\doibase 10.1107/S0567740870003576}
  {\bibfield  {journal} {\bibinfo  {journal} {Acta Crystallographica Section
  B}\ }\textbf {\bibinfo {volume} {26}},\ \bibinfo {pages} {1046--1048}
  (\bibinfo {year} {1970})}\BibitemShut {NoStop}%
\bibitem [{\citenamefont {Withers}\ \emph {et~al.}(2007)\citenamefont
  {Withers}, \citenamefont {Brink}, \citenamefont {Liu},\ and\ \citenamefont
  {Nor{\'e}n}}]{withers2007cluster}%
  \BibitemOpen
  \bibfield  {author} {\bibinfo {author} {\bibfnamefont {Ray~L}\ \bibnamefont
  {Withers}}, \bibinfo {author} {\bibfnamefont {FJ}~\bibnamefont {Brink}},
  \bibinfo {author} {\bibfnamefont {Yun}\ \bibnamefont {Liu}}, \ and\ \bibinfo
  {author} {\bibfnamefont {Lasse}\ \bibnamefont {Nor{\'e}n}},\ }\bibfield
  {title} {\enquote {\bibinfo {title} {Cluster chemistry in the solid state:
  Structured diffuse scattering, oxide/fluoride ordering and polar behaviour in
  transition metal oxyfluorides},}\ }\href@noop {} {\bibfield  {journal}
  {\bibinfo  {journal} {Polyhedron}\ }\textbf {\bibinfo {volume} {26}},\
  \bibinfo {pages} {290--299} (\bibinfo {year} {2007})}\BibitemShut {NoStop}%
\bibitem [{\citenamefont {Kresse}\ and\ \citenamefont
  {Furthm\"uller}(1996)}]{Kresse/Furthmuller:1996a}%
  \BibitemOpen
  \bibfield  {author} {\bibinfo {author} {\bibfnamefont {G.}~\bibnamefont
  {Kresse}}\ and\ \bibinfo {author} {\bibfnamefont {J.}~\bibnamefont
  {Furthm\"uller}},\ }\bibfield  {title} {\enquote {\bibinfo {title} {Efficient
  iterative schemes for \emph{ab initio} total-energy calculations using a
  plane-wave basis set},}\ }\href {\doibase 10.1103/PhysRevB.54.11169}
  {\bibfield  {journal} {\bibinfo  {journal} {Phys. Rev. B}\ }\textbf {\bibinfo
  {volume} {54}},\ \bibinfo {pages} {11169--11186} (\bibinfo {year}
  {1996})}\BibitemShut {NoStop}%
\bibitem [{\citenamefont {Kresse}\ and\ \citenamefont
  {Joubert}(1999)}]{Kresse/Joubert:1999}%
  \BibitemOpen
  \bibfield  {author} {\bibinfo {author} {\bibfnamefont {G.}~\bibnamefont
  {Kresse}}\ and\ \bibinfo {author} {\bibfnamefont {D.}~\bibnamefont
  {Joubert}},\ }\bibfield  {title} {\enquote {\bibinfo {title} {From ultrasoft
  pseudopotentials to the projector augmented-wave method},}\ }\href {\doibase
  10.1103/PhysRevB.59.1758} {\bibfield  {journal} {\bibinfo  {journal} {Phys.
  Rev. B}\ }\textbf {\bibinfo {volume} {59}},\ \bibinfo {pages} {1758--1775}
  (\bibinfo {year} {1999})}\BibitemShut {NoStop}%
\bibitem [{\citenamefont {Bl\"ochl}(1994)}]{Blochl:1994}%
  \BibitemOpen
  \bibfield  {author} {\bibinfo {author} {\bibfnamefont {P.~E.}\ \bibnamefont
  {Bl\"ochl}},\ }\bibfield  {title} {\enquote {\bibinfo {title} {Projector
  augmented-wave method},}\ }\href {\doibase 10.1103/PhysRevB.50.17953}
  {\bibfield  {journal} {\bibinfo  {journal} {Phys. Rev. B}\ }\textbf {\bibinfo
  {volume} {50}},\ \bibinfo {pages} {17953--17979} (\bibinfo {year}
  {1994})}\BibitemShut {NoStop}%
\bibitem [{Sup()}]{Supplmental_Note:Benchmark}%
  \BibitemOpen
  \href@noop {} {\ }\bibinfo {note} {See Supplemental Material at [URL will be
  inserted by publisher] for additional computational details, crystal
  structure information, and GII calculations.}\BibitemShut {Stop}%
\bibitem [{\citenamefont {Monkhorst}\ and\ \citenamefont
  {Pack}(1976)}]{Monkhorst/Pack:1976}%
  \BibitemOpen
  \bibfield  {author} {\bibinfo {author} {\bibfnamefont {Hendrik~J.}\
  \bibnamefont {Monkhorst}}\ and\ \bibinfo {author} {\bibfnamefont {James~D.}\
  \bibnamefont {Pack}},\ }\bibfield  {title} {\enquote {\bibinfo {title}
  {{Special points for Brillouin-zone integrations}},}\ }\href {\doibase
  10.1103/PhysRevB.13.5188} {\bibfield  {journal} {\bibinfo  {journal}
  {Physical Review B}\ }\textbf {\bibinfo {volume} {13}},\ \bibinfo {pages}
  {5188--5192} (\bibinfo {year} {1976})}\BibitemShut {NoStop}%
\bibitem [{\citenamefont {Togo}\ \emph {et~al.}(2008)\citenamefont {Togo},
  \citenamefont {Oba},\ and\ \citenamefont {Tanaka}}]{Phonopy:2008}%
  \BibitemOpen
  \bibfield  {author} {\bibinfo {author} {\bibfnamefont {A}~\bibnamefont
  {Togo}}, \bibinfo {author} {\bibfnamefont {F}~\bibnamefont {Oba}}, \ and\
  \bibinfo {author} {\bibfnamefont {I}~\bibnamefont {Tanaka}},\ }\bibfield
  {title} {\enquote {\bibinfo {title} {{First-principles calculations of the
  ferroelastic transition between rutile-type and CaCl$_2$-type SiO$_2$ at high
  pressures}},}\ }\href@noop {} {\bibfield  {journal} {\bibinfo  {journal}
  {Physical Review B}\ }\textbf {\bibinfo {volume} {78}},\ \bibinfo {pages}
  {134106} (\bibinfo {year} {2008})}\BibitemShut {NoStop}%
\bibitem [{\citenamefont {Fuchs}\ \emph {et~al.}(1998)\citenamefont {Fuchs},
  \citenamefont {Bockstedte}, \citenamefont {Pehlke},\ and\ \citenamefont
  {Scheffler}}]{Fuchs_PAW_LDAvGGA}%
  \BibitemOpen
  \bibfield  {author} {\bibinfo {author} {\bibfnamefont {M.}~\bibnamefont
  {Fuchs}}, \bibinfo {author} {\bibfnamefont {M.}~\bibnamefont {Bockstedte}},
  \bibinfo {author} {\bibfnamefont {E.}~\bibnamefont {Pehlke}}, \ and\ \bibinfo
  {author} {\bibfnamefont {M.}~\bibnamefont {Scheffler}},\ }\bibfield  {title}
  {\enquote {\bibinfo {title} {Pseudopotential study of binding properties of
  solids within generalized gradient approximations: The role of core-valence
  exchange correlation},}\ }\href {\doibase 10.1103/PhysRevB.57.2134}
  {\bibfield  {journal} {\bibinfo  {journal} {Phys. Rev. B}\ }\textbf {\bibinfo
  {volume} {57}},\ \bibinfo {pages} {2134--2145} (\bibinfo {year}
  {1998})}\BibitemShut {NoStop}%
\bibitem [{\citenamefont {Perdew}\ and\ \citenamefont
  {Schmidt}(2001)}]{Jacobsladder}%
  \BibitemOpen
  \bibfield  {author} {\bibinfo {author} {\bibfnamefont {John~P.}\ \bibnamefont
  {Perdew}}\ and\ \bibinfo {author} {\bibfnamefont {Karla}\ \bibnamefont
  {Schmidt}},\ }\bibfield  {title} {\enquote {\bibinfo {title} {Jacob's ladder
  of density functional approximations for the exchange-correlation energy},}\
  }\href {\doibase http://dx.doi.org/10.1063/1.1390175} {\bibfield  {journal}
  {\bibinfo  {journal} {AIP Conference Proceedings}\ }\textbf {\bibinfo
  {volume} {577}},\ \bibinfo {pages} {1--20} (\bibinfo {year}
  {2001})}\BibitemShut {NoStop}%
\bibitem [{\citenamefont {Kohn}\ and\ \citenamefont
  {Sham}(1965)}]{Kohn/Sham:1965}%
  \BibitemOpen
  \bibfield  {author} {\bibinfo {author} {\bibfnamefont {W.}~\bibnamefont
  {Kohn}}\ and\ \bibinfo {author} {\bibfnamefont {L.~J.}\ \bibnamefont
  {Sham}},\ }\bibfield  {title} {\enquote {\bibinfo {title} {Self-consistent
  equations including exchange and correlation effects},}\ }\href@noop {}
  {\bibfield  {journal} {\bibinfo  {journal} {Physical Review}\ }\textbf
  {\bibinfo {volume} {140}},\ \bibinfo {pages} {A1133--A1138} (\bibinfo {year}
  {1965})}\BibitemShut {NoStop}%
\bibitem [{\citenamefont {Wahl}\ \emph {et~al.}(2008)\citenamefont {Wahl},
  \citenamefont {Vogtenhuber},\ and\ \citenamefont {Kresse}}]{Kresse_STO-HSE}%
  \BibitemOpen
  \bibfield  {author} {\bibinfo {author} {\bibfnamefont {Roman}\ \bibnamefont
  {Wahl}}, \bibinfo {author} {\bibfnamefont {Doris}\ \bibnamefont
  {Vogtenhuber}}, \ and\ \bibinfo {author} {\bibfnamefont {Georg}\ \bibnamefont
  {Kresse}},\ }\bibfield  {title} {\enquote {\bibinfo {title} {{SrTiO$_3$ and
  BaTiO$_3$ revisited using the projector augmented wave method: Performance of
  hybrid and semilocal functionals}},}\ }\href {\doibase
  10.1103/PhysRevB.78.104116} {\bibfield  {journal} {\bibinfo  {journal} {Phys.
  Rev. B}\ }\textbf {\bibinfo {volume} {78}},\ \bibinfo {pages} {104116}
  (\bibinfo {year} {2008})}\BibitemShut {NoStop}%
\bibitem [{\citenamefont {Perdew}\ and\ \citenamefont
  {Yue}(1986)}]{Perdew/Wang:1986}%
  \BibitemOpen
  \bibfield  {author} {\bibinfo {author} {\bibfnamefont {John~P.}\ \bibnamefont
  {Perdew}}\ and\ \bibinfo {author} {\bibfnamefont {Wang}\ \bibnamefont
  {Yue}},\ }\bibfield  {title} {\enquote {\bibinfo {title} {Accurate and simple
  density functional for the electronic exchange energy: Generalized gradient
  approximation},}\ }\href {\doibase 10.1103/PhysRevB.33.8800} {\bibfield
  {journal} {\bibinfo  {journal} {Physical Review B}\ }\textbf {\bibinfo
  {volume} {33}},\ \bibinfo {pages} {8800--8802} (\bibinfo {year}
  {1986})}\BibitemShut {NoStop}%
\bibitem [{\citenamefont {Becke}(1992)}]{Becke:GGA1992}%
  \BibitemOpen
  \bibfield  {author} {\bibinfo {author} {\bibfnamefont {Axel~D.}\ \bibnamefont
  {Becke}},\ }\bibfield  {title} {\enquote {\bibinfo {title} {{Density
  functional thermochemistry. I. The effect of the exchange only gradient
  correction}},}\ }\href {\doibase http://dx.doi.org/10.1063/1.462066}
  {\bibfield  {journal} {\bibinfo  {journal} {The Journal of Chemical Physics}\
  }\textbf {\bibinfo {volume} {96}},\ \bibinfo {pages} {2155--2160} (\bibinfo
  {year} {1992})}\BibitemShut {NoStop}%
\bibitem [{\citenamefont {Perdew}\ \emph {et~al.}(1992)\citenamefont {Perdew},
  \citenamefont {Chevary}, \citenamefont {Vosko}, \citenamefont {Jackson},
  \citenamefont {Pederson}, \citenamefont {Singh},\ and\ \citenamefont
  {Fiolhais}}]{PhysRevB.46.6671}%
  \BibitemOpen
  \bibfield  {author} {\bibinfo {author} {\bibfnamefont {John~P.}\ \bibnamefont
  {Perdew}}, \bibinfo {author} {\bibfnamefont {J.~A.}\ \bibnamefont {Chevary}},
  \bibinfo {author} {\bibfnamefont {S.~H.}\ \bibnamefont {Vosko}}, \bibinfo
  {author} {\bibfnamefont {Koblar~A.}\ \bibnamefont {Jackson}}, \bibinfo
  {author} {\bibfnamefont {Mark~R.}\ \bibnamefont {Pederson}}, \bibinfo
  {author} {\bibfnamefont {D.~J.}\ \bibnamefont {Singh}}, \ and\ \bibinfo
  {author} {\bibfnamefont {Carlos}\ \bibnamefont {Fiolhais}},\ }\bibfield
  {title} {\enquote {\bibinfo {title} {Atoms, molecules, solids, and surfaces:
  Applications of the generalized gradient approximation for exchange and
  correlation},}\ }\href {\doibase 10.1103/PhysRevB.46.6671} {\bibfield
  {journal} {\bibinfo  {journal} {Phys. Rev. B}\ }\textbf {\bibinfo {volume}
  {46}},\ \bibinfo {pages} {6671--6687} (\bibinfo {year} {1992})}\BibitemShut
  {NoStop}%
\bibitem [{\citenamefont {Perdew}\ \emph
  {et~al.}(1996{\natexlab{a}})\citenamefont {Perdew}, \citenamefont {Burke},\
  and\ \citenamefont {Ernzerhof}}]{Perdew/Burke/Ernzerhof:1996}%
  \BibitemOpen
  \bibfield  {author} {\bibinfo {author} {\bibfnamefont {John~P.}\ \bibnamefont
  {Perdew}}, \bibinfo {author} {\bibfnamefont {Kieron}\ \bibnamefont {Burke}},
  \ and\ \bibinfo {author} {\bibfnamefont {Matthias}\ \bibnamefont
  {Ernzerhof}},\ }\bibfield  {title} {\enquote {\bibinfo {title} {Generalized
  gradient approximation made simple},}\ }\href {\doibase
  10.1103/PhysRevLett.77.3865} {\bibfield  {journal} {\bibinfo  {journal}
  {Physical Review Letters}\ }\textbf {\bibinfo {volume} {77}},\ \bibinfo
  {pages} {3865--3868} (\bibinfo {year} {1996}{\natexlab{a}})}\BibitemShut
  {NoStop}%
\bibitem [{\citenamefont {Burke}(2012)}]{Perspec:Burke12}%
  \BibitemOpen
  \bibfield  {author} {\bibinfo {author} {\bibfnamefont {K.}~\bibnamefont
  {Burke}},\ }\bibfield  {title} {\enquote {\bibinfo {title} {Perspective on
  density functional theory},}\ }\href
  {http://link.aip.org/link/doi/10.1063/1.4704546} {\bibfield  {journal}
  {\bibinfo  {journal} {J. Chem. Phys.}\ }\textbf {\bibinfo {volume} {136}},\
  \bibinfo {pages} {150901} (\bibinfo {year} {2012})}\BibitemShut {NoStop}%
\bibitem [{\citenamefont {Perdew}\ \emph {et~al.}(2008)\citenamefont {Perdew},
  \citenamefont {Ruzsinszky}, \citenamefont {Csonka}, \citenamefont {Vydrov},
  \citenamefont {Scuseria}, \citenamefont {Constantin}, \citenamefont {Zhou},\
  and\ \citenamefont {Burke}}]{PBEsol:2008}%
  \BibitemOpen
  \bibfield  {author} {\bibinfo {author} {\bibfnamefont {John~P.}\ \bibnamefont
  {Perdew}}, \bibinfo {author} {\bibfnamefont {Adrienn}\ \bibnamefont
  {Ruzsinszky}}, \bibinfo {author} {\bibfnamefont {G\'abor~I.}\ \bibnamefont
  {Csonka}}, \bibinfo {author} {\bibfnamefont {Oleg~A.}\ \bibnamefont
  {Vydrov}}, \bibinfo {author} {\bibfnamefont {Gustavo~E.}\ \bibnamefont
  {Scuseria}}, \bibinfo {author} {\bibfnamefont {Lucian~A.}\ \bibnamefont
  {Constantin}}, \bibinfo {author} {\bibfnamefont {Xiaolan}\ \bibnamefont
  {Zhou}}, \ and\ \bibinfo {author} {\bibfnamefont {Kieron}\ \bibnamefont
  {Burke}},\ }\bibfield  {title} {\enquote {\bibinfo {title} {Restoring the
  density-gradient expansion for exchange in solids and surfaces},}\ }\href
  {\doibase 10.1103/PhysRevLett.100.136406} {\bibfield  {journal} {\bibinfo
  {journal} {Phys. Rev. Lett.}\ }\textbf {\bibinfo {volume} {100}},\ \bibinfo
  {pages} {136406} (\bibinfo {year} {2008})}\BibitemShut {NoStop}%
\bibitem [{\citenamefont {Wu}\ and\ \citenamefont
  {Cohen}(2006)}]{Wu/Cohen:2006}%
  \BibitemOpen
  \bibfield  {author} {\bibinfo {author} {\bibfnamefont {Zhigang}\ \bibnamefont
  {Wu}}\ and\ \bibinfo {author} {\bibfnamefont {R.~E.}\ \bibnamefont {Cohen}},\
  }\bibfield  {title} {\enquote {\bibinfo {title} {More accurate generalized
  gradient approximation for solids},}\ }\href {\doibase
  10.1103/PhysRevB.73.235116} {\bibfield  {journal} {\bibinfo  {journal}
  {Physical Review B}\ }\textbf {\bibinfo {volume} {73}},\ \bibinfo {pages}
  {235116} (\bibinfo {year} {2006})}\BibitemShut {NoStop}%
\bibitem [{\citenamefont {Constantin}\ \emph
  {et~al.}(2016{\natexlab{a}})\citenamefont {Constantin}, \citenamefont
  {Terentjevs}, \citenamefont {Della~Sala}, \citenamefont {Cortona},\ and\
  \citenamefont {Fabiano}}]{SG4_XC:2016}%
  \BibitemOpen
  \bibfield  {author} {\bibinfo {author} {\bibfnamefont {Lucian~A.}\
  \bibnamefont {Constantin}}, \bibinfo {author} {\bibfnamefont {Aleksandrs}\
  \bibnamefont {Terentjevs}}, \bibinfo {author} {\bibfnamefont {Fabio}\
  \bibnamefont {Della~Sala}}, \bibinfo {author} {\bibfnamefont {Pietro}\
  \bibnamefont {Cortona}}, \ and\ \bibinfo {author} {\bibfnamefont {Eduardo}\
  \bibnamefont {Fabiano}},\ }\bibfield  {title} {\enquote {\bibinfo {title}
  {Semiclassical atom theory applied to solid-state physics},}\ }\href
  {\doibase 10.1103/PhysRevB.93.045126} {\bibfield  {journal} {\bibinfo
  {journal} {Phys. Rev. B}\ }\textbf {\bibinfo {volume} {93}},\ \bibinfo
  {pages} {045126} (\bibinfo {year} {2016}{\natexlab{a}})}\BibitemShut
  {NoStop}%
\bibitem [{\citenamefont {Staroverov}\ \emph {et~al.}(2004)\citenamefont
  {Staroverov}, \citenamefont {Scuseria}, \citenamefont {Tao},\ and\
  \citenamefont {Perdew}}]{PhysRevB.69.075102}%
  \BibitemOpen
  \bibfield  {author} {\bibinfo {author} {\bibfnamefont {Viktor~N.}\
  \bibnamefont {Staroverov}}, \bibinfo {author} {\bibfnamefont {Gustavo~E.}\
  \bibnamefont {Scuseria}}, \bibinfo {author} {\bibfnamefont {Jianmin}\
  \bibnamefont {Tao}}, \ and\ \bibinfo {author} {\bibfnamefont {John~P.}\
  \bibnamefont {Perdew}},\ }\bibfield  {title} {\enquote {\bibinfo {title}
  {Tests of a ladder of density functionals for bulk solids and surfaces},}\
  }\href {\doibase 10.1103/PhysRevB.69.075102} {\bibfield  {journal} {\bibinfo
  {journal} {Phys. Rev. B}\ }\textbf {\bibinfo {volume} {69}},\ \bibinfo
  {pages} {075102} (\bibinfo {year} {2004})}\BibitemShut {NoStop}%
\bibitem [{\citenamefont {Sun}\ \emph {et~al.}(2015{\natexlab{a}})\citenamefont
  {Sun}, \citenamefont {Ruzsinszky},\ and\ \citenamefont
  {Perdew}}]{PhysRevLett.115.036402}%
  \BibitemOpen
  \bibfield  {author} {\bibinfo {author} {\bibfnamefont {Jianwei}\ \bibnamefont
  {Sun}}, \bibinfo {author} {\bibfnamefont {Adrienn}\ \bibnamefont
  {Ruzsinszky}}, \ and\ \bibinfo {author} {\bibfnamefont {John~P.}\
  \bibnamefont {Perdew}},\ }\bibfield  {title} {\enquote {\bibinfo {title}
  {Strongly constrained and appropriately normed semilocal density
  functional},}\ }\href {\doibase 10.1103/PhysRevLett.115.036402} {\bibfield
  {journal} {\bibinfo  {journal} {Phys. Rev. Lett.}\ }\textbf {\bibinfo
  {volume} {115}},\ \bibinfo {pages} {036402} (\bibinfo {year}
  {2015}{\natexlab{a}})}\BibitemShut {NoStop}%
\bibitem [{\citenamefont {Haas}\ \emph
  {et~al.}(2009{\natexlab{a}})\citenamefont {Haas}, \citenamefont {Tran},\ and\
  \citenamefont {Blaha}}]{Haas/Tran/Blaha:2009}%
  \BibitemOpen
  \bibfield  {author} {\bibinfo {author} {\bibfnamefont {Philipp}\ \bibnamefont
  {Haas}}, \bibinfo {author} {\bibfnamefont {Fabien}\ \bibnamefont {Tran}}, \
  and\ \bibinfo {author} {\bibfnamefont {Peter}\ \bibnamefont {Blaha}},\
  }\bibfield  {title} {\enquote {\bibinfo {title} {Calculation of the lattice
  constant of solids with semilocal functionals},}\ }\href {\doibase
  10.1103/PhysRevB.79.085104} {\bibfield  {journal} {\bibinfo  {journal}
  {Physical Review B}\ }\textbf {\bibinfo {volume} {79}},\ \bibinfo {pages}
  {085104} (\bibinfo {year} {2009}{\natexlab{a}})}\BibitemShut {NoStop}%
\bibitem [{\citenamefont {Sun}\ \emph {et~al.}(2012)\citenamefont {Sun},
  \citenamefont {Xiao},\ and\ \citenamefont
  {Ruzsinszky}}]{Jianwei_orbitalOverlap}%
  \BibitemOpen
  \bibfield  {author} {\bibinfo {author} {\bibfnamefont {Jianwei}\ \bibnamefont
  {Sun}}, \bibinfo {author} {\bibfnamefont {Bing}\ \bibnamefont {Xiao}}, \ and\
  \bibinfo {author} {\bibfnamefont {Adrienn}\ \bibnamefont {Ruzsinszky}},\
  }\bibfield  {title} {\enquote {\bibinfo {title} {Communication: Effect of the
  orbital-overlap dependence in the meta generalized gradient approximation},}\
  }\href {\doibase http://dx.doi.org/10.1063/1.4742312} {\bibfield  {journal}
  {\bibinfo  {journal} {The Journal of Chemical Physics}\ }\textbf {\bibinfo
  {volume} {137}},\ \bibinfo {eid} {051101} (\bibinfo {year} {2012}),\
  http://dx.doi.org/10.1063/1.4742312}\BibitemShut {NoStop}%
\bibitem [{\citenamefont {Sun}\ \emph {et~al.}(2013{\natexlab{a}})\citenamefont
  {Sun}, \citenamefont {Xiao}, \citenamefont {Fang}, \citenamefont
  {Haunschild}, \citenamefont {Hao}, \citenamefont {Ruzsinszky}, \citenamefont
  {Csonka}, \citenamefont {Scuseria},\ and\ \citenamefont
  {Perdew}}]{MS2_paper}%
  \BibitemOpen
  \bibfield  {author} {\bibinfo {author} {\bibfnamefont {Jianwei}\ \bibnamefont
  {Sun}}, \bibinfo {author} {\bibfnamefont {Bing}\ \bibnamefont {Xiao}},
  \bibinfo {author} {\bibfnamefont {Yuan}\ \bibnamefont {Fang}}, \bibinfo
  {author} {\bibfnamefont {Robin}\ \bibnamefont {Haunschild}}, \bibinfo
  {author} {\bibfnamefont {Pan}\ \bibnamefont {Hao}}, \bibinfo {author}
  {\bibfnamefont {Adrienn}\ \bibnamefont {Ruzsinszky}}, \bibinfo {author}
  {\bibfnamefont {G\'abor~I.}\ \bibnamefont {Csonka}}, \bibinfo {author}
  {\bibfnamefont {Gustavo~E.}\ \bibnamefont {Scuseria}}, \ and\ \bibinfo
  {author} {\bibfnamefont {John~P.}\ \bibnamefont {Perdew}},\ }\bibfield
  {title} {\enquote {\bibinfo {title} {Density functionals that recognize
  covalent, metallic, and weak bonds},}\ }\href {\doibase
  10.1103/PhysRevLett.111.106401} {\bibfield  {journal} {\bibinfo  {journal}
  {Phys. Rev. Lett.}\ }\textbf {\bibinfo {volume} {111}},\ \bibinfo {pages}
  {106401} (\bibinfo {year} {2013}{\natexlab{a}})}\BibitemShut {NoStop}%
\bibitem [{\citenamefont {Sun}\ \emph {et~al.}(2013{\natexlab{b}})\citenamefont
  {Sun}, \citenamefont {Haunschild}, \citenamefont {Xiao}, \citenamefont
  {Bulik}, \citenamefont {Scuseria},\ and\ \citenamefont
  {Perdew}}]{SunKEdepend:2013}%
  \BibitemOpen
  \bibfield  {author} {\bibinfo {author} {\bibfnamefont {Jianwei}\ \bibnamefont
  {Sun}}, \bibinfo {author} {\bibfnamefont {Robin}\ \bibnamefont {Haunschild}},
  \bibinfo {author} {\bibfnamefont {Bing}\ \bibnamefont {Xiao}}, \bibinfo
  {author} {\bibfnamefont {Ireneusz~W.}\ \bibnamefont {Bulik}}, \bibinfo
  {author} {\bibfnamefont {Gustavo~E.}\ \bibnamefont {Scuseria}}, \ and\
  \bibinfo {author} {\bibfnamefont {John~P.}\ \bibnamefont {Perdew}},\
  }\bibfield  {title} {\enquote {\bibinfo {title} {Semilocal and hybrid
  meta-generalized gradient approximations based on the understanding of the
  kinetic-energy-density dependence},}\ }\href {\doibase
  http://dx.doi.org/10.1063/1.4789414} {\bibfield  {journal} {\bibinfo
  {journal} {The Journal of Chemical Physics}\ }\textbf {\bibinfo {volume}
  {138}},\ \bibinfo {eid} {044113} (\bibinfo {year} {2013}{\natexlab{b}}),\
  http://dx.doi.org/10.1063/1.4789414}\BibitemShut {NoStop}%
\bibitem [{\citenamefont {Silvi}\ \emph {et~al.}(1994)\citenamefont {Silvi},
  \citenamefont {Savin} \emph {et~al.}}]{silvi1994classification}%
  \BibitemOpen
  \bibfield  {author} {\bibinfo {author} {\bibfnamefont {Bernard}\ \bibnamefont
  {Silvi}}, \bibinfo {author} {\bibfnamefont {Andreas}\ \bibnamefont {Savin}},
  \emph {et~al.},\ }\bibfield  {title} {\enquote {\bibinfo {title}
  {Classification of chemical bonds based on topological analysis of electron
  localization functions},}\ }\href {http://dx.doi.org/10.1038/371683a0}
  {\bibfield  {journal} {\bibinfo  {journal} {Nature}\ }\textbf {\bibinfo
  {volume} {371}},\ \bibinfo {pages} {683--686} (\bibinfo {year}
  {1994})}\BibitemShut {NoStop}%
\bibitem [{\citenamefont {Becke}\ and\ \citenamefont
  {Edgecombe}(1990)}]{Becke:ELF1990}%
  \BibitemOpen
  \bibfield  {author} {\bibinfo {author} {\bibfnamefont {A.~D.}\ \bibnamefont
  {Becke}}\ and\ \bibinfo {author} {\bibfnamefont {K.~E.}\ \bibnamefont
  {Edgecombe}},\ }\bibfield  {title} {\enquote {\bibinfo {title} {A simple
  measure of electron localization in atomic and molecular systems},}\ }\href
  {\doibase http://dx.doi.org/10.1063/1.458517} {\bibfield  {journal} {\bibinfo
   {journal} {The Journal of Chemical Physics}\ }\textbf {\bibinfo {volume}
  {92}},\ \bibinfo {pages} {5397--5403} (\bibinfo {year} {1990})}\BibitemShut
  {NoStop}%
\bibitem [{\citenamefont {Della~Sala}\ \emph {et~al.}(2015)\citenamefont
  {Della~Sala}, \citenamefont {Fabiano},\ and\ \citenamefont
  {Constantin}}]{PhysRevB.91.035126}%
  \BibitemOpen
  \bibfield  {author} {\bibinfo {author} {\bibfnamefont {Fabio}\ \bibnamefont
  {Della~Sala}}, \bibinfo {author} {\bibfnamefont {Eduardo}\ \bibnamefont
  {Fabiano}}, \ and\ \bibinfo {author} {\bibfnamefont {Lucian~A.}\ \bibnamefont
  {Constantin}},\ }\bibfield  {title} {\enquote {\bibinfo {title} {Kohn-sham
  kinetic energy density in the nuclear and asymptotic regions: Deviations from
  the von weizs\"acker behavior and applications to density functionals},}\
  }\href {\doibase 10.1103/PhysRevB.91.035126} {\bibfield  {journal} {\bibinfo
  {journal} {Phys. Rev. B}\ }\textbf {\bibinfo {volume} {91}},\ \bibinfo
  {pages} {035126} (\bibinfo {year} {2015})}\BibitemShut {NoStop}%
\bibitem [{\citenamefont {Constantin}\ \emph
  {et~al.}(2016{\natexlab{b}})\citenamefont {Constantin}, \citenamefont
  {Fabiano}, \citenamefont {Pitarke},\ and\ \citenamefont
  {Della~Sala}}]{PhysRevB.93.115127}%
  \BibitemOpen
  \bibfield  {author} {\bibinfo {author} {\bibfnamefont {Lucian~A.}\
  \bibnamefont {Constantin}}, \bibinfo {author} {\bibfnamefont {Eduardo}\
  \bibnamefont {Fabiano}}, \bibinfo {author} {\bibfnamefont {J.~M.}\
  \bibnamefont {Pitarke}}, \ and\ \bibinfo {author} {\bibfnamefont {Fabio}\
  \bibnamefont {Della~Sala}},\ }\bibfield  {title} {\enquote {\bibinfo {title}
  {Semilocal density functional theory with correct surface asymptotics},}\
  }\href {\doibase 10.1103/PhysRevB.93.115127} {\bibfield  {journal} {\bibinfo
  {journal} {Phys. Rev. B}\ }\textbf {\bibinfo {volume} {93}},\ \bibinfo
  {pages} {115127} (\bibinfo {year} {2016}{\natexlab{b}})}\BibitemShut
  {NoStop}%
\bibitem [{\citenamefont {Zhao}\ and\ \citenamefont
  {Truhlar}(2006)}]{M06L:2006}%
  \BibitemOpen
  \bibfield  {author} {\bibinfo {author} {\bibfnamefont {Yan}\ \bibnamefont
  {Zhao}}\ and\ \bibinfo {author} {\bibfnamefont {Donald~G.}\ \bibnamefont
  {Truhlar}},\ }\bibfield  {title} {\enquote {\bibinfo {title} {A new local
  density functional for main-group thermochemistry, transition metal bonding,
  thermochemical kinetics, and noncovalent interactions},}\ }\href {\doibase
  http://dx.doi.org/10.1063/1.2370993} {\bibfield  {journal} {\bibinfo
  {journal} {The Journal of Chemical Physics}\ }\textbf {\bibinfo {volume}
  {125}},\ \bibinfo {eid} {194101} (\bibinfo {year} {2006}),\
  http://dx.doi.org/10.1063/1.2370993}\BibitemShut {NoStop}%
\bibitem [{\citenamefont {Tao}\ and\ \citenamefont {Mo}(2016)}]{TM:xc_2016}%
  \BibitemOpen
  \bibfield  {author} {\bibinfo {author} {\bibfnamefont {Jianmin}\ \bibnamefont
  {Tao}}\ and\ \bibinfo {author} {\bibfnamefont {Yuxiang}\ \bibnamefont {Mo}},\
  }\bibfield  {title} {\enquote {\bibinfo {title} {Accurate semilocal density
  functional for condensed-matter physics and quantum chemistry},}\ }\href
  {\doibase 10.1103/PhysRevLett.117.073001} {\bibfield  {journal} {\bibinfo
  {journal} {Phys. Rev. Lett.}\ }\textbf {\bibinfo {volume} {117}},\ \bibinfo
  {pages} {073001} (\bibinfo {year} {2016})}\BibitemShut {NoStop}%
\bibitem [{\citenamefont {Tran}\ \emph {et~al.}(2016)\citenamefont {Tran},
  \citenamefont {Stelzl},\ and\ \citenamefont
  {Blaha}}]{BlahaDFTbenchmark:2016}%
  \BibitemOpen
  \bibfield  {author} {\bibinfo {author} {\bibfnamefont {Fabien}\ \bibnamefont
  {Tran}}, \bibinfo {author} {\bibfnamefont {Julia}\ \bibnamefont {Stelzl}}, \
  and\ \bibinfo {author} {\bibfnamefont {Peter}\ \bibnamefont {Blaha}},\
  }\bibfield  {title} {\enquote {\bibinfo {title} {{Rungs 1 to 4 of DFT Jacob's
  ladder: Extensive test on the lattice constant, bulk modulus, and cohesive
  energy of solids}},}\ }\href {\doibase http://dx.doi.org/10.1063/1.4948636}
  {\bibfield  {journal} {\bibinfo  {journal} {The Journal of Chemical Physics}\
  }\textbf {\bibinfo {volume} {144}},\ \bibinfo {eid} {204120} (\bibinfo {year}
  {2016}),\ http://dx.doi.org/10.1063/1.4948636}\BibitemShut {NoStop}%
\bibitem [{\citenamefont {Becke}(1993)}]{Becke:1993}%
  \BibitemOpen
  \bibfield  {author} {\bibinfo {author} {\bibfnamefont {Axel~D.}\ \bibnamefont
  {Becke}},\ }\bibfield  {title} {\enquote {\bibinfo {title}
  {{Density-functional thermochemistry. III. The role of exact exchange}},}\
  }\href {\doibase 10.1063/1.464913} {\bibfield  {journal} {\bibinfo  {journal}
  {The Journal of Chemical Physics}\ }\textbf {\bibinfo {volume} {98}},\
  \bibinfo {pages} {5648--5652} (\bibinfo {year} {1993})}\BibitemShut {NoStop}%
\bibitem [{\citenamefont {Archer}\ \emph {et~al.}(2011)\citenamefont {Archer},
  \citenamefont {Pemmaraju}, \citenamefont {Sanvito}, \citenamefont
  {Franchini}, \citenamefont {He}, \citenamefont {Filippetti}, \citenamefont
  {Delugas}, \citenamefont {Puggioni}, \citenamefont {Fiorentini},
  \citenamefont {Tiwari},\ and\ \citenamefont {Majumdar}}]{PhysRevB.84.115114}%
  \BibitemOpen
  \bibfield  {author} {\bibinfo {author} {\bibfnamefont {T.}~\bibnamefont
  {Archer}}, \bibinfo {author} {\bibfnamefont {C.~D.}\ \bibnamefont
  {Pemmaraju}}, \bibinfo {author} {\bibfnamefont {S.}~\bibnamefont {Sanvito}},
  \bibinfo {author} {\bibfnamefont {C.}~\bibnamefont {Franchini}}, \bibinfo
  {author} {\bibfnamefont {J.}~\bibnamefont {He}}, \bibinfo {author}
  {\bibfnamefont {A.}~\bibnamefont {Filippetti}}, \bibinfo {author}
  {\bibfnamefont {P.}~\bibnamefont {Delugas}}, \bibinfo {author} {\bibfnamefont
  {D.}~\bibnamefont {Puggioni}}, \bibinfo {author} {\bibfnamefont
  {V.}~\bibnamefont {Fiorentini}}, \bibinfo {author} {\bibfnamefont
  {R.}~\bibnamefont {Tiwari}}, \ and\ \bibinfo {author} {\bibfnamefont
  {P.}~\bibnamefont {Majumdar}},\ }\bibfield  {title} {\enquote {\bibinfo
  {title} {Exchange interactions and magnetic phases of transition metal
  oxides: Benchmarking advanced \textit{ab initio} methods},}\ }\href {\doibase
  10.1103/PhysRevB.84.115114} {\bibfield  {journal} {\bibinfo  {journal} {Phys.
  Rev. B}\ }\textbf {\bibinfo {volume} {84}},\ \bibinfo {pages} {115114}
  (\bibinfo {year} {2011})}\BibitemShut {NoStop}%
\bibitem [{\citenamefont {Perdew}\ \emph
  {et~al.}(1996{\natexlab{b}})\citenamefont {Perdew}, \citenamefont
  {Ernzerhof},\ and\ \citenamefont {Burke}}]{Perdew_et_al:1996}%
  \BibitemOpen
  \bibfield  {author} {\bibinfo {author} {\bibfnamefont {John~P.}\ \bibnamefont
  {Perdew}}, \bibinfo {author} {\bibfnamefont {Matthias}\ \bibnamefont
  {Ernzerhof}}, \ and\ \bibinfo {author} {\bibfnamefont {Kieron}\ \bibnamefont
  {Burke}},\ }\bibfield  {title} {\enquote {\bibinfo {title} {Rationale for
  mixing exact exchange with density functional approximations},}\ }\href
  {\doibase http://dx.doi.org/10.1063/1.472933} {\bibfield  {journal} {\bibinfo
   {journal} {J.\ Chem.\ Phys.}\ }\textbf {\bibinfo {volume} {105}},\ \bibinfo
  {pages} {9982--9985} (\bibinfo {year} {1996}{\natexlab{b}})}\BibitemShut
  {NoStop}%
\bibitem [{\citenamefont {Heyd}\ \emph {et~al.}(2005)\citenamefont {Heyd},
  \citenamefont {Peralta}, \citenamefont {Scuseria},\ and\ \citenamefont
  {Martin}}]{HSE06:2005}%
  \BibitemOpen
  \bibfield  {author} {\bibinfo {author} {\bibfnamefont {Jochen}\ \bibnamefont
  {Heyd}}, \bibinfo {author} {\bibfnamefont {Juan~E.}\ \bibnamefont {Peralta}},
  \bibinfo {author} {\bibfnamefont {Gustavo~E.}\ \bibnamefont {Scuseria}}, \
  and\ \bibinfo {author} {\bibfnamefont {Richard~L.}\ \bibnamefont {Martin}},\
  }\bibfield  {title} {\enquote {\bibinfo {title} {{Energy band gaps and
  lattice parameters evaluated with the Heyd-Scuseria-Ernzerhof screened hybrid
  functional}},}\ }\href {\doibase http://dx.doi.org/10.1063/1.2085170}
  {\bibfield  {journal} {\bibinfo  {journal} {The Journal of Chemical Physics}\
  }\textbf {\bibinfo {volume} {123}},\ \bibinfo {eid} {174101} (\bibinfo {year}
  {2005}),\ http://dx.doi.org/10.1063/1.2085170}\BibitemShut {NoStop}%
\bibitem [{\citenamefont {Marques}\ \emph {et~al.}(2011)\citenamefont
  {Marques}, \citenamefont {Vidal}, \citenamefont {Oliveira}, \citenamefont
  {Reining},\ and\ \citenamefont {Botti}}]{PhysRevB.83.035119}%
  \BibitemOpen
  \bibfield  {author} {\bibinfo {author} {\bibfnamefont {Miguel A.~L.}\
  \bibnamefont {Marques}}, \bibinfo {author} {\bibfnamefont {Julien}\
  \bibnamefont {Vidal}}, \bibinfo {author} {\bibfnamefont {Micael J.~T.}\
  \bibnamefont {Oliveira}}, \bibinfo {author} {\bibfnamefont {Lucia}\
  \bibnamefont {Reining}}, \ and\ \bibinfo {author} {\bibfnamefont {Silvana}\
  \bibnamefont {Botti}},\ }\bibfield  {title} {\enquote {\bibinfo {title}
  {Density-based mixing parameter for hybrid functionals},}\ }\href {\doibase
  10.1103/PhysRevB.83.035119} {\bibfield  {journal} {\bibinfo  {journal} {Phys.
  Rev. B}\ }\textbf {\bibinfo {volume} {83}},\ \bibinfo {pages} {035119}
  (\bibinfo {year} {2011})}\BibitemShut {NoStop}%
\bibitem [{\citenamefont {Garcia-Fernandez}\ \emph {et~al.}(2012)\citenamefont
  {Garcia-Fernandez}, \citenamefont {Ghosh}, \citenamefont {English},\ and\
  \citenamefont {Aramburu}}]{garciabenchmark2012}%
  \BibitemOpen
  \bibfield  {author} {\bibinfo {author} {\bibfnamefont {P.}~\bibnamefont
  {Garcia-Fernandez}}, \bibinfo {author} {\bibfnamefont {S.}~\bibnamefont
  {Ghosh}}, \bibinfo {author} {\bibfnamefont {Niall~J.}\ \bibnamefont
  {English}}, \ and\ \bibinfo {author} {\bibfnamefont {J.~A.}\ \bibnamefont
  {Aramburu}},\ }\bibfield  {title} {\enquote {\bibinfo {title} {Benchmark
  study for the application of density functional theory to the prediction of
  octahedral tilting in perovskites},}\ }\href {\doibase
  10.1103/PhysRevB.86.144107} {\bibfield  {journal} {\bibinfo  {journal} {Phys.
  Rev. B}\ }\textbf {\bibinfo {volume} {86}},\ \bibinfo {pages} {144107}
  (\bibinfo {year} {2012})}\BibitemShut {NoStop}%
\bibitem [{\citenamefont {He}\ and\ \citenamefont
  {Franchini}(2012)}]{FranchiniHSE:2013}%
  \BibitemOpen
  \bibfield  {author} {\bibinfo {author} {\bibfnamefont {Jiangang}\
  \bibnamefont {He}}\ and\ \bibinfo {author} {\bibfnamefont {Cesare}\
  \bibnamefont {Franchini}},\ }\bibfield  {title} {\enquote {\bibinfo {title}
  {Screened hybrid functional applied to
  3${d}^{0}$$\ensuremath{\rightarrow}$3${d}^{8}$ transition-metal perovskites
  {La$M$O$_{3}$ ($M$ = Sc-Cu)}: Influence of the exchange mixing parameter on
  the structural, electronic, and magnetic properties},}\ }\href {\doibase
  10.1103/PhysRevB.86.235117} {\bibfield  {journal} {\bibinfo  {journal} {Phys.
  Rev. B}\ }\textbf {\bibinfo {volume} {86}},\ \bibinfo {pages} {235117}
  (\bibinfo {year} {2012})}\BibitemShut {NoStop}%
\bibitem [{\citenamefont {Finch}\ and\ \citenamefont
  {Fordham}(1936)}]{KF_exp_str}%
  \BibitemOpen
  \bibfield  {author} {\bibinfo {author} {\bibfnamefont {G~I}\ \bibnamefont
  {Finch}}\ and\ \bibinfo {author} {\bibfnamefont {S}~\bibnamefont {Fordham}},\
  }\bibfield  {title} {\enquote {\bibinfo {title} {The effect of crystal-size
  on lattice-dimensions},}\ }\href
  {http://stacks.iop.org/0959-5309/48/i=1/a=312} {\bibfield  {journal}
  {\bibinfo  {journal} {Proceedings of the Physical Society}\ }\textbf
  {\bibinfo {volume} {48}},\ \bibinfo {pages} {85} (\bibinfo {year}
  {1936})}\BibitemShut {NoStop}%
\bibitem [{\citenamefont {Costa}\ and\ \citenamefont
  {de~Almeida}(1987)}]{Costa:VF2}%
  \BibitemOpen
  \bibfield  {author} {\bibinfo {author} {\bibfnamefont {M.~M.~R.}\
  \bibnamefont {Costa}}\ and\ \bibinfo {author} {\bibfnamefont {M.~J.~M.}\
  \bibnamefont {de~Almeida}},\ }\bibfield  {title} {\enquote {\bibinfo {title}
  {{Asphericity effects in the electron density of VF${\sb 2}$}},}\ }\href
  {\doibase 10.1107/S010876818709774X} {\bibfield  {journal} {\bibinfo
  {journal} {Acta Crystallographica Section B}\ }\textbf {\bibinfo {volume}
  {43}},\ \bibinfo {pages} {346--352} (\bibinfo {year} {1987})}\BibitemShut
  {NoStop}%
\bibitem [{\citenamefont {Alte~da Veiga L.~M.}\ and\ \citenamefont
  {W.}(1982)}]{MnF2:Benchmark}%
  \BibitemOpen
  \bibfield  {author} {\bibinfo {author} {\bibfnamefont {Andrade L.~R.}\
  \bibnamefont {Alte~da Veiga L.~M.}}\ and\ \bibinfo {author} {\bibfnamefont
  {Gonschorek}\ \bibnamefont {W.}},\ }\bibfield  {title} {\enquote {\bibinfo
  {title} {The crystal structure of manganese difluoride ({MnF}$_2$) :
  Reliability test of hypothetical intensity variances by means of ${\chi}^2$
  distributions},}\ }\href {\doibase 10.1524/zkri.1982.160.14.171} {\bibfield
  {journal} {\bibinfo  {journal} {Zeitschrift f�r Kristallographie -
  Crystalline Materials}\ }\textbf {\bibinfo {volume} {160}},\ \bibinfo {pages}
  {171--178} (\bibinfo {year} {1982})}\BibitemShut {NoStop}%
\bibitem [{\citenamefont {Scatturin}\ \emph {et~al.}(1961)\citenamefont
  {Scatturin}, \citenamefont {Corliss}, \citenamefont {Elliott},\ and\
  \citenamefont {Hastings}}]{Scatturin:KMnF3}%
  \BibitemOpen
  \bibfield  {author} {\bibinfo {author} {\bibfnamefont {V.}~\bibnamefont
  {Scatturin}}, \bibinfo {author} {\bibfnamefont {L.}~\bibnamefont {Corliss}},
  \bibinfo {author} {\bibfnamefont {N.}~\bibnamefont {Elliott}}, \ and\
  \bibinfo {author} {\bibfnamefont {J.}~\bibnamefont {Hastings}},\ }\bibfield
  {title} {\enquote {\bibinfo {title} {{Magnetic structures of 3{\it d}
  transition metal double fluorides, K{\it Me}F${\sb 3}$}},}\ }\href {\doibase
  10.1107/S0365110X61000036} {\bibfield  {journal} {\bibinfo  {journal} {Acta
  Crystallographica}\ }\textbf {\bibinfo {volume} {14}},\ \bibinfo {pages}
  {19--26} (\bibinfo {year} {1961})}\BibitemShut {NoStop}%
\bibitem [{\citenamefont {Carlson}\ \emph
  {et~al.}(1998{\natexlab{a}})\citenamefont {Carlson}, \citenamefont {Xu},\
  and\ \citenamefont {Norrestam}}]{Carlson1998116}%
  \BibitemOpen
  \bibfield  {author} {\bibinfo {author} {\bibfnamefont {Stefan}\ \bibnamefont
  {Carlson}}, \bibinfo {author} {\bibfnamefont {Yiqiu}\ \bibnamefont {Xu}}, \
  and\ \bibinfo {author} {\bibfnamefont {Rolf}\ \bibnamefont {Norrestam}},\
  }\bibfield  {title} {\enquote {\bibinfo {title} {Single-crystal high-pressure
  studies of {Na$_3$ScF$_6$}},}\ }\href {\doibase
  http://dx.doi.org/10.1006/jssc.1997.7607} {\bibfield  {journal} {\bibinfo
  {journal} {Journal of Solid State Chemistry}\ }\textbf {\bibinfo {volume}
  {135}},\ \bibinfo {pages} {116 -- 120} (\bibinfo {year}
  {1998}{\natexlab{a}})}\BibitemShut {NoStop}%
\bibitem [{\citenamefont {Carlson}\ \emph
  {et~al.}(1998{\natexlab{b}})\citenamefont {Carlson}, \citenamefont {Xu},
  \citenamefont {H{\aa}lenius},\ and\ \citenamefont
  {Norrestam}}]{Carlson/Norrestam:1998}%
  \BibitemOpen
  \bibfield  {author} {\bibinfo {author} {\bibfnamefont {Stefan}\ \bibnamefont
  {Carlson}}, \bibinfo {author} {\bibfnamefont {Yiqiu}\ \bibnamefont {Xu}},
  \bibinfo {author} {\bibfnamefont {Ulf}\ \bibnamefont {H{\aa}lenius}}, \ and\
  \bibinfo {author} {\bibfnamefont {Rolf}\ \bibnamefont {Norrestam}},\
  }\bibfield  {title} {\enquote {\bibinfo {title} {A reversible, isosymmetric,
  high-pressure phase transition in {Na$_3$MnF$_6$}},}\ }\href {\doibase
  10.1021/ic971171g} {\bibfield  {journal} {\bibinfo  {journal} {Inorganic
  Chemistry}\ }\textbf {\bibinfo {volume} {37}},\ \bibinfo {pages} {1486--1492}
  (\bibinfo {year} {1998}{\natexlab{b}})}\BibitemShut {NoStop}%
\bibitem [{\citenamefont {Hawthorne}\ and\ \citenamefont
  {Ferguson}(1975)}]{hawthorne1975refinement}%
  \BibitemOpen
  \bibfield  {author} {\bibinfo {author} {\bibfnamefont {F.~C.}\ \bibnamefont
  {Hawthorne}}\ and\ \bibinfo {author} {\bibfnamefont {R.~B.}\ \bibnamefont
  {Ferguson}},\ }\bibfield  {title} {\enquote {\bibinfo {title} {Refinement of
  the crystal structure of cryolite},}\ }\href
  {http://www.canmin.org/content/13/4/377.short} {\bibfield  {journal}
  {\bibinfo  {journal} {The Canadian Mineralogist}\ }\textbf {\bibinfo {volume}
  {13}},\ \bibinfo {pages} {377--382} (\bibinfo {year} {1975})}\BibitemShut
  {NoStop}%
\bibitem [{\citenamefont {Pinlac}\ \emph {et~al.}(2011)\citenamefont {Pinlac},
  \citenamefont {Stern},\ and\ \citenamefont
  {Poeppelmeier}}]{Pinlac/Poeppelmeier:2011}%
  \BibitemOpen
  \bibfield  {author} {\bibinfo {author} {\bibfnamefont {Rachelle Ann~F.}\
  \bibnamefont {Pinlac}}, \bibinfo {author} {\bibfnamefont {Charlotte~L.}\
  \bibnamefont {Stern}}, \ and\ \bibinfo {author} {\bibfnamefont {Kenneth~R.}\
  \bibnamefont {Poeppelmeier}},\ }\bibfield  {title} {\enquote {\bibinfo
  {title} {{New Layered Oxide-Fluoride Perovskites: KNaNbOF$_5$ and
  KNa$M$O$_2$F$_4$ ($M =$~Mo$^{6+}$, W$^{6+}$)}},}\ }\href {\doibase
  10.3390/cryst1010003} {\bibfield  {journal} {\bibinfo  {journal} {Crystals}\
  }\textbf {\bibinfo {volume} {1}},\ \bibinfo {pages} {3--14} (\bibinfo {year}
  {2011})}\BibitemShut {NoStop}%
\bibitem [{\citenamefont {Bushnell}\ and\ \citenamefont
  {Moss}(1972)}]{CsVOF4:1972}%
  \BibitemOpen
  \bibfield  {author} {\bibinfo {author} {\bibfnamefont {G.~W.}\ \bibnamefont
  {Bushnell}}\ and\ \bibinfo {author} {\bibfnamefont {K.~C.}\ \bibnamefont
  {Moss}},\ }\bibfield  {title} {\enquote {\bibinfo {title} {{The Crystal
  Structure of Caesium Oxotetrafluorovanadate(V)}},}\ }\href {\doibase
  10.1139/v72-585} {\bibfield  {journal} {\bibinfo  {journal} {Canadian Journal
  of Chemistry}\ }\textbf {\bibinfo {volume} {50}},\ \bibinfo {pages}
  {3700--3705} (\bibinfo {year} {1972})}\BibitemShut {NoStop}%
\bibitem [{\citenamefont {Marvel}\ \emph {et~al.}(2007)\citenamefont {Marvel},
  \citenamefont {Lesage}, \citenamefont {Baek}, \citenamefont {Halasyamani},
  \citenamefont {Stern},\ and\ \citenamefont
  {Poeppelmeier}}]{Marvel/Poeppelmeier:2007}%
  \BibitemOpen
  \bibfield  {author} {\bibinfo {author} {\bibfnamefont {Michael~R.}\
  \bibnamefont {Marvel}}, \bibinfo {author} {\bibfnamefont {Julien}\
  \bibnamefont {Lesage}}, \bibinfo {author} {\bibfnamefont {Jaewook}\
  \bibnamefont {Baek}}, \bibinfo {author} {\bibfnamefont {P.~Shiv}\
  \bibnamefont {Halasyamani}}, \bibinfo {author} {\bibfnamefont {Charlotte~L.}\
  \bibnamefont {Stern}}, \ and\ \bibinfo {author} {\bibfnamefont {Kenneth~R.}\
  \bibnamefont {Poeppelmeier}},\ }\bibfield  {title} {\enquote {\bibinfo
  {title} {Cation--anion interactions and polar structures in the solid
  state},}\ }\href {\doibase 10.1021/ja074659h} {\bibfield  {journal} {\bibinfo
   {journal} {Journal of the American Chemical Society}\ }\textbf {\bibinfo
  {volume} {129}},\ \bibinfo {pages} {13963--13969} (\bibinfo {year}
  {2007})}\BibitemShut {NoStop}%
\bibitem [{\citenamefont {Vlasse}\ \emph {et~al.}(1982)\citenamefont {Vlasse},
  \citenamefont {Moutou}, \citenamefont {Cervera-Marzal}, \citenamefont
  {Chaminade},\ and\ \citenamefont {Hagenmuller}}]{Na2WOF:1982}%
  \BibitemOpen
  \bibfield  {author} {\bibinfo {author} {\bibfnamefont {M.}~\bibnamefont
  {Vlasse}}, \bibinfo {author} {\bibfnamefont {J.-M.}\ \bibnamefont {Moutou}},
  \bibinfo {author} {\bibfnamefont {M.}~\bibnamefont {Cervera-Marzal}},
  \bibinfo {author} {\bibfnamefont {J.~P.}\ \bibnamefont {Chaminade}}, \ and\
  \bibinfo {author} {\bibfnamefont {P.}~\bibnamefont {Hagenmuller}},\
  }\bibfield  {title} {\enquote {\bibinfo {title} {{ChemInform Abstract:
  Structure of Disodium Tetrafluorodioxotungstate (Na2WO$_2$F$_4$)}},}\ }\href
  {\doibase 10.1002/chin.198234008} {\bibfield  {journal} {\bibinfo  {journal}
  {Chemischer Informationsdienst}\ }\textbf {\bibinfo {volume} {13}} (\bibinfo
  {year} {1982}),\ 10.1002/chin.198234008}\BibitemShut {NoStop}%
\bibitem [{\citenamefont {Crosnier}\ and\ \citenamefont
  {Fourquet}(1992)}]{BaTiOF4:1992}%
  \BibitemOpen
  \bibfield  {author} {\bibinfo {author} {\bibfnamefont {M.~P.}\ \bibnamefont
  {Crosnier}}\ and\ \bibinfo {author} {\bibfnamefont {J.~L.}\ \bibnamefont
  {Fourquet}},\ }\bibfield  {title} {\enquote {\bibinfo {title} {{ChemInform
  Abstract: Synthesis and Crystal Structure of BaTiOF$_4$.}}}\ }\href {\doibase
  10.1002/chin.199228030} {\bibfield  {journal} {\bibinfo  {journal}
  {ChemInform}\ }\textbf {\bibinfo {volume} {23}} (\bibinfo {year} {1992}),\
  10.1002/chin.199228030}\BibitemShut {NoStop}%
\bibitem [{\citenamefont {Brink}\ \emph {et~al.}(2003)\citenamefont {Brink},
  \citenamefont {Nor�n}, \citenamefont {Goossens}, \citenamefont {Withers},
  \citenamefont {Liu},\ and\ \citenamefont {Xu}}]{Brink2003450}%
  \BibitemOpen
  \bibfield  {author} {\bibinfo {author} {\bibfnamefont {Frank~J.}\
  \bibnamefont {Brink}}, \bibinfo {author} {\bibfnamefont {Lasse}\ \bibnamefont
  {Nor�n}}, \bibinfo {author} {\bibfnamefont {Darren~J.}\ \bibnamefont
  {Goossens}}, \bibinfo {author} {\bibfnamefont {Ray~L.}\ \bibnamefont
  {Withers}}, \bibinfo {author} {\bibfnamefont {Yun}\ \bibnamefont {Liu}}, \
  and\ \bibinfo {author} {\bibfnamefont {Chao-Nan}\ \bibnamefont {Xu}},\
  }\bibfield  {title} {\enquote {\bibinfo {title} {{A combined diffraction
  (XRD, electron and neutron) and electrical study of Na$_3$MoO$_3$F$_3$ }},}\
  }\href {\doibase http://dx.doi.org/10.1016/S0022-4596(03)00303-7} {\bibfield
  {journal} {\bibinfo  {journal} {Journal of Solid State Chemistry}\ }\textbf
  {\bibinfo {volume} {174}},\ \bibinfo {pages} {450 -- 458} (\bibinfo {year}
  {2003})}\BibitemShut {NoStop}%
\bibitem [{\citenamefont {Mattes}\ \emph {et~al.}(1980)\citenamefont {Mattes},
  \citenamefont {Mennemann}, \citenamefont {J�ckel}, \citenamefont
  {Rieskamp},\ and\ \citenamefont {Brockmeyer}}]{Cs3Mo2O6F3:1980}%
  \BibitemOpen
  \bibfield  {author} {\bibinfo {author} {\bibfnamefont {Rainer}\ \bibnamefont
  {Mattes}}, \bibinfo {author} {\bibfnamefont {Karl}\ \bibnamefont
  {Mennemann}}, \bibinfo {author} {\bibfnamefont {Norbert}\ \bibnamefont
  {J�ckel}}, \bibinfo {author} {\bibfnamefont {Helmut}\ \bibnamefont
  {Rieskamp}}, \ and\ \bibinfo {author} {\bibfnamefont {Heinz-Josef}\
  \bibnamefont {Brockmeyer}},\ }\bibfield  {title} {\enquote {\bibinfo {title}
  {{Structure and properties of the fluorine-rich oxofluoromolybdates
  Cs$_3$[Mo$_2$O$_6$F$_3$], (NH$_4$)$_3$[Mo$_2$O$_2$F$_9$] and
  (NH$_4$)$_2$[MoOF$_5$]}},}\ }\href {\doibase
  http://dx.doi.org/10.1016/0022-5088(80)90023-5} {\bibfield  {journal}
  {\bibinfo  {journal} {Journal of the Less Common Metals}\ }\textbf {\bibinfo
  {volume} {76}},\ \bibinfo {pages} {199 -- 212} (\bibinfo {year}
  {1980})}\BibitemShut {NoStop}%
\bibitem [{\citenamefont {Torardi}\ and\ \citenamefont
  {Brixner}(1985)}]{Ba2WO3F4:1985}%
  \BibitemOpen
  \bibfield  {author} {\bibinfo {author} {\bibfnamefont {C.C.}\ \bibnamefont
  {Torardi}}\ and\ \bibinfo {author} {\bibfnamefont {L.H.}\ \bibnamefont
  {Brixner}},\ }\bibfield  {title} {\enquote {\bibinfo {title} {Structure and
  luminescence of {Ba$_2$WO$_3$F$_4$}},}\ }\href {\doibase
  http://dx.doi.org/10.1016/0025-5408(85)90039-X} {\bibfield  {journal}
  {\bibinfo  {journal} {Materials Research Bulletin}\ }\textbf {\bibinfo
  {volume} {20}},\ \bibinfo {pages} {137 -- 145} (\bibinfo {year}
  {1985})}\BibitemShut {NoStop}%
\bibitem [{\citenamefont {McCabe}\ \emph {et~al.}(2007)\citenamefont {McCabe},
  \citenamefont {Jones}, \citenamefont {Zhang}, \citenamefont {Hyatt},\ and\
  \citenamefont {Greaves}}]{Bi2NbO5F:2007}%
  \BibitemOpen
  \bibfield  {author} {\bibinfo {author} {\bibfnamefont {E.~E.}\ \bibnamefont
  {McCabe}}, \bibinfo {author} {\bibfnamefont {I.~P.}\ \bibnamefont {Jones}},
  \bibinfo {author} {\bibfnamefont {D.}~\bibnamefont {Zhang}}, \bibinfo
  {author} {\bibfnamefont {N.~C.}\ \bibnamefont {Hyatt}}, \ and\ \bibinfo
  {author} {\bibfnamefont {C.}~\bibnamefont {Greaves}},\ }\bibfield  {title}
  {\enquote {\bibinfo {title} {{Crystal structure and electrical
  characterisation of Bi$_2$NbO$_5$F: An Aurivillius oxide fluoride}},}\ }\href
  {\doibase 10.1039/B613970A} {\bibfield  {journal} {\bibinfo  {journal} {J.
  Mater. Chem.}\ }\textbf {\bibinfo {volume} {17}},\ \bibinfo {pages}
  {1193--1200} (\bibinfo {year} {2007})}\BibitemShut {NoStop}%
\bibitem [{\citenamefont {Needs}\ and\ \citenamefont
  {Weller}(1995)}]{Ba2InO3F:1995}%
  \BibitemOpen
  \bibfield  {author} {\bibinfo {author} {\bibfnamefont {Richard~L.}\
  \bibnamefont {Needs}}\ and\ \bibinfo {author} {\bibfnamefont {Mark~T.}\
  \bibnamefont {Weller}},\ }\bibfield  {title} {\enquote {\bibinfo {title}
  {{Synthesis and structure of Ba$_2$InO$_3$F: oxide/fluoride ordering in a new
  K$_2$NiF$_4$ superstructure}},}\ }\href {\doibase 10.1039/C39950000353}
  {\bibfield  {journal} {\bibinfo  {journal} {J. Chem. Soc.{,} Chem. Commun.}\
  ,\ \bibinfo {pages} {353--354}} (\bibinfo {year} {1995})}\BibitemShut
  {NoStop}%
\bibitem [{\citenamefont {Kiat}\ and\ \citenamefont
  {Roisnel}(1996)}]{I4mcm:STO1996}%
  \BibitemOpen
  \bibfield  {author} {\bibinfo {author} {\bibfnamefont {J~M}\ \bibnamefont
  {Kiat}}\ and\ \bibinfo {author} {\bibfnamefont {Thierry}\ \bibnamefont
  {Roisnel}},\ }\bibfield  {title} {\enquote {\bibinfo {title} {Rietveld
  analysis of strontium titanate in the m\"uller state},}\ }\href
  {http://stacks.iop.org/0953-8984/8/i=19/a=021} {\bibfield  {journal}
  {\bibinfo  {journal} {Journal of Physics: Condensed Matter}\ }\textbf
  {\bibinfo {volume} {8}},\ \bibinfo {pages} {3471} (\bibinfo {year}
  {1996})}\BibitemShut {NoStop}%
\bibitem [{\citenamefont {Buttner}\ and\ \citenamefont
  {Maslen}(1992)}]{BTO:1992}%
  \BibitemOpen
  \bibfield  {author} {\bibinfo {author} {\bibfnamefont {R.~H.}\ \bibnamefont
  {Buttner}}\ and\ \bibinfo {author} {\bibfnamefont {E.~N.}\ \bibnamefont
  {Maslen}},\ }\bibfield  {title} {\enquote {\bibinfo {title} {{Structural
  parameters and electron difference density in {BaTiO}$_3$}},}\ }\href
  {\doibase 10.1107/S010876819200510X} {\bibfield  {journal} {\bibinfo
  {journal} {Acta Crystallographica Section B}\ }\textbf {\bibinfo {volume}
  {48}},\ \bibinfo {pages} {764--769} (\bibinfo {year} {1992})}\BibitemShut
  {NoStop}%
\bibitem [{\citenamefont {Tressaud}\ and\ \citenamefont
  {Dance}(1982)}]{Tressaud1982}%
  \BibitemOpen
  \bibfield  {author} {\bibinfo {author} {\bibfnamefont {Alain}\ \bibnamefont
  {Tressaud}}\ and\ \bibinfo {author} {\bibfnamefont {Jean-Michel}\
  \bibnamefont {Dance}},\ }\enquote {\bibinfo {title} {Relationships between
  structure and low-dimensional magnetism in fluorides},}\ in\ \href {\doibase
  10.1007/BFb0111297} {\emph {\bibinfo {booktitle} {Structures versus Special
  Properties}}}\ (\bibinfo  {publisher} {Springer Berlin Heidelberg},\ \bibinfo
  {address} {Berlin, Heidelberg},\ \bibinfo {year} {1982})\ pp.\ \bibinfo
  {pages} {87--146}\BibitemShut {NoStop}%
\bibitem [{\citenamefont {Slater}(1951)}]{Slater:Insulator1951}%
  \BibitemOpen
  \bibfield  {author} {\bibinfo {author} {\bibfnamefont {J.~C.}\ \bibnamefont
  {Slater}},\ }\bibfield  {title} {\enquote {\bibinfo {title} {Magnetic effects
  and the hartree-fock equation},}\ }\href {\doibase 10.1103/PhysRev.82.538}
  {\bibfield  {journal} {\bibinfo  {journal} {Phys. Rev.}\ }\textbf {\bibinfo
  {volume} {82}},\ \bibinfo {pages} {538--541} (\bibinfo {year}
  {1951})}\BibitemShut {NoStop}%
\bibitem [{\citenamefont {Saiki}(1972)}]{C-type:KMnF31972}%
  \BibitemOpen
  \bibfield  {author} {\bibinfo {author} {\bibfnamefont {Kunio}\ \bibnamefont
  {Saiki}},\ }\bibfield  {title} {\enquote {\bibinfo {title} {Resonance
  behaviour in canted antiferromagnet {KMnF$_3$}},}\ }\href {\doibase
  10.1143/JPSJ.33.1284} {\bibfield  {journal} {\bibinfo  {journal} {Journal of
  the Physical Society of Japan}\ }\textbf {\bibinfo {volume} {33}},\ \bibinfo
  {pages} {1284--1291} (\bibinfo {year} {1972})},\ \Eprint
  {http://arxiv.org/abs/http://dx.doi.org/10.1143/JPSJ.33.1284}
  {http://dx.doi.org/10.1143/JPSJ.33.1284} \BibitemShut {NoStop}%
\bibitem [{\citenamefont {Charles}\ and\ \citenamefont
  {Rondinelli}(2014)}]{Nenian_PRB_NMF_2014}%
  \BibitemOpen
  \bibfield  {author} {\bibinfo {author} {\bibfnamefont {Nenian}\ \bibnamefont
  {Charles}}\ and\ \bibinfo {author} {\bibfnamefont {James~M.}\ \bibnamefont
  {Rondinelli}},\ }\bibfield  {title} {\enquote {\bibinfo {title} {Microscopic
  origin of pressure-induced isosymmetric transitions in fluoromanganate
  cryolites},}\ }\href {\doibase 10.1103/PhysRevB.90.094114} {\bibfield
  {journal} {\bibinfo  {journal} {Phys. Rev. B}\ }\textbf {\bibinfo {volume}
  {90}},\ \bibinfo {pages} {094114} (\bibinfo {year} {2014})}\BibitemShut
  {NoStop}%
\bibitem [{\citenamefont {Czy\ifmmode~\dot{z}\else \.{z}\fi{}yk}\ and\
  \citenamefont {Sawatzky}(1994)}]{Sawatzky:1994}%
  \BibitemOpen
  \bibfield  {author} {\bibinfo {author} {\bibfnamefont {M.~T.}\ \bibnamefont
  {Czy\ifmmode~\dot{z}\else \.{z}\fi{}yk}}\ and\ \bibinfo {author}
  {\bibfnamefont {G.~A.}\ \bibnamefont {Sawatzky}},\ }\bibfield  {title}
  {\enquote {\bibinfo {title} {{Local-density functional and on-site
  correlations: The electronic structure of La$_2$CuO$_4$ and LaCuO$_3$}},}\
  }\href@noop {} {\bibfield  {journal} {\bibinfo  {journal} {Physical Review
  B}\ }\textbf {\bibinfo {volume} {49}},\ \bibinfo {pages} {14211--14228}
  (\bibinfo {year} {1994})}\BibitemShut {NoStop}%
\bibitem [{\citenamefont {Liechtenstein}\ \emph {et~al.}(1995)\citenamefont
  {Liechtenstein}, \citenamefont {Anisimov},\ and\ \citenamefont
  {Zaanen}}]{Liechtenstein/Anisimov/Zaanen:1995}%
  \BibitemOpen
  \bibfield  {author} {\bibinfo {author} {\bibfnamefont {A.~I.}\ \bibnamefont
  {Liechtenstein}}, \bibinfo {author} {\bibfnamefont {V.~I.}\ \bibnamefont
  {Anisimov}}, \ and\ \bibinfo {author} {\bibfnamefont {J.}~\bibnamefont
  {Zaanen}},\ }\bibfield  {title} {\enquote {\bibinfo {title}
  {Density-functional theory and strong interactions: Orbital ordering in
  mott-hubbard insulators},}\ }\href@noop {} {\bibfield  {journal} {\bibinfo
  {journal} {Physical Review B}\ }\textbf {\bibinfo {volume} {52}},\ \bibinfo
  {pages} {R5467--R5470} (\bibinfo {year} {1995})}\BibitemShut {NoStop}%
\bibitem [{\citenamefont {Sun}\ \emph {et~al.}(2015{\natexlab{b}})\citenamefont
  {Sun}, \citenamefont {Ruzsinszky},\ and\ \citenamefont {Perdew}}]{PRL_SCAN}%
  \BibitemOpen
  \bibfield  {author} {\bibinfo {author} {\bibfnamefont {Jianwei}\ \bibnamefont
  {Sun}}, \bibinfo {author} {\bibfnamefont {Adrienn}\ \bibnamefont
  {Ruzsinszky}}, \ and\ \bibinfo {author} {\bibfnamefont {John~P.}\
  \bibnamefont {Perdew}},\ }\bibfield  {title} {\enquote {\bibinfo {title}
  {Strongly constrained and appropriately normed semilocal density
  functional},}\ }\href {\doibase 10.1103/PhysRevLett.115.036402} {\bibfield
  {journal} {\bibinfo  {journal} {Phys. Rev. Lett.}\ }\textbf {\bibinfo
  {volume} {115}},\ \bibinfo {pages} {036402} (\bibinfo {year}
  {2015}{\natexlab{b}})}\BibitemShut {NoStop}%
\bibitem [{\citenamefont {He}(2004)}]{He2004135}%
  \BibitemOpen
  \bibfield  {author} {\bibinfo {author} {\bibfnamefont {Yi}~\bibnamefont
  {He}},\ }\bibfield  {title} {\enquote {\bibinfo {title} {Heat capacity,
  thermal conductivity, and thermal expansion of barium titanate-based
  ceramics},}\ }\href {\doibase http://dx.doi.org/10.1016/j.tca.2004.02.008}
  {\bibfield  {journal} {\bibinfo  {journal} {Thermochimica Acta}\ }\textbf
  {\bibinfo {volume} {419}},\ \bibinfo {pages} {135 -- 141} (\bibinfo {year}
  {2004})}\BibitemShut {NoStop}%
\bibitem [{\citenamefont {Huang}\ \emph {et~al.}(2016)\citenamefont {Huang},
  \citenamefont {Lu}, \citenamefont {Tennessen},\ and\ \citenamefont
  {Rondinelli}}]{Huang201684}%
  \BibitemOpen
  \bibfield  {author} {\bibinfo {author} {\bibfnamefont {Liang-Feng}\
  \bibnamefont {Huang}}, \bibinfo {author} {\bibfnamefont {Xue-Zeng}\
  \bibnamefont {Lu}}, \bibinfo {author} {\bibfnamefont {Emrys}\ \bibnamefont
  {Tennessen}}, \ and\ \bibinfo {author} {\bibfnamefont {James~M.}\
  \bibnamefont {Rondinelli}},\ }\bibfield  {title} {\enquote {\bibinfo {title}
  {An efficient ab-initio quasiharmonic approach for the thermodynamics of
  solids},}\ }\href {\doibase
  http://dx.doi.org/10.1016/j.commatsci.2016.04.012} {\bibfield  {journal}
  {\bibinfo  {journal} {Computational Materials Science}\ }\textbf {\bibinfo
  {volume} {120}},\ \bibinfo {pages} {84 -- 93} (\bibinfo {year}
  {2016})}\BibitemShut {NoStop}%
\bibitem [{\citenamefont {de~la Flor}\ \emph {et~al.}(2016)\citenamefont {de~la
  Flor}, \citenamefont {Orobengoa}, \citenamefont {Tasci}, \citenamefont
  {Perez-Mato},\ and\ \citenamefont {Aroyo}}]{delaFlor:to5129}%
  \BibitemOpen
  \bibfield  {author} {\bibinfo {author} {\bibfnamefont {Gemma}\ \bibnamefont
  {de~la Flor}}, \bibinfo {author} {\bibfnamefont {Danel}\ \bibnamefont
  {Orobengoa}}, \bibinfo {author} {\bibfnamefont {Emre}\ \bibnamefont {Tasci}},
  \bibinfo {author} {\bibfnamefont {Juan~Manuel}\ \bibnamefont {Perez-Mato}}, \
  and\ \bibinfo {author} {\bibfnamefont {Mois~I.}\ \bibnamefont {Aroyo}},\
  }\bibfield  {title} {\enquote {\bibinfo {title} {{Comparison of structures
  applying the tools available at the Bilbao Crystallographic Server}},}\
  }\href {\doibase 10.1107/S1600576716002569} {\bibfield  {journal} {\bibinfo
  {journal} {Journal of Applied Crystallography}\ }\textbf {\bibinfo {volume}
  {49}},\ \bibinfo {pages} {653--664} (\bibinfo {year} {2016})}\BibitemShut
  {NoStop}%
\bibitem [{\citenamefont {Brown}\ and\ \citenamefont
  {Poeppelmeier}(2014)}]{BondValenceBookPoeppe}%
  \BibitemOpen
  \bibfield  {author} {\bibinfo {author} {\bibfnamefont {I~David}\ \bibnamefont
  {Brown}}\ and\ \bibinfo {author} {\bibfnamefont {Kenneth~R}\ \bibnamefont
  {Poeppelmeier}},\ }\href@noop {} {\emph {\bibinfo {title} {Bond Valences}}},\
  Vol.\ \bibinfo {volume} {158}\ (\bibinfo  {publisher} {Springer},\ \bibinfo
  {year} {2014})\BibitemShut {NoStop}%
\bibitem [{\citenamefont {Adams}\ \emph {et~al.}(2004)\citenamefont {Adams},
  \citenamefont {Moretzki},\ and\ \citenamefont {Canadell}}]{Adams2004281}%
  \BibitemOpen
  \bibfield  {author} {\bibinfo {author} {\bibfnamefont {Stefan}\ \bibnamefont
  {Adams}}, \bibinfo {author} {\bibfnamefont {Olaf}\ \bibnamefont {Moretzki}},
  \ and\ \bibinfo {author} {\bibfnamefont {Enric}\ \bibnamefont {Canadell}},\
  }\bibfield  {title} {\enquote {\bibinfo {title} {Global instability index
  optimizations for the localization of mobile protons},}\ }\href {\doibase
  http://dx.doi.org/10.1016/j.ssi.2003.04.002} {\bibfield  {journal} {\bibinfo
  {journal} {Solid State Ionics}\ }\textbf {\bibinfo {volume} {168}},\ \bibinfo
  {pages} {281 -- 290} (\bibinfo {year} {2004})},\ \bibinfo {note} {proceedings
  of the Workshop on Hydrogen: Ionic, Atomic and Molecular Motion}\BibitemShut
  {NoStop}%
\bibitem [{\citenamefont {Garcia-Fernandez}\ \emph {et~al.}(2010)\citenamefont
  {Garcia-Fernandez}, \citenamefont {Aramburu}, \citenamefont {Barriuso},\ and\
  \citenamefont {Moreno}}]{GarciaFernandezetal:2010}%
  \BibitemOpen
  \bibfield  {author} {\bibinfo {author} {\bibfnamefont {P.}~\bibnamefont
  {Garcia-Fernandez}}, \bibinfo {author} {\bibfnamefont {J.A.}\ \bibnamefont
  {Aramburu}}, \bibinfo {author} {\bibfnamefont {M.T.}\ \bibnamefont
  {Barriuso}}, \ and\ \bibinfo {author} {\bibfnamefont {M.}~\bibnamefont
  {Moreno}},\ }\bibfield  {title} {\enquote {\bibinfo {title} {Key role of
  covalent bonding in octahedral tilting in perovskites},}\ }\href {\doibase
  10.1021/jz900399m} {\bibfield  {journal} {\bibinfo  {journal} {J. Phys. Chem.
  Lett.}\ }\textbf {\bibinfo {volume} {1}},\ \bibinfo {pages} {647--651}
  (\bibinfo {year} {2010})},\ \Eprint
  {http://arxiv.org/abs/http://pubs.acs.org/doi/pdf/10.1021/jz900399m}
  {http://pubs.acs.org/doi/pdf/10.1021/jz900399m} \BibitemShut {NoStop}%
\bibitem [{\citenamefont {Shirane}\ \emph {et~al.}(1957)\citenamefont
  {Shirane}, \citenamefont {Danner},\ and\ \citenamefont
  {Pepinsky}}]{Shirane/BTO:1957}%
  \BibitemOpen
  \bibfield  {author} {\bibinfo {author} {\bibfnamefont {G.}~\bibnamefont
  {Shirane}}, \bibinfo {author} {\bibfnamefont {H.}~\bibnamefont {Danner}}, \
  and\ \bibinfo {author} {\bibfnamefont {R.}~\bibnamefont {Pepinsky}},\
  }\bibfield  {title} {\enquote {\bibinfo {title} {Neutron diffraction study of
  orthorhombic bati${\mathrm{o}}_{3}$},}\ }\href {\doibase
  10.1103/PhysRev.105.856} {\bibfield  {journal} {\bibinfo  {journal} {Phys.
  Rev.}\ }\textbf {\bibinfo {volume} {105}},\ \bibinfo {pages} {856--860}
  (\bibinfo {year} {1957})}\BibitemShut {NoStop}%
\bibitem [{\citenamefont {Bilc}\ \emph {et~al.}(2008)\citenamefont {Bilc},
  \citenamefont {Orlando}, \citenamefont {Shaltaf}, \citenamefont {Rignanese},
  \citenamefont {\'I\~niguez},\ and\ \citenamefont
  {Ghosez}}]{Bilc/Ghosez_et_al:2008}%
  \BibitemOpen
  \bibfield  {author} {\bibinfo {author} {\bibfnamefont {D.~I.}\ \bibnamefont
  {Bilc}}, \bibinfo {author} {\bibfnamefont {R.}~\bibnamefont {Orlando}},
  \bibinfo {author} {\bibfnamefont {R.}~\bibnamefont {Shaltaf}}, \bibinfo
  {author} {\bibfnamefont {G.-M.}\ \bibnamefont {Rignanese}}, \bibinfo {author}
  {\bibfnamefont {Jorge}\ \bibnamefont {\'I\~niguez}}, \ and\ \bibinfo {author}
  {\bibfnamefont {Ph.}\ \bibnamefont {Ghosez}},\ }\bibfield  {title} {\enquote
  {\bibinfo {title} {Hybrid exchange-correlation functional for accurate
  prediction of the electronic and structural properties of ferroelectric
  oxides},}\ }\href {\doibase 10.1103/PhysRevB.77.165107} {\bibfield  {journal}
  {\bibinfo  {journal} {Physical Review B}\ }\textbf {\bibinfo {volume} {77}},\
  \bibinfo {pages} {165107} (\bibinfo {year} {2008})}\BibitemShut {NoStop}%
\bibitem [{\citenamefont {Wu}\ \emph {et~al.}(2004)\citenamefont {Wu},
  \citenamefont {Cohen},\ and\ \citenamefont {Singh}}]{Wu/CohenSingh:2004}%
  \BibitemOpen
  \bibfield  {author} {\bibinfo {author} {\bibfnamefont {Zhigang}\ \bibnamefont
  {Wu}}, \bibinfo {author} {\bibfnamefont {R.~E.}\ \bibnamefont {Cohen}}, \
  and\ \bibinfo {author} {\bibfnamefont {D.~J.}\ \bibnamefont {Singh}},\
  }\bibfield  {title} {\enquote {\bibinfo {title} {Comparing the weighted
  density approximation with the lda and gga for ground-state properties of
  ferroelectric perovskites},}\ }\href {\doibase 10.1103/PhysRevB.70.104112}
  {\bibfield  {journal} {\bibinfo  {journal} {Phys. Rev. B}\ }\textbf {\bibinfo
  {volume} {70}},\ \bibinfo {pages} {104112} (\bibinfo {year}
  {2004})}\BibitemShut {NoStop}%
\bibitem [{\citenamefont {Fox}(2001)}]{fox2001optical}%
  \BibitemOpen
  \bibfield  {author} {\bibinfo {author} {\bibfnamefont {Anthony~Mark}\
  \bibnamefont {Fox}},\ }\href@noop {} {\emph {\bibinfo {title} {Optical
  properties of solids}}},\ Vol.~\bibinfo {volume} {3}\ (\bibinfo  {publisher}
  {Oxford University Press, USA},\ \bibinfo {year} {2001})\BibitemShut
  {NoStop}%
\bibitem [{\citenamefont {van Benthem}\ \emph {et~al.}(2001)\citenamefont {van
  Benthem}, \citenamefont {Els\"{a}sser},\ and\ \citenamefont
  {French}}]{Benthem/French:2001}%
  \BibitemOpen
  \bibfield  {author} {\bibinfo {author} {\bibfnamefont {K.}~\bibnamefont {van
  Benthem}}, \bibinfo {author} {\bibfnamefont {C.}~\bibnamefont
  {Els\"{a}sser}}, \ and\ \bibinfo {author} {\bibfnamefont {R.~H.}\
  \bibnamefont {French}},\ }\bibfield  {title} {\enquote {\bibinfo {title}
  {{Bulk electronic structure of SrTiO$_3$: Experiment and theory}},}\ }\href
  {\doibase 10.1063/1.1415766} {\bibfield  {journal} {\bibinfo  {journal}
  {Journal of Applied Physics}\ }\textbf {\bibinfo {volume} {90}},\ \bibinfo
  {pages} {6156--6164} (\bibinfo {year} {2001})}\BibitemShut {NoStop}%
\bibitem [{\citenamefont {Wemple}(1970)}]{BTOBandgap:1970}%
  \BibitemOpen
  \bibfield  {author} {\bibinfo {author} {\bibfnamefont {S.~H.}\ \bibnamefont
  {Wemple}},\ }\bibfield  {title} {\enquote {\bibinfo {title} {Polarization
  fluctuations and the optical-absorption edge in {BaTiO}$_3$},}\ }\href
  {\doibase 10.1103/PhysRevB.2.2679} {\bibfield  {journal} {\bibinfo  {journal}
  {Phys. Rev. B}\ }\textbf {\bibinfo {volume} {2}},\ \bibinfo {pages}
  {2679--2689} (\bibinfo {year} {1970})}\BibitemShut {NoStop}%
\bibitem [{\citenamefont {Yang}\ \emph {et~al.}(2016)\citenamefont {Yang},
  \citenamefont {Peng}, \citenamefont {Sun},\ and\ \citenamefont
  {Perdew}}]{Zeng-hui:metaGGAgaps2016}%
  \BibitemOpen
  \bibfield  {author} {\bibinfo {author} {\bibfnamefont {Zeng-hui}\
  \bibnamefont {Yang}}, \bibinfo {author} {\bibfnamefont {Haowei}\ \bibnamefont
  {Peng}}, \bibinfo {author} {\bibfnamefont {Jianwei}\ \bibnamefont {Sun}}, \
  and\ \bibinfo {author} {\bibfnamefont {John~P.}\ \bibnamefont {Perdew}},\
  }\bibfield  {title} {\enquote {\bibinfo {title} {{More realistic band gaps
  from meta-generalized gradient approximations: Only in a generalized
  Kohn-Sham scheme}},}\ }\href {\doibase 10.1103/PhysRevB.93.205205} {\bibfield
   {journal} {\bibinfo  {journal} {Phys. Rev. B}\ }\textbf {\bibinfo {volume}
  {93}},\ \bibinfo {pages} {205205} (\bibinfo {year} {2016})}\BibitemShut
  {NoStop}%
\bibitem [{\citenamefont {Kapusta}\ \emph {et~al.}(1999)\citenamefont
  {Kapusta}, \citenamefont {Daniel},\ and\ \citenamefont
  {Ratuszna}}]{KMnF3_revised_str}%
  \BibitemOpen
  \bibfield  {author} {\bibinfo {author} {\bibfnamefont {Joanna}\ \bibnamefont
  {Kapusta}}, \bibinfo {author} {\bibfnamefont {Philippe}\ \bibnamefont
  {Daniel}}, \ and\ \bibinfo {author} {\bibfnamefont {Alicja}\ \bibnamefont
  {Ratuszna}},\ }\bibfield  {title} {\enquote {\bibinfo {title} {Revised
  structural phase transitions in the archetype {KMnF}$_{3}$ perovskite
  crystal},}\ }\href {\doibase 10.1103/PhysRevB.59.14235} {\bibfield  {journal}
  {\bibinfo  {journal} {Phys. Rev. B}\ }\textbf {\bibinfo {volume} {59}},\
  \bibinfo {pages} {14235--14245} (\bibinfo {year} {1999})}\BibitemShut
  {NoStop}%
\bibitem [{\citenamefont {Islam}\ \emph {et~al.}(2013)\citenamefont {Islam},
  \citenamefont {Rondinelli},\ and\ \citenamefont
  {Spanier}}]{Rondinelli/Spanier:2013}%
  \BibitemOpen
  \bibfield  {author} {\bibinfo {author} {\bibfnamefont {Mohammad~A}\
  \bibnamefont {Islam}}, \bibinfo {author} {\bibfnamefont {James~M}\
  \bibnamefont {Rondinelli}}, \ and\ \bibinfo {author} {\bibfnamefont
  {Jonathan~E}\ \bibnamefont {Spanier}},\ }\bibfield  {title} {\enquote
  {\bibinfo {title} {Normal mode determination of perovskite crystal structures
  with octahedral rotations: theory and applications},}\ }\href
  {http://stacks.iop.org/0953-8984/25/i=17/a=175902} {\bibfield  {journal}
  {\bibinfo  {journal} {Journal of Physics: Condensed Matter}\ }\textbf
  {\bibinfo {volume} {25}},\ \bibinfo {pages} {175902} (\bibinfo {year}
  {2013})}\BibitemShut {NoStop}%
\bibitem [{\citenamefont {Axe}\ and\ \citenamefont
  {Pettit}(1967)}]{KMnF3_Phonondata_cubic}%
  \BibitemOpen
  \bibfield  {author} {\bibinfo {author} {\bibfnamefont {J.~D.}\ \bibnamefont
  {Axe}}\ and\ \bibinfo {author} {\bibfnamefont {G.~D.}\ \bibnamefont
  {Pettit}},\ }\bibfield  {title} {\enquote {\bibinfo {title} {Infrared
  dielectric dispersion of several fluoride perovskites},}\ }\href {\doibase
  10.1103/PhysRev.157.435} {\bibfield  {journal} {\bibinfo  {journal} {Phys.
  Rev.}\ }\textbf {\bibinfo {volume} {157}},\ \bibinfo {pages} {435--437}
  (\bibinfo {year} {1967})}\BibitemShut {NoStop}%
\bibitem [{\citenamefont {Krylov}\ \emph {et~al.}(2014)\citenamefont {Krylov},
  \citenamefont {Sofronova}, \citenamefont {Kolesnikova}, \citenamefont
  {Ivanov}, \citenamefont {Sukhovsky}, \citenamefont {Goryainov}, \citenamefont
  {Ivanenko}, \citenamefont {Shestakov}, \citenamefont {Kocharova},\ and\
  \citenamefont {Vtyurin}}]{Krylov201432}%
  \BibitemOpen
  \bibfield  {author} {\bibinfo {author} {\bibfnamefont {A.S.}\ \bibnamefont
  {Krylov}}, \bibinfo {author} {\bibfnamefont {S.N.}\ \bibnamefont
  {Sofronova}}, \bibinfo {author} {\bibfnamefont {E.M.}\ \bibnamefont
  {Kolesnikova}}, \bibinfo {author} {\bibfnamefont {Yu.N.}\ \bibnamefont
  {Ivanov}}, \bibinfo {author} {\bibfnamefont {A.A.}\ \bibnamefont
  {Sukhovsky}}, \bibinfo {author} {\bibfnamefont {S.V.}\ \bibnamefont
  {Goryainov}}, \bibinfo {author} {\bibfnamefont {A.A.}\ \bibnamefont
  {Ivanenko}}, \bibinfo {author} {\bibfnamefont {N.P.}\ \bibnamefont
  {Shestakov}}, \bibinfo {author} {\bibfnamefont {A.G.}\ \bibnamefont
  {Kocharova}}, \ and\ \bibinfo {author} {\bibfnamefont {A.N.}\ \bibnamefont
  {Vtyurin}},\ }\bibfield  {title} {\enquote {\bibinfo {title} {Experimental
  and theoretical methods to study structural phase transition mechanisms in
  {K$_3$WO$_3$F$_3$} oxyfluoride},}\ }\href {\doibase
  http://dx.doi.org/10.1016/j.jssc.2014.05.028} {\bibfield  {journal} {\bibinfo
   {journal} {Journal of Solid State Chemistry}\ }\textbf {\bibinfo {volume}
  {218}},\ \bibinfo {pages} {32 -- 37} (\bibinfo {year} {2014})}\BibitemShut
  {NoStop}%
\bibitem [{\citenamefont {Voit}\ \emph {et~al.}(2006)\citenamefont {Voit},
  \citenamefont {Voit}, \citenamefont {Mashkovskii}, \citenamefont {Laptash},\
  and\ \citenamefont {Kavun}}]{Voit2006}%
  \BibitemOpen
  \bibfield  {author} {\bibinfo {author} {\bibfnamefont {E.~I.}\ \bibnamefont
  {Voit}}, \bibinfo {author} {\bibfnamefont {A.~V.}\ \bibnamefont {Voit}},
  \bibinfo {author} {\bibfnamefont {A.~A.}\ \bibnamefont {Mashkovskii}},
  \bibinfo {author} {\bibfnamefont {N.~M.}\ \bibnamefont {Laptash}}, \ and\
  \bibinfo {author} {\bibfnamefont {V.~Ya.}\ \bibnamefont {Kavun}},\ }\bibfield
   {title} {\enquote {\bibinfo {title} {Dynamic disorder in ammonium
  oxofluorotungstates {(NH$_4$)$_2$WO$_2$F$_4$} and
  {(NH$_4$)$_3$WO$_3$F$_3$}},}\ }\href {\doibase 10.1007/s10947-006-0351-3}
  {\bibfield  {journal} {\bibinfo  {journal} {Journal of Structural Chemistry}\
  }\textbf {\bibinfo {volume} {47}},\ \bibinfo {pages} {642--650} (\bibinfo
  {year} {2006})}\BibitemShut {NoStop}%
\bibitem [{\citenamefont {Jr.}(1963)}]{doi:10.1021/ic50008a029}%
  \BibitemOpen
  \bibfield  {author} {\bibinfo {author} {\bibfnamefont {O.~L.~Keller}\
  \bibnamefont {Jr.}},\ }\bibfield  {title} {\enquote {\bibinfo {title}
  {Identification of complex ions of niobium({V}) in hydrofluoric acid
  solutions by raman and infrared spectroscopy},}\ }\href {\doibase
  10.1021/ic50008a029} {\bibfield  {journal} {\bibinfo  {journal} {Inorganic
  Chemistry}\ }\textbf {\bibinfo {volume} {2}},\ \bibinfo {pages} {783--787}
  (\bibinfo {year} {1963})},\ \Eprint
  {http://arxiv.org/abs/http://dx.doi.org/10.1021/ic50008a029}
  {http://dx.doi.org/10.1021/ic50008a029} \BibitemShut {NoStop}%
\bibitem [{\citenamefont {Pausewang}\ and\ \citenamefont
  {R{\"U}dorff}(1969)}]{OxF6-x_Vibrational_spec}%
  \BibitemOpen
  \bibfield  {author} {\bibinfo {author} {\bibfnamefont {G.}~\bibnamefont
  {Pausewang}}\ and\ \bibinfo {author} {\bibfnamefont {W.}~\bibnamefont
  {R{\"U}dorff}},\ }\bibfield  {title} {\enquote {\bibinfo {title} {{\"U}ber
  alkali-oxofluorometallate der {\"u}bergangsmetalle.
  {A$^I_3$MeO$_x$F$_{6-x}$}-verbindungen mit x = 1, 2, 3},}\ }\href {\doibase
  10.1002/zaac.19693640107} {\bibfield  {journal} {\bibinfo  {journal}
  {Zeitschrift f�r anorganische und allgemeine Chemie}\ }\textbf {\bibinfo
  {volume} {364}},\ \bibinfo {pages} {69--87} (\bibinfo {year}
  {1969})}\BibitemShut {NoStop}%
\bibitem [{\citenamefont {Krylov}\ \emph {et~al.}(2012)\citenamefont {Krylov},
  \citenamefont {Merkushova}, \citenamefont {Vtyurin},\ and\ \citenamefont
  {Isaenko}}]{Krylov2012}%
  \BibitemOpen
  \bibfield  {author} {\bibinfo {author} {\bibfnamefont {A.~S.}\ \bibnamefont
  {Krylov}}, \bibinfo {author} {\bibfnamefont {E.~M.}\ \bibnamefont
  {Merkushova}}, \bibinfo {author} {\bibfnamefont {A.~N.}\ \bibnamefont
  {Vtyurin}}, \ and\ \bibinfo {author} {\bibfnamefont {L.~I.}\ \bibnamefont
  {Isaenko}},\ }\bibfield  {title} {\enquote {\bibinfo {title} {Raman
  spectroscopic study of the lattice dynamics in the {Rb$_2$KMoO$_3$F$_3$}
  oxyfluoride},}\ }\href {\doibase 10.1134/S1063783412060170} {\bibfield
  {journal} {\bibinfo  {journal} {Physics of the Solid State}\ }\textbf
  {\bibinfo {volume} {54}},\ \bibinfo {pages} {1275--1280} (\bibinfo {year}
  {2012})}\BibitemShut {NoStop}%
\bibitem [{\citenamefont {Dehnicke}\ \emph {et~al.}(1969)\citenamefont
  {Dehnicke}, \citenamefont {Pausewang},\ and\ \citenamefont
  {R�dorff}}]{ZAAC19693660107}%
  \BibitemOpen
  \bibfield  {author} {\bibinfo {author} {\bibfnamefont {K.}~\bibnamefont
  {Dehnicke}}, \bibinfo {author} {\bibfnamefont {G.}~\bibnamefont {Pausewang}},
  \ and\ \bibinfo {author} {\bibfnamefont {W.}~\bibnamefont {R�dorff}},\
  }\bibfield  {title} {\enquote {\bibinfo {title} {Die ir-spektren der
  oxofluorokomplexe {TiOF$_5$$^{3-}$}, {VOF$_5$$^{3-}$}, {NbO$_2$F$_4$$^{3-}$},
  {MoO$_3$F$_3$$^{3-}$} und {WO$_3$F$_3$$^{3-}$}},}\ }\href {\doibase
  10.1002/zaac.19693660107} {\bibfield  {journal} {\bibinfo  {journal}
  {Zeitschrift f�r anorganische und allgemeine Chemie}\ }\textbf {\bibinfo
  {volume} {366}},\ \bibinfo {pages} {64--72} (\bibinfo {year}
  {1969})}\BibitemShut {NoStop}%
\bibitem [{\citenamefont {Catlow}\ and\ \citenamefont
  {Stoneham}(1983)}]{IonicityReview:83}%
  \BibitemOpen
  \bibfield  {author} {\bibinfo {author} {\bibfnamefont {C~R~A}\ \bibnamefont
  {Catlow}}\ and\ \bibinfo {author} {\bibfnamefont {A~M}\ \bibnamefont
  {Stoneham}},\ }\bibfield  {title} {\enquote {\bibinfo {title} {Ionicity in
  solids},}\ }\href {http://stacks.iop.org/0022-3719/16/i=22/a=010} {\bibfield
  {journal} {\bibinfo  {journal} {Journal of Physics C: Solid State Physics}\
  }\textbf {\bibinfo {volume} {16}},\ \bibinfo {pages} {4321} (\bibinfo {year}
  {1983})}\BibitemShut {NoStop}%
\bibitem [{\citenamefont {Duffy}(1986)}]{DUFFY1986145}%
  \BibitemOpen
  \bibfield  {author} {\bibinfo {author} {\bibfnamefont {J.A.}\ \bibnamefont
  {Duffy}},\ }\bibfield  {title} {\enquote {\bibinfo {title} {Chemical bonding
  in the oxides of the elements: {A} new appraisal},}\ }\href {\doibase
  http://dx.doi.org/10.1016/0022-4596(86)90225-2} {\bibfield  {journal}
  {\bibinfo  {journal} {Journal of Solid State Chemistry}\ }\textbf {\bibinfo
  {volume} {62}},\ \bibinfo {pages} {145 -- 157} (\bibinfo {year}
  {1986})}\BibitemShut {NoStop}%
\bibitem [{\citenamefont {Meister}\ and\ \citenamefont
  {Schwarz}(1994)}]{Prin/ion:1994}%
  \BibitemOpen
  \bibfield  {author} {\bibinfo {author} {\bibfnamefont {J.}~\bibnamefont
  {Meister}}\ and\ \bibinfo {author} {\bibfnamefont {W.~H.~E.}\ \bibnamefont
  {Schwarz}},\ }\bibfield  {title} {\enquote {\bibinfo {title} {Principal
  components of ionicity},}\ }\href {\doibase 10.1021/j100084a048} {\bibfield
  {journal} {\bibinfo  {journal} {The Journal of Physical Chemistry}\ }\textbf
  {\bibinfo {volume} {98}},\ \bibinfo {pages} {8245--8252} (\bibinfo {year}
  {1994})}\BibitemShut {NoStop}%
\bibitem [{\citenamefont {Cammarata}\ and\ \citenamefont
  {Rondinelli}(2014)}]{AntonioCovalency}%
  \BibitemOpen
  \bibfield  {author} {\bibinfo {author} {\bibfnamefont {Antonio}\ \bibnamefont
  {Cammarata}}\ and\ \bibinfo {author} {\bibfnamefont {James~M.}\ \bibnamefont
  {Rondinelli}},\ }\bibfield  {title} {\enquote {\bibinfo {title} {Covalent
  dependence of octahedral rotations in orthorhombic perovskite oxides},}\
  }\href {\doibase http://dx.doi.org/10.1063/1.4895967} {\bibfield  {journal}
  {\bibinfo  {journal} {The Journal of Chemical Physics}\ }\textbf {\bibinfo
  {volume} {141}},\ \bibinfo {eid} {114704} (\bibinfo {year} {2014}),\
  http://dx.doi.org/10.1063/1.4895967}\BibitemShut {NoStop}%
\bibitem [{\citenamefont {Cohen}(1992)}]{Cohen:1992}%
  \BibitemOpen
  \bibfield  {author} {\bibinfo {author} {\bibfnamefont {Ronald~E.}\
  \bibnamefont {Cohen}},\ }\bibfield  {title} {\enquote {\bibinfo {title}
  {Origin of ferroelectricity in perovskite oxides},}\ }\href
  {http://dx.doi.org/10.1038/358136a0} {\bibfield  {journal} {\bibinfo
  {journal} {Nature}\ }\textbf {\bibinfo {volume} {358}},\ \bibinfo {pages}
  {136--138} (\bibinfo {year} {1992})}\BibitemShut {NoStop}%
\bibitem [{\citenamefont {Van~Vechten}(1969)}]{PhysRev.182.891}%
  \BibitemOpen
  \bibfield  {author} {\bibinfo {author} {\bibfnamefont {J.~A.}\ \bibnamefont
  {Van~Vechten}},\ }\bibfield  {title} {\enquote {\bibinfo {title} {Quantum
  dielectric theory of electronegativity in covalent systems. {I.} electronic
  dielectric constant},}\ }\href {\doibase 10.1103/PhysRev.182.891} {\bibfield
  {journal} {\bibinfo  {journal} {Phys. Rev.}\ }\textbf {\bibinfo {volume}
  {182}},\ \bibinfo {pages} {891--905} (\bibinfo {year} {1969})}\BibitemShut
  {NoStop}%
\bibitem [{\citenamefont {Whiteside}\ \emph {et~al.}(2011)\citenamefont
  {Whiteside}, \citenamefont {Xantheas},\ and\ \citenamefont
  {Gutowski}}]{Whiteside:2011}%
  \BibitemOpen
  \bibfield  {author} {\bibinfo {author} {\bibfnamefont {Alexander}\
  \bibnamefont {Whiteside}}, \bibinfo {author} {\bibfnamefont {Sotiris~S.}\
  \bibnamefont {Xantheas}}, \ and\ \bibinfo {author} {\bibfnamefont {Maciej}\
  \bibnamefont {Gutowski}},\ }\bibfield  {title} {\enquote {\bibinfo {title}
  {Is electronegativity a useful descriptor for the pseudo-alkali metal
  {NH}$_4$?}}\ }\href {\doibase 10.1002/chem.201101949} {\bibfield  {journal}
  {\bibinfo  {journal} {Chemistry ? A European Journal}\ }\textbf {\bibinfo
  {volume} {17}},\ \bibinfo {pages} {13197--13205} (\bibinfo {year}
  {2011})}\BibitemShut {NoStop}%
\bibitem [{\citenamefont {Pauling}(1932)}]{PaulingElectro}%
  \BibitemOpen
  \bibfield  {author} {\bibinfo {author} {\bibfnamefont {Linus}\ \bibnamefont
  {Pauling}},\ }\bibfield  {title} {\enquote {\bibinfo {title} {The nature of
  the chemical bond. {IV.} the energy of single bonds and the relative
  electronegativity of atoms},}\ }\href {\doibase 10.1021/ja01348a011}
  {\bibfield  {journal} {\bibinfo  {journal} {Journal of the American Chemical
  Society}\ }\textbf {\bibinfo {volume} {54}},\ \bibinfo {pages} {3570--3582}
  (\bibinfo {year} {1932})},\ \Eprint
  {http://arxiv.org/abs/http://dx.doi.org/10.1021/ja01348a011}
  {http://dx.doi.org/10.1021/ja01348a011} \BibitemShut {NoStop}%
\bibitem [{\citenamefont {Xiao}\ \emph {et~al.}(2013)\citenamefont {Xiao},
  \citenamefont {Sun}, \citenamefont {Ruzsinszky}, \citenamefont {Feng},
  \citenamefont {Haunschild}, \citenamefont {Scuseria},\ and\ \citenamefont
  {Perdew}}]{Xiao:Sun2013SiO}%
  \BibitemOpen
  \bibfield  {author} {\bibinfo {author} {\bibfnamefont {Bing}\ \bibnamefont
  {Xiao}}, \bibinfo {author} {\bibfnamefont {Jianwei}\ \bibnamefont {Sun}},
  \bibinfo {author} {\bibfnamefont {Adrienn}\ \bibnamefont {Ruzsinszky}},
  \bibinfo {author} {\bibfnamefont {Jing}\ \bibnamefont {Feng}}, \bibinfo
  {author} {\bibfnamefont {Robin}\ \bibnamefont {Haunschild}}, \bibinfo
  {author} {\bibfnamefont {Gustavo~E.}\ \bibnamefont {Scuseria}}, \ and\
  \bibinfo {author} {\bibfnamefont {John~P.}\ \bibnamefont {Perdew}},\
  }\bibfield  {title} {\enquote {\bibinfo {title} {Testing density functionals
  for structural phase transitions of solids under pressure: Si, sio${}_{2}$,
  and zr},}\ }\href {\doibase 10.1103/PhysRevB.88.184103} {\bibfield  {journal}
  {\bibinfo  {journal} {Phys. Rev. B}\ }\textbf {\bibinfo {volume} {88}},\
  \bibinfo {pages} {184103} (\bibinfo {year} {2013})}\BibitemShut {NoStop}%
\bibitem [{\citenamefont {Perdew}\ \emph
  {et~al.}(1996{\natexlab{c}})\citenamefont {Perdew}, \citenamefont {Burke},\
  and\ \citenamefont {Wang}}]{Perdew/Burke/Wang:1996}%
  \BibitemOpen
  \bibfield  {author} {\bibinfo {author} {\bibfnamefont {John~P.}\ \bibnamefont
  {Perdew}}, \bibinfo {author} {\bibfnamefont {Kieron}\ \bibnamefont {Burke}},
  \ and\ \bibinfo {author} {\bibfnamefont {Yue}\ \bibnamefont {Wang}},\
  }\bibfield  {title} {\enquote {\bibinfo {title} {Generalized gradient
  approximation for the exchange-correlation hole of a many-electron system},}\
  }\href {\doibase 10.1103/PhysRevB.54.16533} {\bibfield  {journal} {\bibinfo
  {journal} {Physical Review B}\ }\textbf {\bibinfo {volume} {54}},\ \bibinfo
  {pages} {16533--16539} (\bibinfo {year} {1996}{\natexlab{c}})}\BibitemShut
  {NoStop}%
\bibitem [{\citenamefont {Haas}\ \emph
  {et~al.}(2009{\natexlab{b}})\citenamefont {Haas}, \citenamefont {Tran},
  \citenamefont {Blaha}, \citenamefont {Schwarz},\ and\ \citenamefont
  {Laskowski}}]{CoreVal:Blaha:2009}%
  \BibitemOpen
  \bibfield  {author} {\bibinfo {author} {\bibfnamefont {Philipp}\ \bibnamefont
  {Haas}}, \bibinfo {author} {\bibfnamefont {Fabien}\ \bibnamefont {Tran}},
  \bibinfo {author} {\bibfnamefont {Peter}\ \bibnamefont {Blaha}}, \bibinfo
  {author} {\bibfnamefont {Karlheinz}\ \bibnamefont {Schwarz}}, \ and\ \bibinfo
  {author} {\bibfnamefont {Robert}\ \bibnamefont {Laskowski}},\ }\bibfield
  {title} {\enquote {\bibinfo {title} {Insight into the performance of {GGA}
  functionals for solid-state calculations},}\ }\href {\doibase
  10.1103/PhysRevB.80.195109} {\bibfield  {journal} {\bibinfo  {journal} {Phys.
  Rev. B}\ }\textbf {\bibinfo {volume} {80}},\ \bibinfo {pages} {195109}
  (\bibinfo {year} {2009}{\natexlab{b}})}\BibitemShut {NoStop}%
\bibitem [{\citenamefont {Csonka}\ \emph {et~al.}(2009)\citenamefont {Csonka},
  \citenamefont {Perdew}, \citenamefont {Ruzsinszky}, \citenamefont
  {Philipsen}, \citenamefont {Leb\`egue}, \citenamefont {Paier}, \citenamefont
  {Vydrov},\ and\ \citenamefont {\'Angy\'an}}]{PhysRevB.79.155107}%
  \BibitemOpen
  \bibfield  {author} {\bibinfo {author} {\bibfnamefont {G\'abor~I.}\
  \bibnamefont {Csonka}}, \bibinfo {author} {\bibfnamefont {John~P.}\
  \bibnamefont {Perdew}}, \bibinfo {author} {\bibfnamefont {Adrienn}\
  \bibnamefont {Ruzsinszky}}, \bibinfo {author} {\bibfnamefont {Pier H.~T.}\
  \bibnamefont {Philipsen}}, \bibinfo {author} {\bibfnamefont {S\'ebastien}\
  \bibnamefont {Leb\`egue}}, \bibinfo {author} {\bibfnamefont {Joachim}\
  \bibnamefont {Paier}}, \bibinfo {author} {\bibfnamefont {Oleg~A.}\
  \bibnamefont {Vydrov}}, \ and\ \bibinfo {author} {\bibfnamefont {J\'anos~G.}\
  \bibnamefont {\'Angy\'an}},\ }\bibfield  {title} {\enquote {\bibinfo {title}
  {Assessing the performance of recent density functionals for bulk solids},}\
  }\href {\doibase 10.1103/PhysRevB.79.155107} {\bibfield  {journal} {\bibinfo
  {journal} {Phys. Rev. B}\ }\textbf {\bibinfo {volume} {79}},\ \bibinfo
  {pages} {155107} (\bibinfo {year} {2009})}\BibitemShut {NoStop}%
\bibitem [{Note1()}]{Note1}%
  \BibitemOpen
  \bibinfo {note} {The exchange energy density, is $n(\protect \mathbf
  {r})\times E_x (n)$ in LDA or $n(\protect \mathbf {r})\times E_x (n)\times
  F_x(s,\alpha )$ in GGA or meta-GGA. $E_x(n)$ is the exchange energy per
  particle of LDA, which is a function of $r_s$.}\BibitemShut {Stop}%
\bibitem [{\citenamefont {Perdew}\ \emph {et~al.}(2009)\citenamefont {Perdew},
  \citenamefont {Ruzsinszky}, \citenamefont {Csonka}, \citenamefont
  {Constantin},\ and\ \citenamefont {Sun}}]{PhysRevLett.103.026403}%
  \BibitemOpen
  \bibfield  {author} {\bibinfo {author} {\bibfnamefont {John~P.}\ \bibnamefont
  {Perdew}}, \bibinfo {author} {\bibfnamefont {Adrienn}\ \bibnamefont
  {Ruzsinszky}}, \bibinfo {author} {\bibfnamefont {G\'abor~I.}\ \bibnamefont
  {Csonka}}, \bibinfo {author} {\bibfnamefont {Lucian~A.}\ \bibnamefont
  {Constantin}}, \ and\ \bibinfo {author} {\bibfnamefont {Jianwei}\
  \bibnamefont {Sun}},\ }\bibfield  {title} {\enquote {\bibinfo {title}
  {Workhorse semilocal density functional for condensed matter physics and
  quantum chemistry},}\ }\href {\doibase 10.1103/PhysRevLett.103.026403}
  {\bibfield  {journal} {\bibinfo  {journal} {Phys. Rev. Lett.}\ }\textbf
  {\bibinfo {volume} {103}},\ \bibinfo {pages} {026403} (\bibinfo {year}
  {2009})}\BibitemShut {NoStop}%
\end{thebibliography}%
 
\end{document}